\documentclass[rnote]{aa} 

\usepackage{tablefootnote}
\usepackage{graphicx}	
\usepackage{subfigure}
\usepackage{aalongtable}
\usepackage{amsmath}
\usepackage{natbib}
\usepackage{lscape}
\usepackage{txfonts}
\shortcites{Alibert05}
\shortcites{Allard01}
\shortcites{Baraffe03}
\shortcites{Bejar08}
\shortcites{Benedict06}
\shortcites{Bihain06}
\shortcites{Biller06}
\shortcites{Biller07}
\shortcites{Bouvier08}
\shortcites{Burrows97}
\shortcites{Chabrier00}
\shortcites{Chauvin04}
\shortcites{Chauvin05}
\shortcites{Close03}
\shortcites{Downes08}
\shortcites{Duchene07}
\shortcites{Epchtein99}
\shortcites{Guenther01}
\shortcites{Guenther05}
\shortcites{Hatzes00}
\shortcites{Itoh08}
\shortcites{Janson07}
\shortcites{Janson08}
\shortcites{Kalas08}
\shortcites{King03}
\shortcites{Koenig02}
\shortcites{Kraus08}
\shortcites{Lagrange08}
\shortcites{Leggett02}
\shortcites{Leggett08}
\shortcites{Lenzen98}
\shortcites{Lenzen03}
\shortcites{Liu08}
\shortcites{Lloyd06}
\shortcites{Machida08}
\shortcites{Marcy03}
\shortcites{Marengo06}
\shortcites{Marois08}
\shortcites{Martin00}
\shortcites{Masciadri05}
\shortcites{Montes01a}
\shortcites{Nakajima95}
\shortcites{Neuhaeuser00}
\shortcites{Neuhaeuser03}
\shortcites{Neuhaeuser05}
\shortcites{Neuhaeuser07}
\shortcites{Nielsen08}
\shortcites{Perryman97}
\shortcites{Persson98}
\shortcites{Potter02}
\shortcites{Raghavan06}
\shortcites{Rousset00}
\shortcites{Rousset03}
\shortcites{Sato07}
\shortcites{Schmidt08}
\shortcites{Skrutskie06}
\shortcites{York00}
\shortcites{Zapatero04}

\def \fdeg {\hbox{$.\!^{\circ}$}}
\def \farcs {\hbox{$.\!''$}}

\def \imgwidth {8cm}
\def \Ks {K_\mathrm{s}}
\def \Mjup {M_\mathrm{Jup}}
\def \Msun {M_\odot}

\begin{document}

   
   \title{A coronagraphic search for wide sub-stellar companions among members of the Ursa Major moving group\thanks{Based on observations collected at the European Southern Observatory, Chile, in programmes 72.C-0485, 73.C-0225, 76.C-0777, 77.C-0268, 384.C-0245A}}

   \author{M. Ammler - von Eiff\inst{1,2,3,4,5,6} \and A. Bedalov\inst{3,7} \and C. Kranhold\inst{2} \and M. Mugrauer\inst{3} \and T.O.B. Schmidt\inst{3,8} \and R. Neuh\"auser\inst{3} \and R. Errmann\inst{3,9}}
   \offprints{M. Ammler - von Eiff}

   \institute{Max-Planck-Institut f\"ur Sonnensystemforschung, Justus-von-Liebig-Weg 3, 37077 G\"ottingen, Germany, \email{ammler@mps.mpg.de}
  \and Th\"uringer Landessternwarte, Sternwarte 5, 07778 Tautenburg, Germany
  \and Astrophysikalisches Institut und Universit\"ats-Sternwarte Jena, Schillerg\"a{\ss}chen 2-3, 07745 Jena, Germany 
  \and Centro de Astronomia e Astrof\'{\i}sica da Universidade de Lisboa, Observat\'orio Astron\'omico de Lisboa, Tapada da Ajuda, 1349-018 Lisboa, Portugal
  \and Centro de Astrof\'{\i}sica da Universidade do Porto, Rua das Estrelas, 4150-762 Porto, Portugal 
  \and Georg-August-Universit\"at, Institut f\"ur Astrophysik. Friedrich-Hund-Platz 1, 37077 G\"ottingen, Germany
  \and Faculty of Natural Sciences, University of Split, Teslina 12. 21000 Split, Croatia
  \and Hamburger Sternwarte, Gojenbergsweg 112, 21029 Hamburg, Germany 
  \and Abbe Center of Photonics, Friedrich-Schiller-Universit\"at Jena, Max-Wien-Platz 1, 07743 Jena, Germany}
   \date{}

 
  \abstract
   {We present the results of a survey to detect low-mass companions of UMa group members, carried out in 2003-2006 with NACO at the ESO VLT. While many extra-solar planets and planetary candidates have been found in close orbits around stars by the radial velocity and the transit method, direct detections at wider orbits are rare. The Ursa Major (UMa) group, a young stellar association at an age of about 200-600\,Myr and an average distance of 25\,pc, has not yet been addressed as a whole although its members represent a very interesting sample to search for and characterize sub-stellar companions by direct imaging.}
   {Our goal was to find or to provide detection limits on wide sub-stellar companions around nearby UMa group members using high-resolution imaging. 
   }
   {We searched for faint companions around 20 UMa group members within 30\,pc. The primaries were placed below a semi-transparent coronagraph, a rather rarely used mode of NACO, to increase the dynamic range of the images. In most cases, second epoch images of companion candidates were taken to check whether they share common proper motion with the primary. 
   }
   {Our coronagraphic images rule out sub-stellar companions around the stars of the sample. A dynamical range of typically 13-15\,mag in the $\Ks$ band was achieved at separations beyond 3" from the star. Candidates as faint as $\Ks\approx20$ were securely identified and measured. The survey is most sensitive between separations of 100 and 200\,au but only on average because of the very different target distance. Field coverage reaches about 650\,au for the most distant targets. Most of the 200 candidates are visible in two epochs. All of those were rejected being distant background objects. 
   }
   {}

   \keywords{Stars: binaries: visual -- Stars: imaging -- Stars: brown dwarfs -- Galaxy: open clusters and associations: individual: UMa group -- Galaxy: solar neighborhood}

   \titlerunning{Multiplicity of UMa group members}
   \authorrunning{Ammler-von Eiff et al.}

   \maketitle
%

\section{Introduction}

By now almost 1,300 brown dwarfs have been found, classified with spectral types L, T, and Y \citep{Kirkpatrick08,2011ApJ...743...50C}\footnote{See http://www.DwarfArchives.org for a full account.}.
This number has been surpassed by indirect detections in the planetary regime -- by now more than 600 by radial velocity variations and more than 1,200 by the transit method.\footnote{See http://www.exoplanet.eu for further details and updates.}

While indirect methods are most sensitive to objects in very close orbits around the targets, direct imaging detects objects in wider orbits and thus is complementary. Ideally, direct imaging is combined with simultaneous measurement of the radial velocity variation \citep{Guenther05}.
Young sub-stellar objects still contract and are self-luminous \citep{Burrows97} so 
that their direct detection is less difficult up to ages of a few hundred Myr \citep{Malkov98,Neuhauser+2012}.
In contrast to high-resolution imaging in space \citep[e.g.][]{Marengo06}, ground-based observations need to encompass the seeing which can be done with adaptive optics \citep[e.g.][]{Neuhaeuser03,Duchene07}. Coronagraphy with intransparent (e.g. \citealp{McCarthy04}, \citealp{2005A&A...438L..29C}) or semi-transparent coronagraphs (e.g. \citealp{2010ApJ...720L..82B,2013A&A...551L..14B}, \citealp{Guenther05,Itoh06,Itoh08,Neuhaeuser07}, \citealp{2011ApJ...729..139W}) shade the bright star so that the exposure time and thus the sensitivity of the images can be increased. A number of very sophisticated techniques have been developed recently (see the review by \citealp{2014prpl.conf..715F}).

The frequency of low-mass companions to stars gives important clues regarding our understanding of the formation of brown dwarfs and planets \citep{Ida04,Alibert05,Broeg07}. At wide orbits, where direct imaging surveys are sensitive, the frequency of brown dwarfs is of the order of several percent for host stars with spectral types A-M and is less constrained for giant planets \citep{2013A&A...553A..60R,2015ApJS..216....7B}. In young nearby associations \citet{Neuhaeuser04} measure a frequency of $6\pm4\,\%$ of sub-stellar companions which is not very different from the value of $1\pm1\,\%$ obtained for isolated late-type stars \citep{McCarthy04}. The frequency of brown dwarfs around Hyades members does not turn out very different from the latter (\citealp{Guenther05,Bouvier08}, \citealp{2014MNRAS.445.3908L})
although the Hyades are still young with an age of $\approx$600\,Myr. 
In the younger Pleiades (125\,Myr), \citet{2013PASJ...65...90Y} confirmed two brown dwarf companions in a sample of 20 stars.


Young moving groups of an age intermediate between the Pleiades and the Hyades offer interesting opportunities to study homogeneous samples  of common age and origin. No sub-stellar companions have been found in the Her-Lyr assiociation (\citealp{Eisenbeiss+2007}, \citealp{2013ApJ...777..160B}) which has an age of $\approx$250\,Myr, similar to the UMa group \citep{2013A&A...556A..53E}. Although a systematic survey of the UMa group at high resolution has been missing, low-mass companions have been detected: GJ 569 Ba, Bb \citep{Martin00,Zapatero04}, HD 130948\,B \& C, \citep{Potter02}, and ${\chi}^1$ Ori B \citep{Koenig02}\footnote{It is worth noting that \citet{2007MNRAS.378L..24B} identified one T dwarf and three L dwarfs in the UMa group using the moving cluster method.}. The only known planet around a probable UMa group member, $\epsilon$\,Eri, was found by radial velocity variations \citep{Hatzes00,Benedict06}. Although there are still doubts about its existence \citep{2013A&A...552A..78Z}, there are even suspections of a second planet \citep{Quillen02,Deller05}. The planet(s) of $\epsilon$\,Eri have been subject to many, yet unsuccessful, attempts of direct detection (\citealp{Itoh06,Janson07,Marengo06,Neuhauser+2012}, and the present work).

\begin{figure}
\subfigure{\includegraphics[width=5cm]{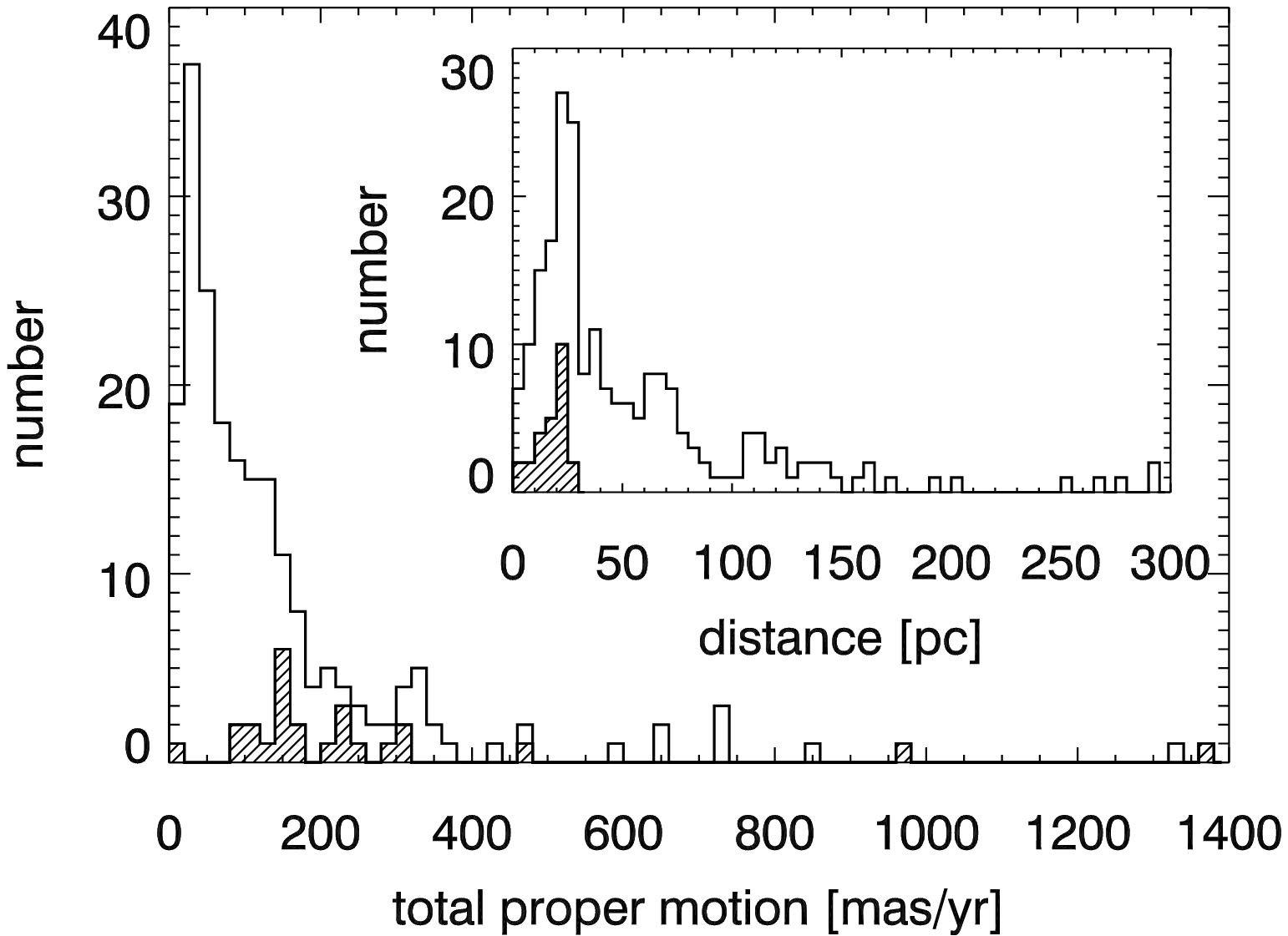}}
\subfigure{\includegraphics[width=3cm]{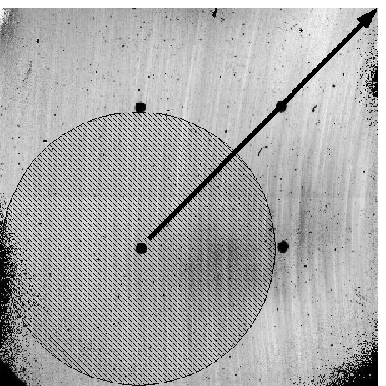}}
\caption{\label{fig:hist_coro} Left: distribution of distance and proper motion. The major concentration of the UMa group is the UMa open cluster in the Big Dipper constellation at distances of about $25\,$pc. All known members and candidates (see references in the text) are included in the histogram (solid line; distances mostly from \citealp{Montes01a} and \citealp{2007A&A...474..653V}). We add the distribution of the southern stars within 30\,pc addressed in the present work (hatched; distances from \citealp{2007A&A...474..653V}). Right: an image of the coronagraphic mask with the four coronagraphs (black filled circles) taken with the S27 camera. Each of the coronagraphs has an angular diameter of $0\farcs7$. We placed the star below the lower left (south-eastern) coronagraph, so that a field with a radius of $\approx$9" is completely covered outside the coronagraph (hatched circle). Incomplete coverage is achieved up to separations of $\approx$25" (arrow).}
\end{figure}

The goal of the present work is to find additional sub-stellar companions in the UMa group by direct imaging.
The search for close and faint companions in the UMa group benefits from a relatively young age of 200-600\,Myrs \footnote{The Ursa Major (UMa) has been studied extensively under various aspects \citep[e.g.][]{Eggen94,Montes01a,King03,Fuhrmann04,2006PhDT........21A} but the precise age and the list of members are still a matter of debate. Age estimates range from 200 to 600 Myrs \citep{Koenig02,King03,Fuhrmann04,King05,2015arXiv150104404B}.
}, the small distance of members ($\approx$50\,pc on average; Fig.~\ref{fig:hist_coro}), and thus the availability of precise Hipparcos astrometry \citep{Perryman97,2007A&A...474..653V}.
Furthermore, the proper motion of UMa group members is high on average ($\approx$50\,mas/yr) because of their proximity (Fig.~\ref{fig:hist_coro}) and their peculiar space motion \citep{2006PhDT........21A}. 
Hence, co-moving companions can be identified already after a short epoch difference. 
Preliminary results of the present study have been published earlier \citep{Ammler+2009}. 

\section{Coronagraphic observations and data reduction}
\label{sect:obs}

As the definition of the UMa group and the list of members is controversial, we compiled the targets in the following way to get a meaningful number of reliable UMa group members closer than $30\,$pc and observable with NACO at the ESO VLT:
\begin{itemize}
\item stars found by \citet{Montes01a} to fulfil at least one of Eggen's kinematic criteria \citep{Eggen95}. 
\item certain or probable members compiled by \citet{King03} based on photometric, kinematic, and spectroscopic criteria. 
\item HD\,135599 which is an UMa group member according to \citet{Fuhrmann04}. 
\end{itemize}

One system, GJ\,569 (=HIP\,72944), we did not observe since it has been extensively studied before (e.g. \citealp{Simon+2006}). Some objects were excluded because of known bright secondaries in the field of view (\object{HD\,24916}, \object{HD\,29875}, \object{HD\,98712}, and \object{HD\,134083)}. \object{HIP\,104383\,A} is a close binary, too, but fully fits below the coronagraph so that we did observe it (Fig.~\ref{fig:HIP104383}), i.e. 20 targets in total (Tables~\ref{tab:sample_mem} and \ref{tab:sample_all}).

The targets have been observed with NACO, the adaptive optics imager at the Nasmyth platform of the ESO VLT UT4 (Yepun)\footnote{NACO consists of the adaptive optics system NAOS \citep{Lenzen03} and the camera CONICA \citep{Rousset03}. The instrument was decommissioned in August 2013 and recommissioned on UT1 in January 2015.}. The stars themselves were used as adaptive optics reference stars for the visual wavefront sensor of NAOS (VIS). First epoch images were taken in 2003/2004 (programmes 072.C-0485, 073.C-0225) and second epoch imaging followed in 2005/2006 (programmes 076.C-777, 077.C-0268) for those stars with faint companion candidates identified. 

The semi-transparent coronagraph is a rarely used NACO mode. It has a diameter of $0\farcs7$ and dims the incoming stellar light by about 6 magnitudes in the $K$ band (see Fig.~\ref{fig:hist_coro}). Hence, it blocks most light of the star but still allows one to do precise astrometry with the stellar point spread function (PSF). 


\begin{figure}
\subfigure{\includegraphics[width=4.3cm]{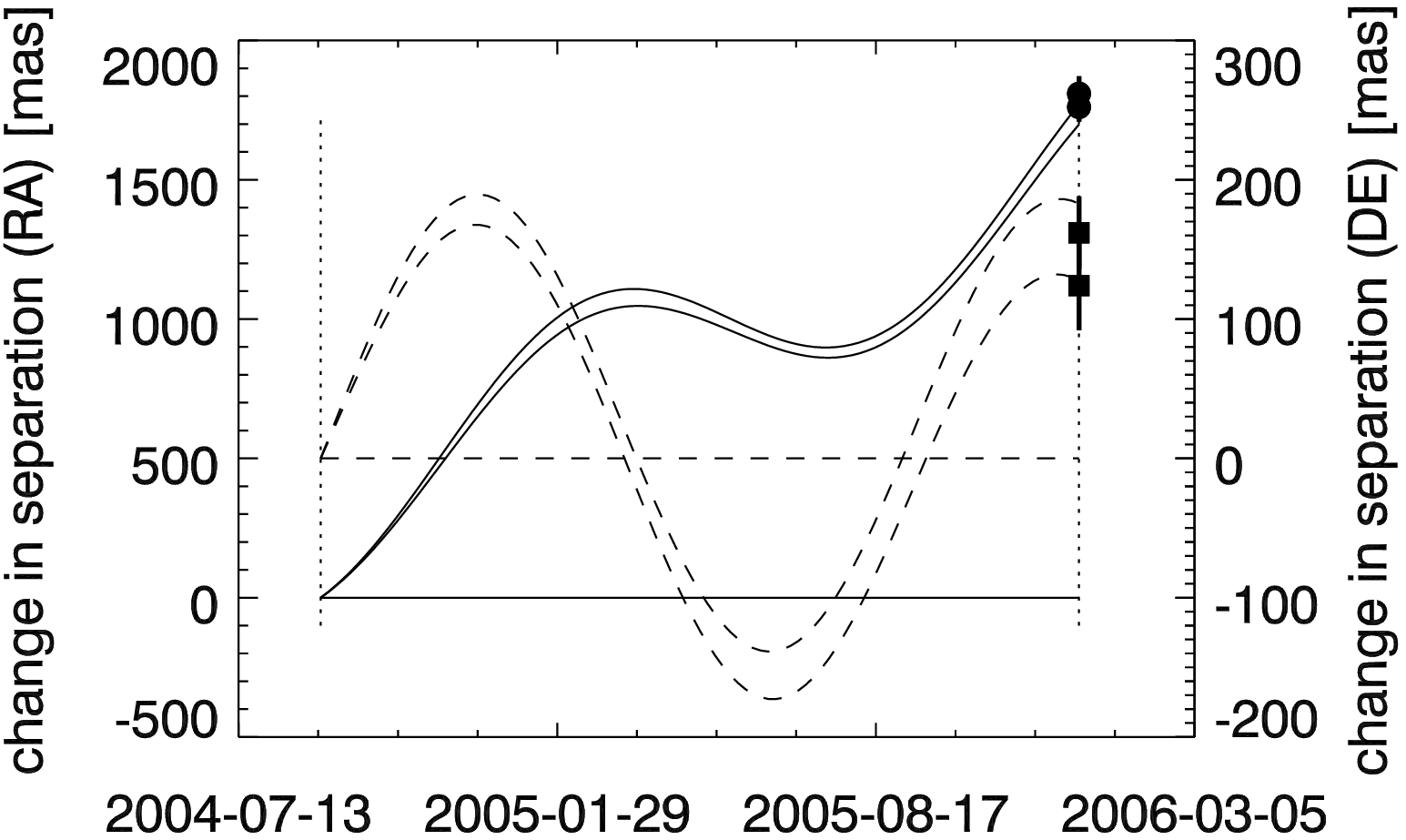}}
\subfigure{\includegraphics[width=4.3cm]{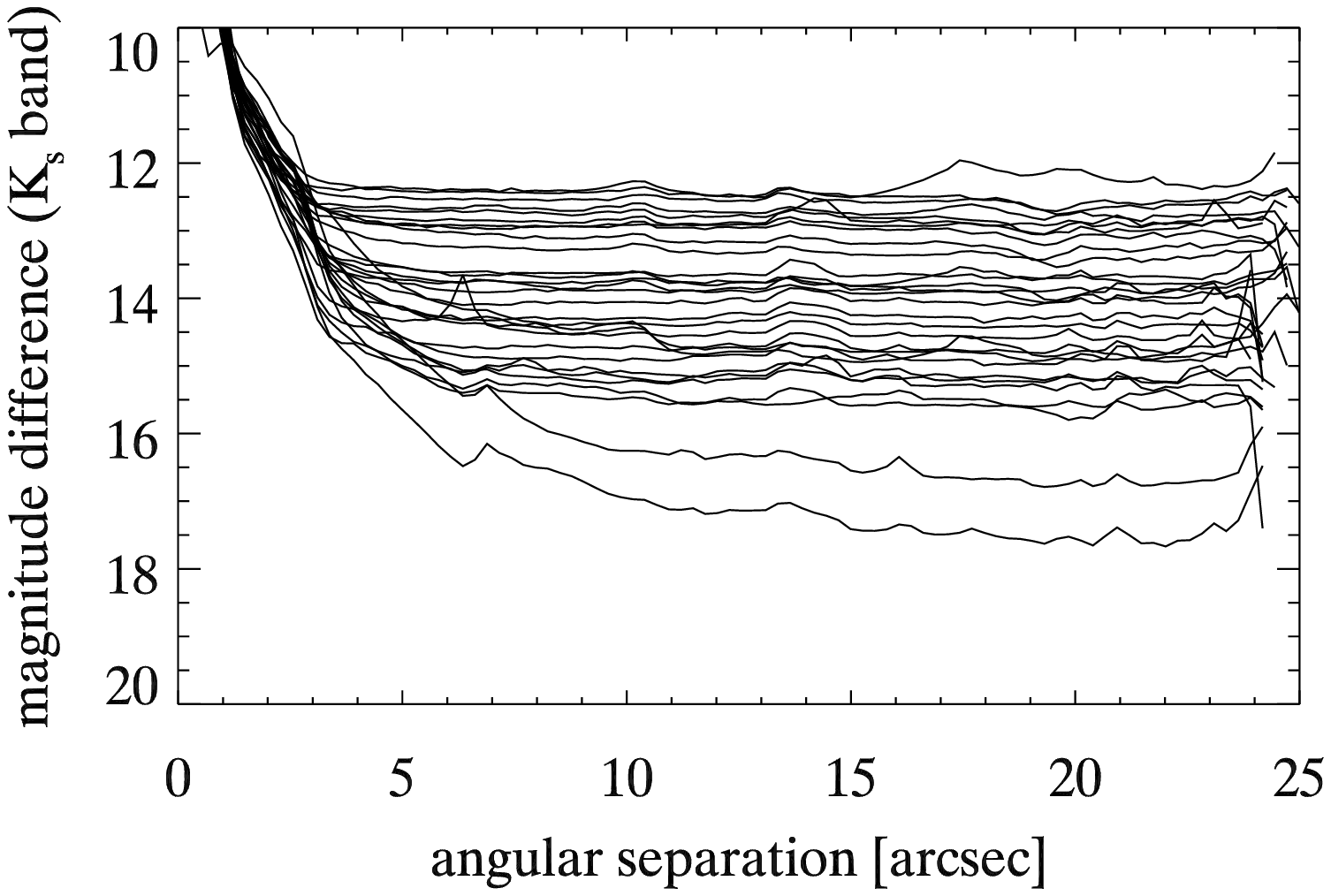}}
\caption{\label{fig:curves_radec} Left: based on Hipparcos data (cf. Tables~ \ref{tab:par_motion}, \ref{tab:sample_mem}, and \ref{tab:sample_all}), this example of HD\,22049 shows how the relative motion of non-moving background stars in right ascension (solid lines) and declination (dashed lines) will differ from a comoving companion (respective horizontal lines). The lines encompass the Hipparcos uncertainty, scaled by a factor of 100 for visibility. Possible orbital motion of a hypothetical companion has been neglected. The vertical dotted lines highlight the dates of the $1^\mathrm{st}$ and $2^\mathrm{nd}$ epoch exposures, respectively. Obviously, the two candidates found  are background objects (circles for right ascension and squares for declination, resp.; see Table~\ref{tab:all_cand_formatted} and Fig.~\ref{fig:images1}). Right: dynamic range curves for a $10\,\sigma$ detection as a function of angular separation for all exposures of the survey. }
\end{figure}

The observations have been obtained in the $\Ks$ band \citep{Persson98,Tokunaga02} further reducing the brightness difference of stars and any low-mass companions. 
 The sensitivity of the observations has been further improved by not using the S13 camera with the smallest pixel scale but instead observing with the S27 camera (27.15 mas/pix, FOV 28"x28"). This way, a larger field is covered (Fig.~\ref{fig:hist_coro}) and more light is collected in a single pixel. The astrometric precision using the S27 camera is sufficient since the average proper motion of the UMa group of 150\,mas (Fig.~\ref{fig:hist_coro}) per year corresponds to 5.5 pixels on the detector.

The data reduction follows the usual steps of sky subtraction, flat-field correction, bad pixel correction, shift and add\footnote{using own scripts and tools provided by ESO, including \textsl{ESO-eclipse} v5.0.0, the \textsl{jitter} recipe  \citep{1997Msngr..87...19D,2011ascl.soft12001D}, and \textsl{esorex} using the recipe \textsl{naco\_img\_twflat} to derive bad pixel frames.}. 
The sky has been subtracted using jittered exposures of nearby sky positions free of bright stars. 
The flat-field correction is based on coronagraphic night-time flat-field exposures to correct for the variable transmittance of the coronagraphic substrate and standard day-time exposures to correct for pixel-to-pixel variations of the detector below the coronagraph used.
Bad pixel frames have been taken from standard calibrations or derived individually with ESO tools and standard twilight flat-field exposures.

\section{Identification and characterisation of candidates}
\label{sect:cand}


The stellar PSF has been subtracted in a box of 500x500 pixels centred on the star to facilitate the visual identification of faint companion candidates. The PSF has been derived by calculating the average of a set of rotated frames (in steps of 2$^\circ$) and using sigma-clipping to get rid of bright features, in particular the diffraction spikes. A smoothed image has been subtracted to further facilitate detection (Figs~\ref{fig:images1}-\ref{fig:images5})\footnote{Smoothing has been done by the application of a Gaussian filter ({\it FILTER/GAUSS} in ESO MIDAS version 13SEP; \citealp{Banse83,1992ASPC...25..115W}) using all adjacent pixels within a radius of 18 pixels weighted by a Gaussian with a width of $\sigma=3\,$pixels.}.

We noticed that artefacts due to dust particles on the coronagraphic substrate can be mistaken for companion candidates. The flat-field correction fails in removing those since occasional displacements of the coronagraphic substrate can occur between calibration and target exposures. We removed those artefacts from the list of companion candidates guided by a visual cross-match of the scientific exposures with the flat-field exposures. In the most extreme cases (exposures of HD\,11171 and HD\,22049), several tens of artefacts had to be removed. 

A wide comoving companion will not change its position relative to the star while a distant and thus non-moving background object will reflect the stellar parallactic and proper motion (Fig.~\ref{fig:curves_radec}). Our assessment of astrometric measurement uncertainties is based on \citet{Chauvin+2010} who give a long-term average of the S27 pixel scale of $27.012\pm0.004$\,mas/pix and of the detector position angle (true North) of $-0\fdeg04\pm0\fdeg14$. In the present work, we corrected the position angles of all candidates for the mean deviation of $-0\fdeg04$ from the true North. From the measurements obtained by \citet{Chauvin+2010}, we expect uncertainties of 0.016\,mas/pix on separation and $0\fdeg15$ on position angle. We neglected the errors in fitting Gaussian profiles to measure the positions of the stars and the candidates. It is as precise as a few milliarcseconds and of the order of the {\it Hipparcos} errors (Table~\ref{tab:par_motion}).

The instrumental magnitudes of the target stars and the candidates have all been measured in apertures of 13 pixels across which is more than double the typical FWHM of the PSF. Then, the $\Ks$ band magnitude of the candidates has been derived from the measured flux ratios taking into account the brightness of the target star (Table~\ref{tab:sample_all}) and the attenuation of the coronagraph.  While the brightness of the star has been measured in the reduced frames, the brightness of the candidates has been measured in the PSF-subtracted frames constructed in the way described above.

To calibrate the transmission of the coronagraph, a standard star (\object{HD\,1274}) was taken during the programme 384.C-0245A, once outside and once below the coronagraph, and reduced in the same way as the exposures of the science targets. The flux values have been integrated within apertures of 19\,pixels radius which fits well inside the coronagraph. Relating the measurements inside and outside the coronagraph gives a transmission of $0.47\pm0.03$\,\% in the $\Ks$ band which translates to a dimming of $5.83\pm0.06$\,mag. The error bar accounts for uncertainties involved in the aperture photometry but not for other systematics like cross-talk, light-leaks, temporal variation of the PSF, improper/variable placement of the star below the coronagraph, or variable residual absorption by the transparent substrate which carries the coronagraphs. The new measurement gives a dimming weaker by half a magnitude than the value given in the NACO manual ($\Ks=6.3\pm0.1\,$mag). The latter measurement was done in a way similar to the present work but using frames reduced in a different way and using peak counts instead of aperture photometry (ESO, priv. comm.). A conservative estimate of $6.1\pm0.3$ has been adopted in the present work since the discrepancy cannot be explained with the data available.

For each one of the companion candidates, the average magnitude of two epochs  has been adopted and the uncertainty is given by half their difference plus the square-added error bar of 0.3\,mag of the transmissivity of the coronagraph. For single-epoch images, the transmissivity is the only well-known dominant source of error.

\section{Assessment of field coverage and detection limits}
\label{sect:coverage}
We could not take advantage of the full field of view of the NACO mode applied (28''x28'') since the corners are affected by obstructions. Furthermore, the coronagraphs themselves are insensitive patches in the field of view.

The field of view is covered completely between angular separations of $0\farcs35$ (coronagraphic radius) and $9\farcs0$ which correspond to different linear scales depending on the distance of the star. The parallaxes of the sample stars vary from 37\,mas (HIP\,104383) to 311\,mas (HD\,22049) so that the field of view covers very different parts of the stellar environments, e.g. the closest view in the case of $\epsilon$\,Eri (=HD\,22049; $1.1-29$\,au) and the farthest view for HIP\,104383 ($9.2-237$\,AU) (Table~\ref{tab:fov}). The expected widest bound orbit of low-mass companions has been estimated based on the linear law given in \citealp{Close+2003} given the central mass of the target systems.

Although the field of view is complete only up to separations of $\sim$9'', it samples separations of up to $\sim$25'' corresponding to as much as $\sim$650\,au in the case of HD\,125451 and HIP\,104383. Still the field of view never reaches the widest possible bound orbit in the present sample. We note, however, that inclination is a free parameter in this whole consideration and that measured separations are projected separations. Stars with exoplanets are known for which very wide (up to one third of a parsec) companions with common proper motion have been detected \citep{2014MNRAS.439.1063M}.

To understand the detection limits in the wings of the PSF, we built noise maps from the reduced and PSF-subtracted images. We measured the standard deviation at each pixel using 7x7 adjacent pixels. This value has been multiplied by the square root of the number of pixels in an aperture of 13 pixels across in order to compare to the photometric flux measurements described above.

Based on the noise analysis, the limiting magnitudes have been derived relative to the magnitude of the primary star  in the same way as by \citet{2006ApJ...652.1572B}. We found that a source can be detected at a given location if the flux measured is larger than a detection limit of $5-10\,\sigma$ of the local noise level and adopted the conservative and common value of 10$\sigma$. The limits derived have been corrected for the coronagraphic attenuation of $6.1$mag inferred above. For each exposure, the right panel of Fig.~\ref{fig:curves_radec} shows the average value of pixels at same angular distance.

When it comes to the determination of the mass of a companion that is still detectable, uncertain age is the largest contribution to the error budget. Using the stellar $\Ks$ band magnitude and the measured dynamic range, we calculated the $\Ks$ band magnitude of faint objects which could still have been detected at a given separation.
The magnitudes have been interpolated in evolutionary models to obtain a mass estimate. We used the COND03 models \citep{Baraffe03} for effective temperatures lower than 1,300\,K\footnote{According to \citet{Baraffe03}, the COND03 models are more appropriate below 1300\,K to predict infrared colours, for methane dwarfs and extrasolar giant planets at large orbital separation. The difference between the $\Ks$ and the $K$ band has been neglected in the present study.} and DUSTY00 \citep{Chabrier00} models for temperatures higher than 1,300\,K. For ages of 100\,Myr, 500\,Myr, and 1\,Gyr, this temperature corresponds to a mass of 10, 31, and 42 \,$\Mjup$, respectively, and is accessible to the present survey. In this range of mass, the difference in $K$ band magnitude between 100\,Myr and 1\,Gyr varies from 4 at the high-mass end to 8 at the low-mass end. Even if we avoid interpolating the true age range of the UMa group in the evolutionary models, we can assess that the age uncertainty of the UMa group implies an error of several $K$ band magnitudes which is the dominant source of error in the present study.

\section{Results and Discussion}
\label{sect:results}

The noise level is observed to decrease strongly with increasing separation from the primary star (Fig.~\ref{fig:curves_radec}). The longest on-target exposure time was spent on the brightest target, HD\,22049, resulting in the highest dynamic range of the present work ($\gtrsim$17\,mag for a $10\,\sigma$ detection). Typically, the dynamic range for a $10\,\sigma$ detection limit ranges between $\Delta\Ks=13$ and 15 at a separation beyond 3" (Figs.~\ref{fig:hist_coro}, \ref{fig:candidates_2epochs1}, \ref{fig:candidates_2epochs2}, and \ref{fig:more_detlimits}). Given the $K$ band magnitude of the central star, this corresponds to about $12-20\,\Mjup$ provided an age of 500\,Myr (Figs.~\ref{fig:mass_of_sep1} and \ref{fig:mass_of_sep2}). 
Objects with 12\,$M_\mathrm{Jup}$ and younger than 1\,Gyr could have been detected  at separations of less than $10\,$au in the case of HIP\,57548 with the deepest exposure of the present work, and objects with 20\,$M_\mathrm{Jup}$ closer than $3\,$au (Table~\ref{tab:det_limits2}). 

More than 200 candidates have been identified (Figs.~\ref{fig:images1}-\ref{fig:images5}, Table~\ref{tab:all_cand_formatted}). While in some fields, not a single object has been detected around the star (\object{HD\,11171}, \object{HD\,26923}, \object{HD\,38393}, \object{HD\,63433}, \object{HIP\,57548}, \object{HD\,95650}, \object{HD\,125451}, \object{HD\,139006}, and \object{HD\,217813}), some 160 have been found around \object{HD\,165185} which is located in the galactic plane. A number of additional candidates have been found close to \object{HD\,11131}, \object{HD\,22049}, \object{HD\,26913}, \object{HD\,41593}, \object{HD\,60491}, \object{HD\,61606\,A}, \object{HD\,135599}, \object{HD\,147584}, \object{HD\,175742}, and \object{HIP\,104383\,A}\footnote{
HIP\,104383\,A is a special case since it appears as a visual binary below the coronagraph (Fig.~\ref{fig:HIP104383}). According to \citet{Balega04}, it is a binary with a separation of 0\farcs3 and a magnitude difference of 1.67 in the $R^\prime$ band. In the $\Ks$ band we measured a magnitude difference of 0.45. The astrometric measurements presented here have been done w.r.t. the brighter component while the co-added signal has been used for photometric measurements.
}.

\begin{figure}
\center	
\subfigure{\includegraphics[width=3.5cm]{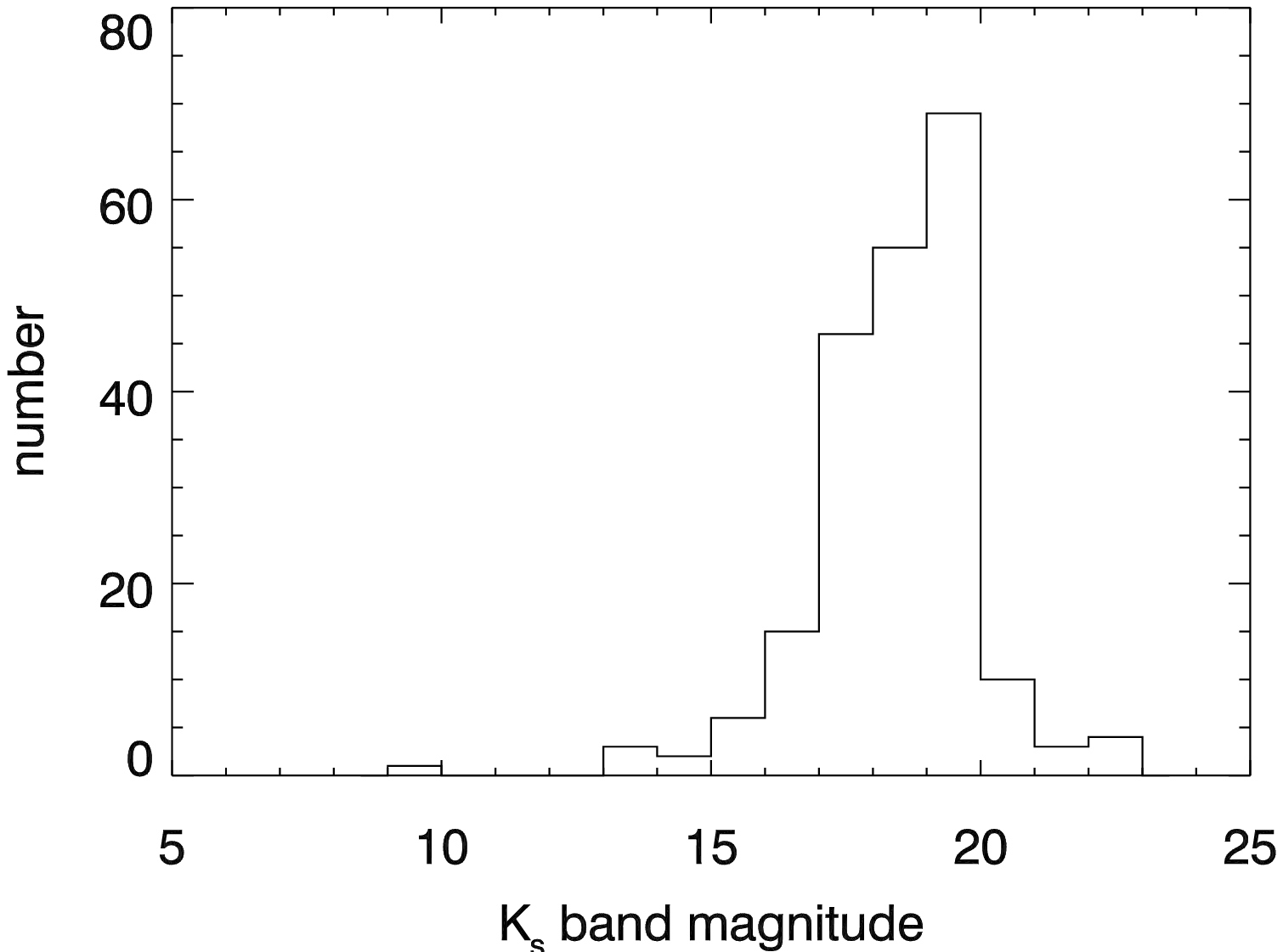}}
\subfigure{\includegraphics[width=4.5cm]{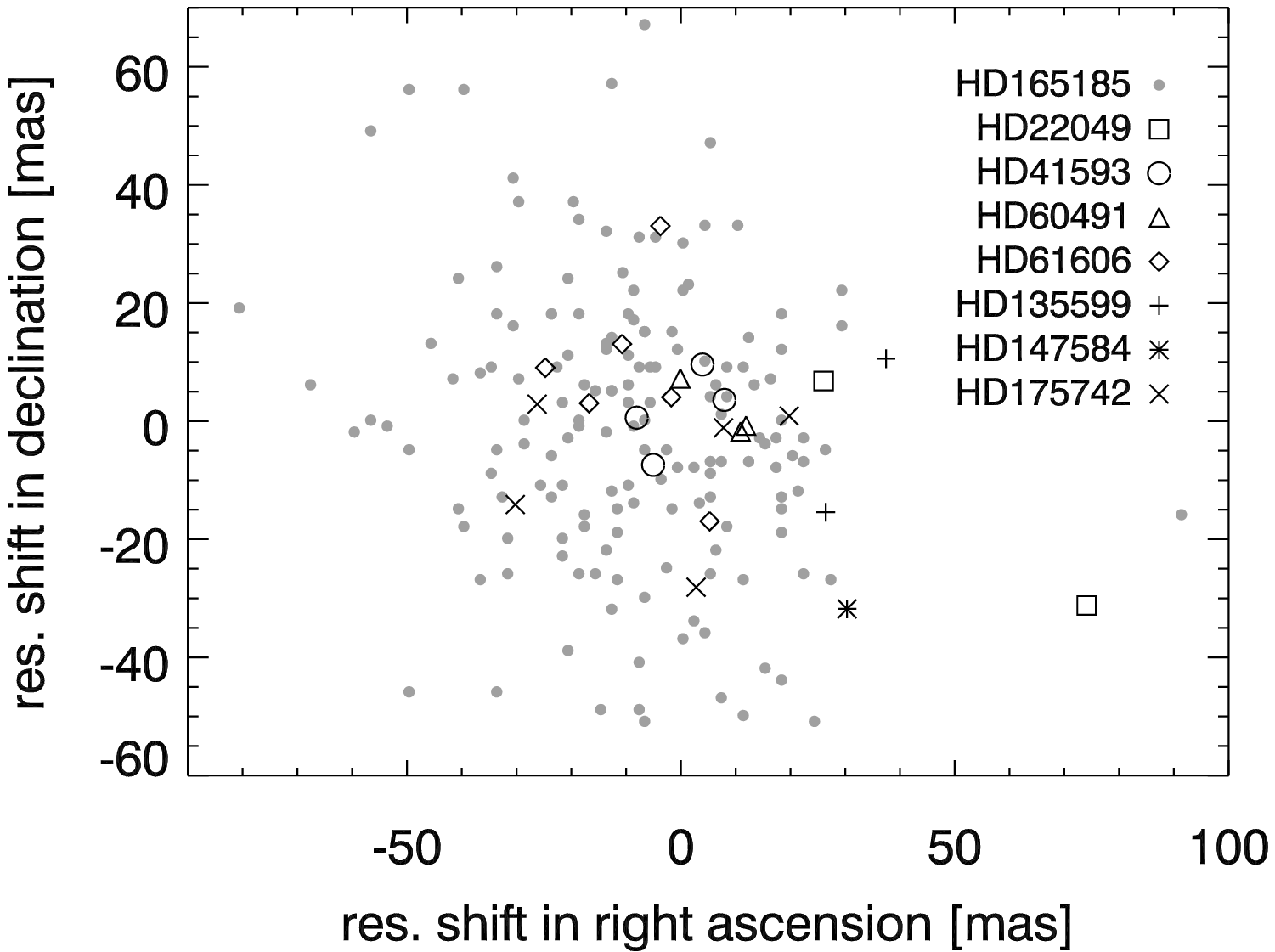}}
\caption{\label{fig:all_cand}Left: $\Ks$ magnitude distribution of candidates. Based on Table~\ref{tab:all_cand_formatted}. Errors in magnitude have not been considered. Right: residual motion of candidates. For all candidates detected in two epochs, the figure shows the residual shifts in right ascension and declination over the time elapsed between the epochs (Table~\ref{tab:all_cand_formatted}) after correcting for the proper and parallactic motion of the central stars (Table~\ref{tab:par_motion}). The symbols discern the candidates according to their central stars. Error bars have been omitted for clarity.}
\end{figure}

The relative shift of a companion w.r.t. the star has been measured when a second epoch was available (Table~\ref{tab:all_cand_formatted}). All those cases are non-moving background stars when comparing to the predicted shift due to the stellar parallactic and proper motion (Table~\ref{tab:par_motion} and Fig.~\ref{fig:all_cand})\footnote{The parallactic motion has been calculated assuming a value of the obliquity of the ecliptic plane of $23\fdeg4$ and expressing the solar longitude by $L=279\fdeg697+36,000\fdeg770\,T$ with the time $T$ given by Julian centuries since 1900, January 0, 12\,h \citep{1995moas.book.....K}. The eccentricity of Earth's orbit has been neglected.}. The deviation from these expectations is less than 0\farcs1 in all cases and of the order of the astrometric error bars (Table~\ref{tab:all_cand_formatted}). This shows that the true astrometric uncertainties agree with the assessments by \citet{Chauvin+2010} implemented here (Sect.~\ref{sect:cand}).

The detection limits can be compared to the brightness of faint background objects detected in the field of view. The uncertainties of the photometric measurements are below 1\,mag in almost all cases. Usually, they are close to 0.5\,mag and increasing towards the detection limit while the photometric measurements deviate by less than a magnitude from epoch to epoch. Enhanced error bars can be ascribed to smearing, noise, flux missed by the aperture photometry at the edges of the frame, or background features like diffraction spikes, ghosts, and reflections (indicated in Table~\ref{tab:all_cand_formatted}).

The faintest object identified (HD 165185, ID0101) advocates an empirical detection limit of $\Ks$$\approx$22 for HD\,165185. However, the aperture photometry of this object is affected by insufficient background correction so that it must be somewhat brighter (Fig.~\ref{fig:images3}). Moreover, the lower envelope to the distribution of most candidates (Fig.~\ref{fig:candidates_2epochs2}) is a bit higher as is also indicated by the brightness distribution of all candidates (cf. Fig.~\ref{fig:all_cand}). It covers magnitudes in the range $\Ks=13-23$  with a single brighter object next to HD\,11131. As can be expected for a distribution of field stars, there are a few bright objects and a larger number of faint objects. The number of objects decreases rapidly at a brightness level fainter than $\Ks=20$ indicating that the census is complete down to this value. Since the distribution is almost entirely constituted of objects next to HD\,165185, this completeness limit is certainly valid for this subset.

\begin{table}
\centering
\caption{\label{tab:literature}Identification of candidates with previous detections.}
\begin{tabular}{lrrrl}
\hline	
\hline
target/cand.&\multicolumn{2}{l}{epoch/position}&\multicolumn{2}{l}{reference/magnitude/ID}\\
\hline
HD 22049&\multicolumn{2}{l}{2002-08-20}&\multicolumn{2}{l}{(1)}\\
ID0001&$4\farcs5$&$17\farcs0$&$K'=17.3$&2\\
ID0002&$-9\farcs6$&$14\farcs2$&$K'=17.3$&1\\
\hline
HD 41593&\multicolumn{2}{l}{2002-2004}&\multicolumn{2}{l}{(2)}\\
ID0002&$11\farcs10$&$328\fdeg40$&$H=15.88$&cc1\\
\hline
HD 175742&\multicolumn{2}{l}{2004-06-28}&\multicolumn{2}{l}{(3)}\\
ID0001&$9\farcs45$&$308\fdeg5$&$\Ks=16.99\pm0.09$&3\\
ID0003&$7\farcs57$&$335\fdeg5$&$\Ks=19.13\pm0.23$&4\\
ID0004&$2\farcs64$&$89^\circ$&$\Ks=16.88\pm0.09$&1\\
ID0006&$9\farcs36$&$199^\circ$&$\Ks=17.34\pm0.09$&2\\
\hline
HD 175742&\multicolumn{2}{l}{2011-05-23}&\multicolumn{2}{l}{(4)}\\
ID0004&$1\farcs72$&$1\farcs97$&...&...\\
\hline
\end{tabular}\\
\tablefoot{For each target with previous imaging, the candidates recovered are listed and the corresponding previous astrometric and photometric measurements are given together with the reference ID assigned in the previous work. Relative positions in right ascension and declination are given for \citet{Macintosh+2003} (1) and \citet{2013ApJ...773...73J} (4) while total separation and position angle are given for \citet{2008PASJ...60..209I} (2) and \citet{2009ApJS..181...62M} (3).}
\nocite{Macintosh+2003}
\nocite{2009ApJS..181...62M}
\nocite{2008PASJ...60..209I}
\nocite{2013ApJ...773...73J}
\end{table}

We compared our candidates with previous work for targets in common with other surveys which have not necessarily been dedicated to the UMa group \citep{Macintosh+2003,2008PASJ...60..209I,2009ApJS..181...62M}.We recovered all of the previously known candidates (Table~\ref{tab:literature}) at the positions expected as far as they are covered by our field of view.   Photometric measurements agree well within the error bars as long as they have been measured in the same or a similar photometric band. The single exception is ID 4 next to HD\,175742 \citep[][object 1]{2009ApJS..181...62M}. Most certainly, the difference of one $\Ks$ magnitude can be explained by the location of ID 4 in the PSF of HD\,175742 where the photometric measurement is very sensitive to imperfections of the PSF subtraction. The same object was found by \citet{2013ApJ...773...73J} but no photometric measurement is given there. The southern component of HD\,165185 listed by \citet{2014yCat....102026M} is out of the field of view of the present work but displayed in Table~\ref{tab:fov}.

\begin{figure}
\center	
\includegraphics[width=6cm]{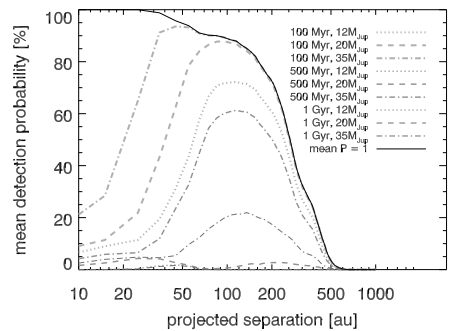}
\caption{\label{fig:prob_freq_sep} Mean survey detection probability for companions of 12, 20, and 35\,$\Mjup$ as a function of separation and for different age of 100\,Myr, 500\,Myr, and 1\,Gyr as indicated in the legend. The solid line gives the limiting case of a detection probability of 1 in each frame.}
\end{figure}

The deepest exposure has been chosen for each target to derive a survey upper limit on the frequency of stars with a companion of given mass and age. This has been done in a basic way similar to \citet{2007ApJ...670.1367L} without assuming any prior knowledge on the mass-period distribution of companions.

For this purpose, maps of 10\,$\sigma$ detection probability have been derived from the maps of limiting magnitude presented in Sect.~\ref{sect:coverage} by comparing them to the signal of a 12, 20, and 35$\,\Mjup$ companion at different ages of 100\,Myr, 500\,Myr, and 1\,Gyr (using the evolutionary models described in Sect.~\ref{sect:coverage})\footnote{The probability has been computed that a signal exceeds the 10\,$\sigma$ detection threshold in the presence of noise, assuming a normal distribution.}. The mean survey detection probability has been assessed by centering, rescaling, and averaging those probability maps (Fig.~\ref{fig:prob_freq_sep}). In the limiting case, where the detection probability equals 1 throughout each frame, the average detection probability is only limited by the respective field of view and approaches zero at separations beyond $\sim$500\,au which are not covered by any exposure of the survey. The age range considered has a stronger effect than the mass range. On average, the survey is most sensitive at separations between 100 and 200\,au since the distance of the targets and thus the field of view covered is very different.

We derived the upper limit on the frequency of stars with a companion where the mean $10\,\sigma$ detection probability is high. We followed the Bayesian approach described by \citet{2007ApJ...670.1367L} using the Poisson approximation and a confidence level of $95\,\%$. In the most sensitive range between 100 and 200\,au, we can place an upper limit of $\sim$25\,\% on the frequency of UMa group members with low-mass companions more massive than 35\,$\Mjup$ if the age of the UMa group is close to 500\,Myr or younger. The limit is never below 15\,\%, defined by the number of stars, the confidence level, and the case of a detection probability of one everywhere (solid line in Fig.~\ref{fig:prob_freq_sep}).

When interpreting the outcome of the present study one has to be aware of specific formation environments represented by the significant fraction of targets with stellar companions (cf. Tables~\ref{tab:sample_all} and \ref{tab:fov}). In addition we note that the known companions of GJ\,569 have not been considered for the frequency estimate since they are closer to the star than the field covered by the present work and we did not account for the companions of HD\,130948 since they have masses close to the Deuterium burning limit of $12-13\,M_\mathrm{Jup}$ which we are not sensitive to on average. Although the results are not at variance with previous work, it is obvious that the study of the UMa group would benefit from a larger sample size. So far, northern UMa group members have not been studied systematically. Those actually comprise the largest part of the UMa group. Although its densest part on the northern hemisphere is within 30\,pc, the group extends far beyond.


\begin{acknowledgements}
M.A. thanks Eike W. Guenther for fruitful discussions. We thank the anonymous referee for the constructive comments. AB acknowledges support from DFG in grants NE 515/13-1 and 13-2. M.A. was supported by a graduate scholarship of the Cusanuswerk, one of the national student elite programs of Germany, and an individual fellowship granted by the Funda\c{c}\~ao para a Ci\^{e}ncia e a Tecnologia (FCT), Portugal (reference SFRH/BPD/26817/2006). M.A. acknowledges research funding granted by the Deutsche Forschungsgemeinschaft (DFG) under the project RE 1664/4-1. M.A. further acknowledges support by DLR under the projects 50OW0204 and 50OO1501. RN acknowledges general support from the German National Science Foundation (Deutsche Forschungsgemeinschaft, DFG)  in grants NE 515/13-1, 13-2, and 23-1. TOBS would like to thank Evangelisches Studienwerk e. V. Villigst, the state of Thuringia as well as DFG for support in program NE 515/30-1. RN and RE would like to thank DFG for support in the Priority Programme SPP 1385 on the {\em First ten Million years of the Solar System} in project NE 515 / 34-1 and 34-2. Use was made of the CDS services SIMBAD, Vizier, and the NASA/ADS abstract service. This research has made use of the Washington Double Star Catalog maintained at the U.S. Naval Observatory. This publication makes use of data products from the Two Micron All Sky Survey, which is a joint project of the University of Massachusetts and the Infrared Processing and Analysis Center/California Institute of Technology, funded by the National Aeronautics and Space Administration and the National Science Foundation.
\end{acknowledgements}

\bibliographystyle{aa}
\bibliography{./non_detect}


\Online

\begin{appendix}

\section{Sample}
\label{app:sample}

\begin{table*}
\centering
\caption{\label{tab:sample_mem} The sample of UMa group members.}
\begin{tabular}{lcclllcrrr}
\hline
\hline
name&\multicolumn{2}{c}{\citeauthor{Montes01a}}&\multicolumn{3}{c}{\citet{King03}}&\citeauthor{Fuhrmann04}&\multicolumn{2}{c}{proper motion [mas/yr]}&number of\\
&\multicolumn{2}{c}{\citeyearpar{Montes01a}}&&&&\citeyearpar{Fuhrmann04}&&\\
&$V_\mathrm{pec}$&$\rho_\mathrm{C}$&kin.&phot.&final&&$\mu_{\alpha}\cos\delta$&$\mu_{\delta}$&NACO epochs \\
\hline
HD 11131            & Y & Y &?   & Y?&Y?  & Y & -124.54$\,\pm\,$  3.03& -105.82$\,\pm\,$  2.93&1\\                      
HD 11171            & Y & Y &?/Y?& Y?&Y?  & Y & -148.11$\,\pm\,$  0.25&  -93.43$\,\pm\,$  0.24&2\\                      
HD 22049            &...&...& ...&...& ...& Y & -975.17$\,\pm\,$  0.21&   19.49$\,\pm\,$  0.20&2\\                      
HD 26913            & N & Y &?   & ? &?   & Y & -102.64$\,\pm\,$  0.66& -113.30$\,\pm\,$  0.59&2\\                      
HD 26923            & Y & Y &Y?  & Y &Y?  & Y & -109.46$\,\pm\,$  0.48& -108.25$\,\pm\,$  0.43&1\\                      
HD 38393            & Y & Y &?/Y?& Y?&Y?  &...& -291.67$\,\pm\,$  0.14& -368.97$\,\pm\,$  0.15&1\\                      
HD 41593            & Y & Y &N?/?& Y &N?/?& Y & -120.46$\,\pm\,$  0.71& -103.21$\,\pm\,$  0.43&2\\                      
HD 60491            & N & Y &N?/?& Y?&N?/?&...&  -81.17$\,\pm\,$  1.26&  -42.66$\,\pm\,$  0.66&2\\                      
HD 61606            & N & Y &N?  & Y?&N?  &...&   69.90$\,\pm\,$  0.71& -278.33$\,\pm\,$  0.31&2\\                      
HD 63433            & Y & N &Y?  & ? &?   & Y &   -8.31$\,\pm\,$  0.65&  -10.48$\,\pm\,$  0.46&1\\                      
HD 95650            & N & Y &Y   & Y?&Y   &...&  142.30$\,\pm\,$  1.16&  -51.69$\,\pm\,$  0.79&1\\                      
HIP 57548           & N & Y &N?  & N?&N?  &...&  605.26$\,\pm\,$  2.32&-1219.28$\,\pm\,$  1.97&1\\                      
HD 125451           & N & Y & ...&...& ...&...&  105.95$\,\pm\,$  0.23&  -31.80$\,\pm\,$  0.21&1\\                      
HD 135599           &...&...&?   & Y &?   & Y &  178.35$\,\pm\,$  0.66& -137.52$\,\pm\,$  0.62&2\\                      
HD 139006           &...&...&Y   & Y &Y   & Y &  120.27$\,\pm\,$  0.19&  -89.58$\,\pm\,$  0.20&2\\                      
HD 147584           &...&...&Y   & Y &Y   &...&  199.97$\,\pm\,$  0.25&  110.97$\,\pm\,$  0.43&2\\                      
HD 165185           & N & Y &Y   & Y?&Y   &...&  105.05$\,\pm\,$  0.60&    7.95$\,\pm\,$  0.32&2\\                      
HD 175742           &...&...&N?  & Y?&?   &...&  131.31$\,\pm\,$  0.49& -283.72$\,\pm\,$  0.63&2\\                      
HIP 104383          & N & Y & ...&...& ...&...&  -77.56$\,\pm\,$  2.42&  -32.89$\,\pm\,$  0.87&2\\                      
HD 217813           & N & Y &?   & ? &?   & Y & -117.70$\,\pm\,$  0.59&  -27.66$\,\pm\,$  0.49&1\\                      

\hline
\end{tabular}
\tablefoot{UMa membership information, proper motion \citep{2007A&A...474..653V}, and number of epochs of NACO observations obtained for the present work. We reiterate the assignments by \citet{Montes01a} and \citet{King03} concerning the match of Eggen's criteria or, respectively, the decision on kinematic, photometric, and final membership ('Y' = yes; 'N' = no). Uncertainty is expressed by '?' and we refer the reader to \citet{King03} for more information. Some of the stars were not observed (zero NACO epochs taken) as is explained in the text and omitted here.}
\end{table*}

\begin{table*}
\centering
\caption{\label{tab:sample_all} Basic stellar data of the sample of UMa group members.} 
\begin{tabular}{lrrrrlrr}
\hline
object&HIP number&$\alpha$ (2000.0)&$\delta$ (2000.0)&parallax [mas]&spec. type&$V$ mag.&$Ks$ mag.\\
\hline
HD 11131  &  8486&01 49 23.36&-10 42 12.8&  44.32$\,\pm\,$ 3.02&G1V     & 6.727& 5.149\\                                
HD 11171  &  8497&01 49 35.10&-10 41 11.1&  43.13$\,\pm\,$ 0.26&F3III   & 4.664& 3.872\\                                
HD 22049  & 16537&03 32 55.84&-09 27 29.7& 310.94$\,\pm\,$ 0.16&K2V     & 3.730& 1.776\\                                
HD 26913  & 19855&04 15 25.79&+06 11 58.7&  47.49$\,\pm\,$ 0.68&G5IV    & 6.960& 5.271\\                                
HD 26923  & 19859&04 15 28.80&+06 11 12.7&  46.88$\,\pm\,$ 0.47&G0IV    & 6.314& 4.903\\                                
HD 38393  & 27072&05 44 27.79&-22 26 54.2& 112.02$\,\pm\,$ 0.18&F7V     & 3.586& 2.508\\                                
HD 41593  & 28954&06 06 40.48&+15 32 31.6&  65.48$\,\pm\,$ 0.67&K0V     & 6.762& 4.822\\                                
HD 60491  & 36827&07 34 26.17&-06 53 48.0&  40.73$\,\pm\,$ 1.00&K2V     & 8.160& 6.019\\                                
HD 61606  & 37349&07 39 59.33&-03 35 51.0&  70.37$\,\pm\,$ 0.64&K2V     & 7.200& 4.885\\                                
HD 63433  & 38228&07 49 55.06&+27 21 47.5&  45.45$\,\pm\,$ 0.53&G5IV    & 6.930& 5.258\\                                
HD 95650  & 53985&11 02 38.34&+21 58 01.7&  84.95$\,\pm\,$ 1.05&M0      & 9.690& 5.688\\                                
HIP 57548 & 57548&11 47 44.40&+00 48 16.4& 298.04$\,\pm\,$ 2.30&M4V     &11.080& 5.654\\                                
HD 125451 & 69989&14 19 16.28&+13 00 15.5&  38.32$\,\pm\,$ 0.28&F5IV    & 5.400& 4.394\\                                
HD 135599 & 74702&15 15 59.17&+00 47 46.9&  63.11$\,\pm\,$ 0.70&K0      & 7.000& 4.958\\                                
HD 139006 & 76267&15 34 41.27&+26 42 52.9&  43.46$\,\pm\,$ 0.28&A0V+G5V\tablefootmark{a}  & 2.210& 2.206\\                                
HD 147584 & 80686&16 28 28.14&-70 05 03.8&  82.53$\,\pm\,$ 0.52&F9V+M4V\tablefootmark{b}  & 4.910& 3.661\\                                
HD 165185 & 88694&18 06 23.72&-36 01 11.2&  56.97$\,\pm\,$ 0.48&G5V     & 5.949& 4.469\\                                
HD 175742 & 92919&18 55 53.22&+23 33 23.9&  46.74$\,\pm\,$ 0.85&K0V\tablefootmark{c}     & 8.090& 5.637\\                                
HIP 104383&104383&21 08 45.47&-04 25 36.9&  36.92$\,\pm\,$ 1.53&K6+M1\tablefootmark{d}    & 9.430& 6.398\\                                
HD 217813 &113829&23 03 04.98&+20 55 06.9&  40.46$\,\pm\,$ 0.57&G5V     & 6.652& 5.148\\                                

\hline
\end{tabular}\\
\tablefoot{All parallaxes are {\it HIPPARCOS} parallaxes \citep{2007A&A...474..653V}. $\Ks$ band magnitudes were taken from 2MASS \citep{Skrutskie06} and are combined magnitudes for tight binaries. For spectroscopic binaries, the spectral type is displayed for both components. 

{\bf References for spectral type.}
\tablefoottext{a}{\citet{2007MNRAS.378..179B}} \tablefoottext{b}{HD 147584: mean value of M4 of the range given by \citet{Skuljan+2004} (M1V-M7V)} \tablefoottext{c}{HD 175742: SB1 \citep{Tokovinin06}; literature not unanimous on the parameters \citep{Eggleton+2008}} \tablefoottext{d}{HIP 104383: \citet{Balega04}}.
}
\end{table*}

\section{Field of view and nearby visual companions}
\label{app:coverage}
\begin{table*}
\centering
\caption{\label{tab:fov}Characterisation of the NACO field of view according to Fig.~\ref{fig:hist_coro}.}
\begin{tabular}{lrrrrrllrlr}
\hline\hline
name&\multicolumn{2}{c}{FOV}&\multicolumn{1}{c}{max.}&\multicolumn{4}{c}{nearby stellar companions}&central&ref.&\multicolumn{1}{c}{expected approx.}\\
          &inner&outer                          &\multicolumn{1}{c}{FOV}&\#1&\#2&ref.&WDS entry  &mass&&\multicolumn{1}{c}{widest bound}\\
          &[au] &[au]        &[au]          & [au]&[au]&&                                                                    &$[\Msun]$&&\multicolumn{1}{c}{orbit [au]}\\
\hline
         HD 11131   &    8.3& 213& 590&4340& ...& 1&01496-1041&1.00& 6&  5400\\                                                                                                                         
         HD 11171   &    8.3& 213& 590&4340& ...& 1&01496-1041&1.52& 3&  8200\\                                                                                                                         
    HD 22049        &    1.1&  29&  80& ...& ...&..&       ...&0.79& 6&  4300\\                                                                                                                         
         HD 26913   &    7.3& 188& 522&1350& ...& 1&04155+0611&0.96& 6&  5200\\                                                                                                                         
         HD 26923   &    7.4& 191& 530&1360& ...& 1&04155+0611&1.07& 6&  5800\\                                                                                                                         
    HD 38393        &    3.1&  81& 224& 852&1250& 2&05445-2227&1.23& 4&  6600\\                                                                                                                         
    HD 41593        &    5.4& 139& 386& ...& ...&..&       ...&0.89& 6&  4800\\                                                                                                                         
    HD 60491        &    8.7& 223& 620& ...& ...&..&       ...&0.86&10&  4600\\                                                                                                                         
         HD 61606   &    5.0& 128& 355& 822& ...& 1&07400-0336&1.43& 5&  7700\\                                                                                                                         
         HD 63433   &    7.6& 196& 545& ...& ...&..&       ...&0.99& 6&  5400\\                                                                                                                         
        HIP 57548   &    1.2&  30&  83& ...& ...&..&       ...&0.30&12&  1600\\                                                                                                                         
         HD 95650   &    4.1& 105& 292& ...& ...&..&       ...&0.50&12&  2700\\                                                                                                                         
        HD 125451   &    9.1& 235& 652&4280& ...& 2&14193+1300&1.40& 3&  7600\\                                                                                                                         
   HD 135599        &    5.5& 140& 389& ...& ...&..&       ...&0.85& 6&  4600\\                                                                                                                         
        HD 139006   &    8.0& 206& 573& ...& ...&..&       ...&3.50&11& 18900\\                                                                                                                         
        HD 147584   &    4.2& 109& 303& ...& ...&..&       ...&1.39& 8&  7500\\                                                                                                                         
   HD 165185        &    6.1& 156& 434& 214& ...& 2&18064-3601&1.10&12&  5900\\                                                                                                                         
        HD 175742   &    7.5& 193& 536& ...& ...&..&       ...&1.10& 9&  5900\\                                                                                                                         
   HD 217813        &    8.5& 218& 607& ...& ...&..&       ...&1.05& 6&  5700\\                                                                                                                         
       HIP 104383   &    9.2& 237& 659&   7& 554& 2&21088-0426&1.10&12&  5900\\                                                                                                                         

\hline
\end{tabular}\\
\tablefoot{\textbf{(2, 3)} Range of separations which are completely covered around the coronograph (from inner radius, i.e. outer bound of the coronagraph of $0\farcs35$ to the outer radius of $9''$ shown in Fig.~\ref{fig:hist_coro}).  -- \textbf{(4)} maximum (incomplete) field of view ($25''$) -- \textbf{(5-8)} separation of up to two known nearby visual companions, reference, and WDS catalogue entry. -- \textbf{(9, 10)} Mass of the central star/binary and reference -- \textbf{(11)} Separation of the approximate widest expected bound orbit of very low-mass companions according to the approximation by \citet{Close+2003} for masses greater than $0.185\,\Msun$. Very close (spectroscopic) components are not listed but accounted for in the central mass. Masses have been derived separately for the binary components of HIP\,104383 ($0.6\,\Msun$ and $0.5\,\Msun$, respectively) and then added up. HD\,147584 is a spectroscopic binary with components of $1.12\,\Msun$ and $0.09-0.45\,\Msun$ \citep[][an average of $0.27\,\Msun$ has been adopted for the secondary]{Skuljan+2004}. HD\,175742 is a single-lined spectroscopic binary with a primary mass of $0.75\,\Msun$ and a minimum mass of the secondary of $0.35\,\Msun$ \citep{Tokovinin06}. The maximum expected orbital separation is thus a lower limit in this particular case.
\tablebib{(1)~\citet{2002A&A...384..180F}; (2)~\citet{2001AJ....122.3466M,2014yCat....102026M}; (3)~\citet{Allende+1999a}; (4)~\citet{Ammler+2009a}; (5)~\citet{Bonavita+2007}; (6)~\citet{Fuhrmann04}; (7)~\citet{Simon+2006}; (8)~\citet{Skuljan+2004}; (9)~\citet{Tokovinin06}; (10)~\citet{Valenti+2005}; (11)~\citet{2007MNRAS.378..179B}; (12)~based on spectral type (Table~\ref{tab:sample_all}) and \citet[][table 1]{Gray2005}.}
}
\end{table*}

\section{Images with candidates}
\label{app:images}

\begin{figure*}
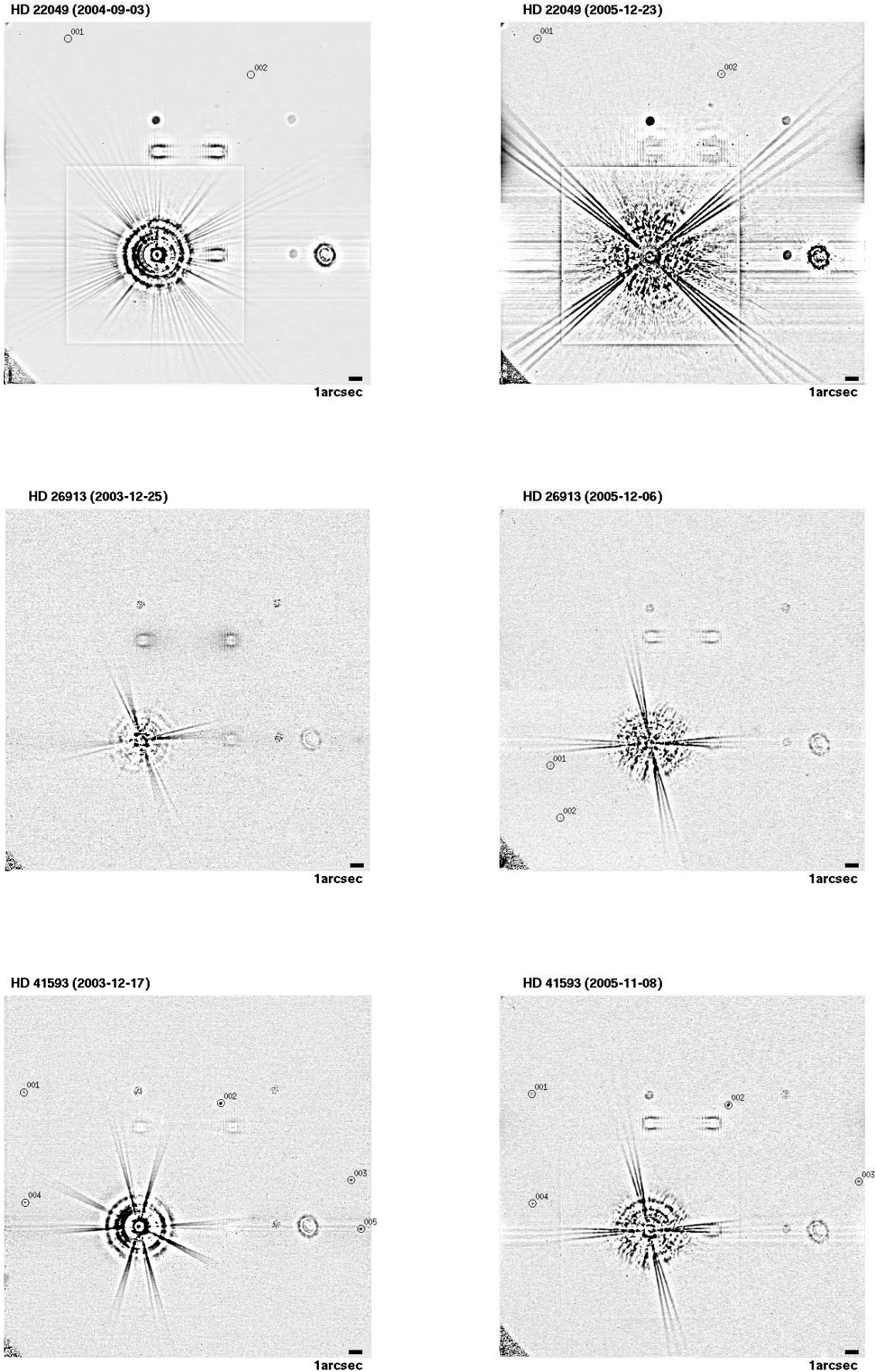

\center
\begin{tabular}{cc}
\subfigure{\includegraphics[bb=0 0 800 800,trim=100 300 200 0,width=\imgwidth]{./figures/rHD22049_073_C_0225A_cand}}&
\subfigure{\includegraphics[bb=0 0 800 800,trim=100 300 200 0,width=\imgwidth]{./figures/rHD22049_051223_cand}}\\     
\subfigure{\includegraphics[bb=0 0 800 800,trim=100 300 200 0,width=\imgwidth]{./figures/rHD26913_031225_cand}}&    
\subfigure{\includegraphics[bb=0 0 800 800,trim=100 300 200 0,width=\imgwidth]{./figures/rHD26913_051206_cand}}\\       
\subfigure{\includegraphics[bb=0 0 800 800,trim=100 300 200 0,width=\imgwidth]{./figures/rHD41593_072_C_0485A_cand}}&  
\subfigure{\includegraphics[bb=0 0 800 800,trim=100 300 200 0,width=\imgwidth]{./figures/rHD41593_051108_cand}}\\   
\end{tabular}
\caption{\label{fig:images1}Candidates for stars with images in two epochs: HD\,22049, HD\,26913, and HD\,41593. The images display the central star under the coronagraph. Faint companion candidates in the field of view are indicated by circles and are enumerated (cf. Table.~\ref{tab:all_cand_formatted}). The scale is indicated in the image. North is at the top and East is to the left. The PSF of the central star has been removed in a rectangular area centred on the star and a smoothed image has been subtracted. The residual speckle pattern and the diffraction spikes of the mount of the secondary mirror remain visible (several of those due to the addition of several exposures). In addition, there are reflections, ghosts, a vignetted region to the lower left, and the shadows of the other three coronagraphs.}
\end{figure*}

\begin{figure*}
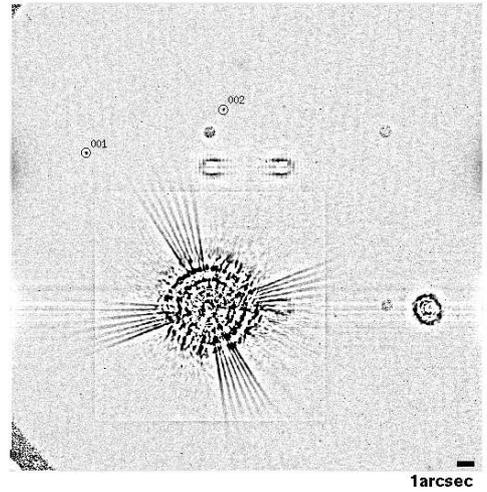

\center
\begin{tabular}{cc}
\subfigure{\includegraphics[bb=0 0 800 800,trim=100 300 200 0,width=\imgwidth]{./figures/rHD60491_072_C_0485A_cand}}&
\subfigure{\includegraphics[bb=0 0 800 800,trim=100 300 200 0,width=\imgwidth]{./figures/rHD60491_051108_cand}}\\
\subfigure{\includegraphics[bb=0 0 800 800,trim=100 300 200 0,width=\imgwidth]{./figures/rHD61606_040110_cand}}&
\subfigure{\includegraphics[bb=0 0 800 800,trim=100 300 200 0,width=\imgwidth]{./figures/rHD61606_051108_cand}}\\    
\subfigure{\includegraphics[bb=0 0 800 800,trim=100 300 200 0,width=\imgwidth]{./figures/rHD135599_073_C_0225A_cand}}& 
\subfigure{\includegraphics[bb=0 0 800 800,trim=100 300 200 0,width=\imgwidth]{./figures/rHD135599_060406_cand}}\\  
\end{tabular}
\caption{\label{fig:images2} Candidates for stars with images in two epochs: HD\,60491, HD\,61606, and HD\,135599. The layout is the same as in Fig.~\ref{fig:images1}.}
\end{figure*}

\begin{figure*}
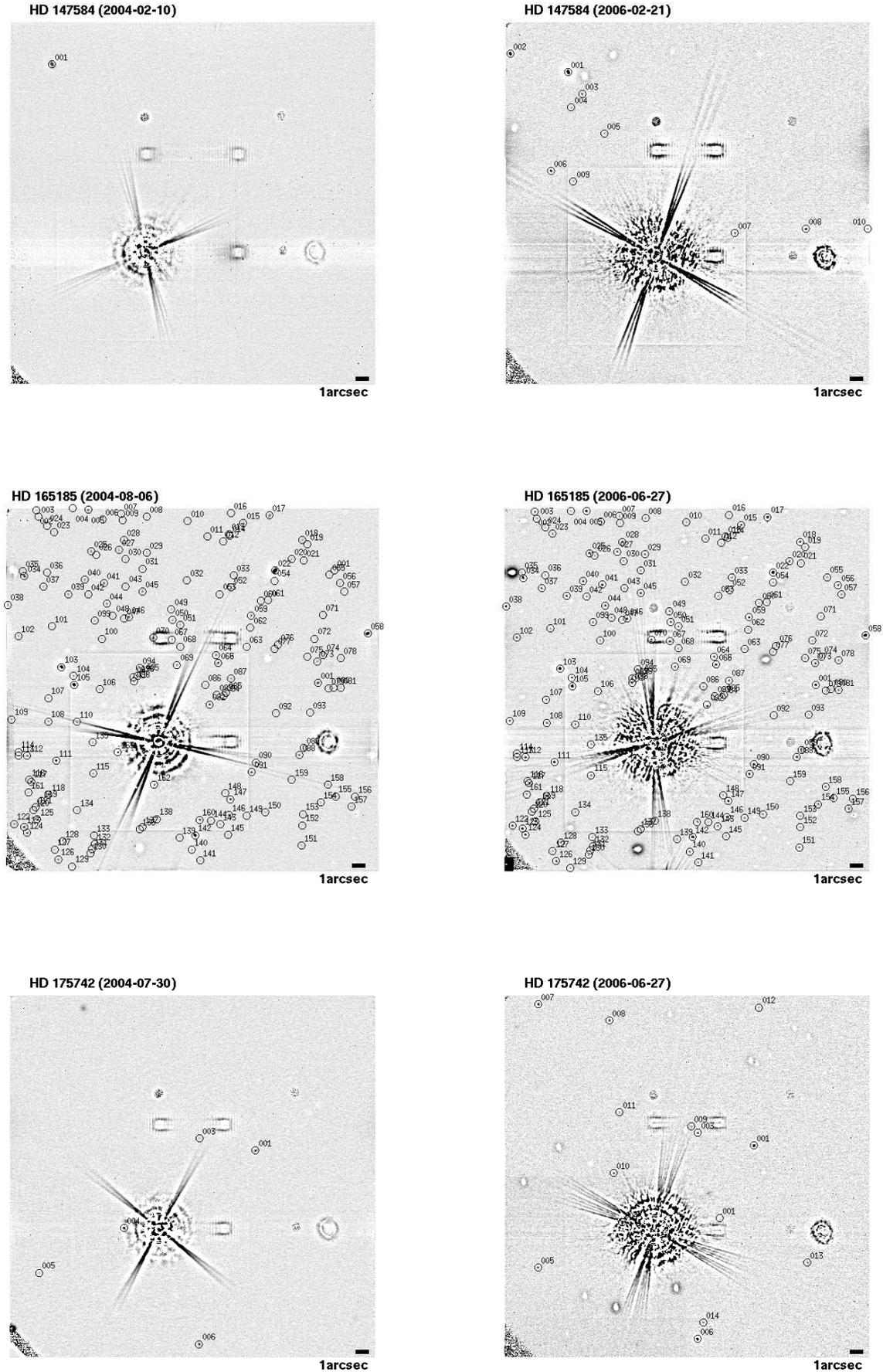

\center
\begin{tabular}{cc}
\subfigure{\includegraphics[bb=0 0 800 800,trim=100 300 200 0,width=\imgwidth]{./figures/rHD147584_040210_cand}}&
\subfigure{\includegraphics[bb=0 0 800 800,trim=100 300 200 0,width=\imgwidth]{./figures/rHD147584_060221_cand}}\\
\subfigure{\includegraphics[bb=0 0 800 800,trim=100 300 200 0,width=\imgwidth]{./figures/rHD165185_073_C_0225A_cand}}&
\subfigure{\includegraphics[bb=0 0 800 800,trim=100 300 200 0,width=\imgwidth]{./figures/rHD165185_060627_cand}}\\   
\subfigure{\includegraphics[bb=0 0 800 800,trim=100 300 200 0,width=\imgwidth]{./figures/rHD175742_040730_cand}}&      
\subfigure{\includegraphics[bb=0 0 800 800,trim=100 300 200 0,width=\imgwidth]{./figures/rHD175742_060627_cand}}    
\end{tabular}
\caption{\label{fig:images3} Candidates for stars with images in two epochs: HD\,147584, HD\,165185, and HD\,175742. 
The layout is the same as in Fig.~\ref{fig:images1}. Stellar residuals are present in the second epochs since jittering of the sky exposures did not work in those cases.}
\end{figure*}

\begin{figure*}
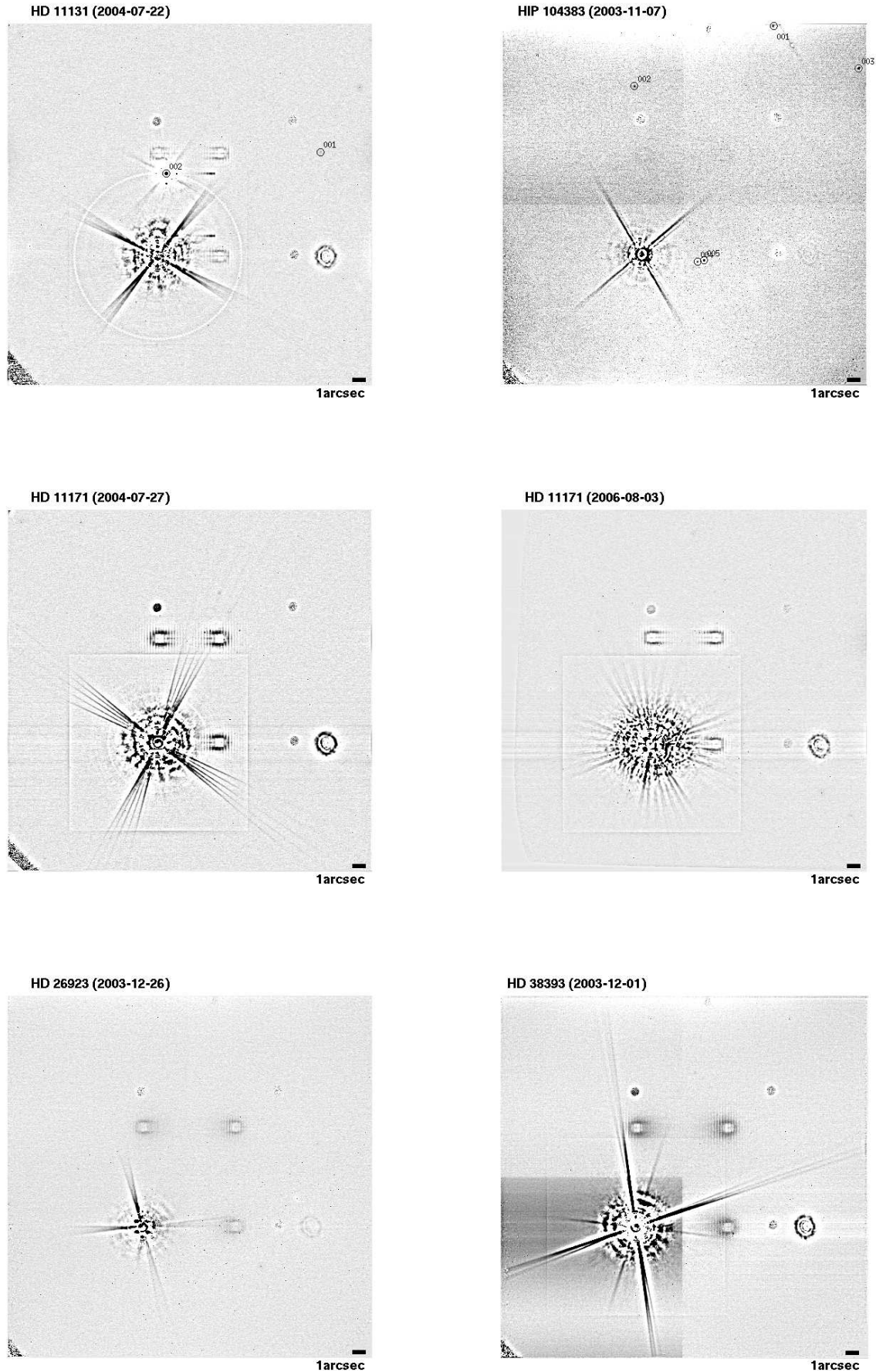

\center
\begin{tabular}{cc}
\subfigure{\includegraphics[bb=0 0 800 800,trim=100 300 200 0,width=\imgwidth]{./figures/rHD11131_040722_cand}}&       
\subfigure{\includegraphics[bb=0 0 800 800,trim=100 300 200 0,width=\imgwidth]{./figures/rHIP104383_031107_cand}}\\
\subfigure{\includegraphics[bb=0 0 800 800,trim=100 300 200 0,width=\imgwidth]{./figures/rHD11171_040727_cand}}&
\subfigure{\includegraphics[bb=0 0 800 800,trim=100 300 200 0,width=\imgwidth]{./figures/rHD11171_060803_cand}}\\
\subfigure{\includegraphics[bb=0 0 800 800,trim=100 300 200 0,width=\imgwidth]{./figures/rHD26923_031226_cand}}&  
\subfigure{\includegraphics[bb=0 0 800 800,trim=100 300 200 0,width=\imgwidth]{./figures/rHD38393_072_C_0485A_cand}}
\end{tabular}
\caption{\label{fig:images4} Candidates found in single-epoch observations (HD\,11131 and HIP\,104383) and observations of stars without any candidates (HD\,11171, HD\,26923, HD\,38393). The layout is the same as in Fig.~\ref{fig:images1}. HIP\,104383\,A is a double star below the coronagraph (see Fig.~\ref{fig:HIP104383} and footnote to Sect.~\ref{sect:results}).}
\end{figure*}

\begin{figure*}
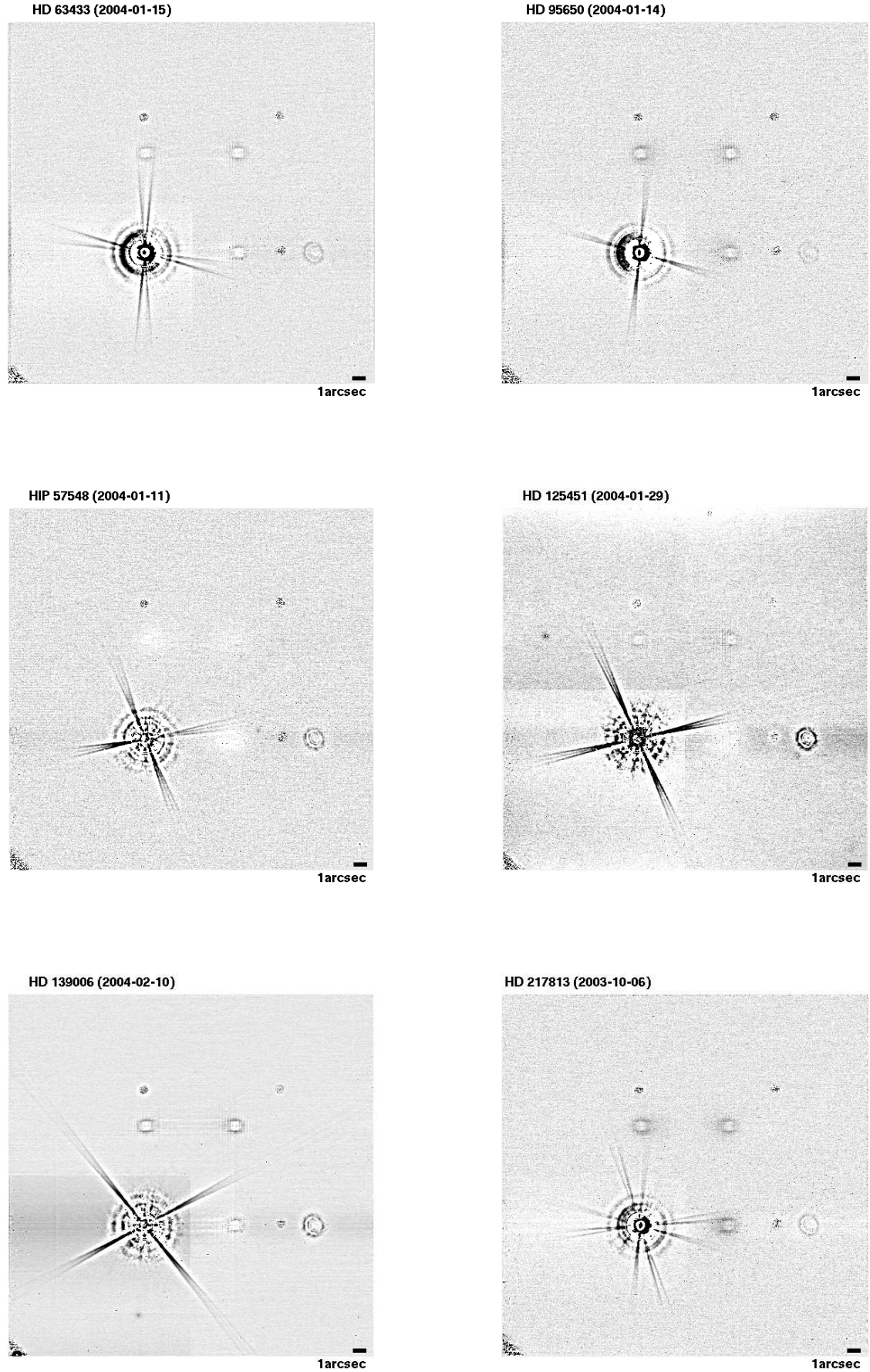

\center
\begin{tabular}{cc}
\subfigure{\includegraphics[bb=0 0 800 800,trim=100 300 200 0,width=\imgwidth]{./figures/rHD63433_040115_cand}}&
\subfigure{\includegraphics[bb=0 0 800 800,trim=100 300 200 0,width=\imgwidth]{./figures/rHD95650_040114_cand}}\\ 
\subfigure{\includegraphics[bb=0 0 800 800,trim=100 300 200 0,width=\imgwidth]{./figures/rHIP57548_040111_cand}}&  
\subfigure{\includegraphics[bb=0 0 800 800,trim=100 300 200 0,width=\imgwidth]{./figures/rHD125451_040129_cand}}\\   
\subfigure{\includegraphics[bb=0 0 800 800,trim=100 300 200 0,width=\imgwidth]{./figures/rHD139006_040210_cand}}&   
\subfigure{\includegraphics[bb=0 0 800 800,trim=100 300 200 0,width=\imgwidth]{./figures/rHD217813_072_C_0485A_cand}}
\end{tabular}
\caption{\label{fig:images5}Observations of stars without any candidates:  HD\,63433, HD\,95650, HIP\,57548, HD\,125451, HD\,139006, and HD\,217813. The layout is the same as in Fig.~\ref{fig:images1}.}
\end{figure*}


\section{List of candidates}
\onllongtab{1}{
\begin{center}
\begin{landscape}
\label{app:cand}
\small  
\setlength\LTleft{-4cm}
\setlength\LTright{-4cm plus 1 fill}
\begin{longtable}{l|rrrrrr|rrrrrr|rrrrrrrrr}
\caption{\label{tab:all_cand_formatted}Astrometry and photometry of candidates for each target and each epoch of observation.}\\
\hline\hline
ID&\multicolumn{3}{c}{$\Delta\alpha["]$}&\multicolumn{3}{c|}{$\Delta\delta["]$}&\multicolumn{3}{c}{$\Delta\alpha["]$}&\multicolumn{3}{c|}{$\Delta\delta["]$}&\multicolumn{3}{c}{$\Ks$ [mag]}&\multicolumn{3}{c}{$\delta\Delta\alpha[\mathrm{mas}]$}&\multicolumn{3}{c}{$\delta\Delta\delta[\mathrm{mas}]$}\\
\hline
\endfirsthead
\caption{continued.}\\
\hline\hline
ID&\multicolumn{3}{c}{$\Delta\alpha["]$}&\multicolumn{3}{c|}{$\Delta\delta["]$}&\multicolumn{3}{c}{$\Delta\alpha["]$}&\multicolumn{3}{c|}{$\Delta\delta["]$}&\multicolumn{3}{c}{$\Ks$ [mag]}&\multicolumn{3}{c}{$\delta\Delta\alpha[\mathrm{mas}]$}&\multicolumn{3}{c}{$\delta\Delta\delta[\mathrm{mas}]$}\\
\hline
\endhead
\hline
\endfoot
\hline
\endlastfoot
         HD 11131             &\multicolumn{6}{l|}{2004-07-22     }&\multicolumn{6}{l|}{}&&&&&&&&&\\
\hline
ID0001&-12.437&$\pm$&  0.022&  7.981&$\pm$&  0.033&\multicolumn{3}{c}{...}&\multicolumn{3}{c|}{...}& 18.8&$\pm$&  0.3&\multicolumn{3}{c}{...}&\multicolumn{3}{c}{...}\\
ID0002& -0.680&$\pm$&  0.017&  6.335&$\pm$&  0.004&\multicolumn{3}{c}{...}&\multicolumn{3}{c|}{...}&  9.5&$\pm$&  0.3&\multicolumn{3}{c}{...}&\multicolumn{3}{c}{...}\\
\hline
    HD 22049                  &\multicolumn{6}{l|}{2004-09-03     }&\multicolumn{6}{l|}{2005-12-23     }&&&&&&&&&\\
\hline
ID0001&  6.749&$\pm$&  0.044& 16.607&$\pm$&  0.020&  8.558&$\pm$&  0.044& 16.731&$\pm$&  0.025& 17.2&$\pm$&  0.5&1809&$\pm$& 62& 124&$\pm$& 32\\                            
ID0002& -7.215&$\pm$&  0.036& 13.812&$\pm$&  0.021& -5.454&$\pm$&  0.037& 13.974&$\pm$&  0.016& 17.3&$\pm$&  0.4&1761&$\pm$& 52& 162&$\pm$& 26\\                            
\hline
         HD 26913             &\multicolumn{6}{l|}{2003-12-25     }&\multicolumn{6}{l|}{2005-12-06     }&&&&&&&&&\\
\hline
ID0001&\multicolumn{3}{c}{...}&\multicolumn{3}{c|}{...}&  7.626&$\pm$&  0.006& -1.747&$\pm$&  0.020& 18.3&$\pm$&  0.3&\multicolumn{3}{c}{...}&\multicolumn{3}{c}{...}\\
ID0002&\multicolumn{3}{c}{...}&\multicolumn{3}{c|}{...}&  6.864&$\pm$&  0.016& -5.755&$\pm$&  0.018& 18.6&$\pm$&  0.3&\multicolumn{3}{c}{...}&\multicolumn{3}{c}{...}\\
\hline
    HD 41593                  &\multicolumn{6}{l|}{2003-12-17     }&\multicolumn{6}{l|}{2005-11-08     }&&&&&&&&&\\
\hline
ID0001&  8.784&$\pm$&  0.027& 10.313&$\pm$&  0.024&  8.966&$\pm$&  0.028& 10.507&$\pm$&  0.024& 17.8&$\pm$&  0.4& 182&$\pm$& 39& 194&$\pm$& 34\\                            
ID0002& -6.208&$\pm$&  0.025&  9.453&$\pm$&  0.017& -6.023&$\pm$&  0.026&  9.639&$\pm$&  0.017& 15.4&$\pm$&  0.4& 185&$\pm$& 36& 186&$\pm$& 24\\                            
ID0003&-16.149&$\pm$&  0.013&  3.558&$\pm$&  0.042&-15.951&$\pm$&  0.014&  3.755&$\pm$&  0.042& 16.8&$\pm$&  0.5& 198&$\pm$& 19& 197&$\pm$& 59\\                            
ID0004&  8.700&$\pm$&  0.007&  1.857&$\pm$&  0.023&  8.894&$\pm$&  0.008&  2.060&$\pm$&  0.023& 17.6&$\pm$&  0.4& 194&$\pm$& 11& 203&$\pm$& 33\\                            
ID0005&-16.878&$\pm$&  0.010& -0.197&$\pm$&  0.044&\multicolumn{3}{c}{...}&\multicolumn{3}{c|}{...}& 16.2&$\pm$&  0.3&\multicolumn{3}{c}{...}&\multicolumn{3}{c}{...}\\
\hline
    HD 60491                  &\multicolumn{6}{l|}{2003-12-26     }&\multicolumn{6}{l|}{2005-11-08     }&&&&&&&&&\\
\hline
ID0001&-15.721&$\pm$&  0.018&  5.847&$\pm$&  0.041&-15.582&$\pm$&  0.018&  5.918&$\pm$&  0.041& 17.4&$\pm$&  0.3& 139&$\pm$& 25&  71&$\pm$& 58\\                            
ID0002& -1.695&$\pm$&  0.045& 17.052&$\pm$&  0.011& -1.567&$\pm$&  0.045& 17.132&$\pm$&  0.011& 15.6&$\pm$&  0.3& 128&$\pm$& 64&  80&$\pm$& 16\\                            
ID0003&  2.382&$\pm$&  0.015& -5.890&$\pm$&  0.007&  2.522&$\pm$&  0.015& -5.818&$\pm$&  0.007& 18.5&$\pm$&  0.3& 140&$\pm$& 21&  72&$\pm$& 10\\                            
ID0004&\multicolumn{3}{c}{...}&\multicolumn{3}{c|}{...}& 11.312&$\pm$&  0.013& -4.255&$\pm$&  0.030& 18.5&$\pm$&  0.3&\multicolumn{3}{c}{...}&\multicolumn{3}{c}{...}\\
\hline
         HD 61606             &\multicolumn{6}{l|}{2004-01-10     }&\multicolumn{6}{l|}{2005-11-08     }&&&&&&&&&\\
\hline
ID0001& -1.114&$\pm$&  0.035& 13.389&$\pm$&  0.008& -1.324&$\pm$&  0.036& 13.898&$\pm$&  0.009& 17.7&$\pm$&  0.3&-210&$\pm$& 50& 509&$\pm$& 12\\                            
ID0002&  1.021&$\pm$&  0.017&  6.328&$\pm$&  0.005&  0.819&$\pm$&  0.018&  6.831&$\pm$&  0.005& 19.0&$\pm$&  0.9\tablefootmark{a}&-202&$\pm$& 25& 503&$\pm$&  7\\                            
ID0003&  4.780&$\pm$&  0.004&  1.135&$\pm$&  0.013&  4.584&$\pm$&  0.005&  1.648&$\pm$&  0.012& 16.9&$\pm$&  0.3&-196&$\pm$&  6& 513&$\pm$& 18\\                            
ID0004& -1.017&$\pm$&  0.012& -4.491&$\pm$&  0.004& -1.204&$\pm$&  0.010& -3.987&$\pm$&  0.004& 17.8&$\pm$&  0.9\tablefootmark{a}&-187&$\pm$& 16& 504&$\pm$&  6\\                            
ID0005&-11.218&$\pm$&  0.017& -5.886&$\pm$&  0.030&-11.398&$\pm$&  0.016& -5.403&$\pm$&  0.030& 18.1&$\pm$&  0.3&-180&$\pm$& 23& 483&$\pm$& 42\\                            
ID0006&  9.264&$\pm$&  0.026& -9.712&$\pm$&  0.025&  9.075&$\pm$&  0.025& -9.179&$\pm$&  0.024& 18.2&$\pm$&  0.3&-189&$\pm$& 36& 533&$\pm$& 35\\                            
\hline
   HD 135599                  &\multicolumn{6}{l|}{2004-07-13     }&\multicolumn{6}{l|}{2006-04-06     }&&&&&&&&&\\
\hline
ID0001&  7.610&$\pm$&  0.024&  8.923&$\pm$&  0.021&  7.238&$\pm$&  0.024&  9.163&$\pm$&  0.020& 17.1&$\pm$&  0.3&-372&$\pm$& 34& 240&$\pm$& 29\\                            
ID0002& -0.490&$\pm$&  0.030& 11.476&$\pm$&  0.007& -0.851&$\pm$&  0.031& 11.742&$\pm$&  0.007& 17.9&$\pm$&  0.3&-361&$\pm$& 43& 266&$\pm$& 10\\                            
\hline
        HD 147584             &\multicolumn{6}{l|}{2004-02-10     }&\multicolumn{6}{l|}{2006-02-21     }&&&&&&&&&\\
\hline
ID0001&  7.082&$\pm$&  0.038& 14.442&$\pm$&  0.020&  6.702&$\pm$&  0.037& 14.197&$\pm$&  0.020& 13.4&$\pm$&  0.8\tablefootmark{b}&-380&$\pm$& 53&-245&$\pm$& 28\\                            
ID0002&\multicolumn{3}{c}{...}&\multicolumn{3}{c|}{...}& 11.121&$\pm$&  0.041& 15.617&$\pm$&  0.031& 15.1&$\pm$&  0.3&\multicolumn{3}{c}{...}&\multicolumn{3}{c}{...}\\
ID0003&\multicolumn{3}{c}{...}&\multicolumn{3}{c|}{...}&  5.645&$\pm$&  0.033& 12.509&$\pm$&  0.017& 17.7&$\pm$&  0.3&\multicolumn{3}{c}{...}&\multicolumn{3}{c}{...}\\
ID0004&\multicolumn{3}{c}{...}&\multicolumn{3}{c|}{...}&  6.463&$\pm$&  0.030& 11.456&$\pm$&  0.018& 18.5&$\pm$&  0.3&\multicolumn{3}{c}{...}&\multicolumn{3}{c}{...}\\
ID0005&\multicolumn{3}{c}{...}&\multicolumn{3}{c|}{...}&  3.956&$\pm$&  0.025&  9.465&$\pm$&  0.012& 19.5&$\pm$&  0.3&\multicolumn{3}{c}{...}&\multicolumn{3}{c}{...}\\
ID0006&\multicolumn{3}{c}{...}&\multicolumn{3}{c|}{...}&  8.022&$\pm$&  0.018&  6.592&$\pm$&  0.021& 16.5&$\pm$&  0.3&\multicolumn{3}{c}{...}&\multicolumn{3}{c}{...}\\
ID0007&\multicolumn{3}{c}{...}&\multicolumn{3}{c|}{...}& -6.014&$\pm$&  0.006&  1.809&$\pm$&  0.016& 17.7&$\pm$&  0.3&\multicolumn{3}{c}{...}&\multicolumn{3}{c}{...}\\
ID0008&\multicolumn{3}{c}{...}&\multicolumn{3}{c|}{...}&-11.427&$\pm$&  0.009&  2.109&$\pm$&  0.030& 16.4&$\pm$&  0.3&\multicolumn{3}{c}{...}&\multicolumn{3}{c}{...}\\
ID0009&\multicolumn{3}{c}{...}&\multicolumn{3}{c|}{...}&  6.303&$\pm$&  0.016&  5.777&$\pm$&  0.017& 19.2&$\pm$&  0.3&\multicolumn{3}{c}{...}&\multicolumn{3}{c}{...}\\
ID0010&\multicolumn{3}{c}{...}&\multicolumn{3}{c|}{...}&-16.132&$\pm$&  0.011&  2.120&$\pm$&  0.042& 21.4&$\pm$&  0.3&\multicolumn{3}{c}{...}&\multicolumn{3}{c}{...}\\
\hline
   HD 165185                  &\multicolumn{6}{l|}{2004-08-06     }&\multicolumn{6}{l|}{2006-06-27     }&&&&&&&&&\\
\hline
ID0001&-12.109&$\pm$&  0.014&  4.500&$\pm$&  0.032&-12.367&$\pm$&  0.014&  4.475&$\pm$&  0.032& 16.7&$\pm$&  0.4&-258&$\pm$& 20& -25&$\pm$& 45\\                            
ID0002&  9.264&$\pm$&  0.047& 17.735&$\pm$&  0.027&  9.035&$\pm$&  0.047& 17.770&$\pm$&  0.026& 18.6&$\pm$&  0.6&-229&$\pm$& 66&  35&$\pm$& 37\\                            
ID0003&  9.114&$\pm$&  0.045& 17.253&$\pm$&  0.026&  8.881&$\pm$&  0.045& 17.264&$\pm$&  0.025& 19.8&$\pm$&  0.9\tablefootmark{c}&-233&$\pm$& 64&  11&$\pm$& 36\\                            
ID0004&  6.524&$\pm$&  0.047& 17.879&$\pm$&  0.020&  6.301&$\pm$&  0.047& 17.840&$\pm$&  0.020& 21.8&$\pm$&  2.1\tablefootmark{c}&-223&$\pm$& 66& -39&$\pm$& 28\\                            
ID0005&  5.351&$\pm$&  0.047& 17.779&$\pm$&  0.018&  5.110&$\pm$&  0.047& 17.834&$\pm$&  0.017& 17.0&$\pm$&  0.8\tablefootmark{c}&-241&$\pm$& 66&  55&$\pm$& 25\\                            
ID0006&  4.259&$\pm$&  0.045& 17.036&$\pm$&  0.015&  4.019&$\pm$&  0.045& 17.027&$\pm$&  0.015& 19.9&$\pm$&  0.4&-240&$\pm$& 64&  -9&$\pm$& 21\\                            
ID0007&  2.842&$\pm$&  0.046& 17.473&$\pm$&  0.013&  2.595&$\pm$&  0.046& 17.429&$\pm$&  0.012& 19.5&$\pm$&  0.6&-247&$\pm$& 65& -44&$\pm$& 18\\                            
ID0008&  0.875&$\pm$&  0.045& 17.248&$\pm$&  0.010&  0.610&$\pm$&  0.045& 17.277&$\pm$&  0.010& 19.4&$\pm$&  0.5&-265&$\pm$& 64&  29&$\pm$& 14\\                            
ID0009&  2.753&$\pm$&  0.044& 16.955&$\pm$&  0.012&  2.534&$\pm$&  0.044& 16.901&$\pm$&  0.012& 22.3&$\pm$&  1.5\tablefootmark{d}&-219&$\pm$& 62& -54&$\pm$& 17\\                            
ID0010& -2.224&$\pm$&  0.044& 16.927&$\pm$&  0.012& -2.515&$\pm$&  0.044& 16.964&$\pm$&  0.012& 19.5&$\pm$&  0.3&-291&$\pm$& 62&  37&$\pm$& 17\\                            
ID0011& -3.699&$\pm$&  0.041& 15.713&$\pm$&  0.013& -3.968&$\pm$&  0.041& 15.710&$\pm$&  0.014& 19.2&$\pm$&  0.5&-269&$\pm$& 58&  -3&$\pm$& 19\\                            
ID0012& -4.902&$\pm$&  0.040& 15.352&$\pm$&  0.016& -5.177&$\pm$&  0.040& 15.364&$\pm$&  0.016& 20.7&$\pm$&  0.8\tablefootmark{d}&-275&$\pm$& 57&  12&$\pm$& 23\\                            
ID0013& -5.338&$\pm$&  0.042& 15.915&$\pm$&  0.017& -5.481&$\pm$&  0.042& 15.887&$\pm$&  0.017& 19.1&$\pm$&  0.4&-143&$\pm$& 59& -28&$\pm$& 24\\                            
ID0014& -5.415&$\pm$&  0.041& 15.784&$\pm$&  0.017& -5.673&$\pm$&  0.041& 15.790&$\pm$&  0.018& 18.5&$\pm$&  0.6&-258&$\pm$& 58&   6&$\pm$& 25\\                            
ID0015& -6.450&$\pm$&  0.044& 16.762&$\pm$&  0.020& -6.697&$\pm$&  0.044& 16.764&$\pm$&  0.020& 18.1&$\pm$&  0.4&-247&$\pm$& 62&   2&$\pm$& 28\\                            
ID0016& -5.497&$\pm$&  0.046& 17.530&$\pm$&  0.018& -5.785&$\pm$&  0.046& 17.517&$\pm$&  0.018& 21.9&$\pm$&  1.9\tablefootmark{d}&-288&$\pm$& 65& -13&$\pm$& 25\\                            
ID0017& -8.445&$\pm$&  0.046& 17.338&$\pm$&  0.024& -8.713&$\pm$&  0.046& 17.352&$\pm$&  0.025& 16.9&$\pm$&  0.5&-268&$\pm$& 65&  14&$\pm$& 35\\                            
ID0018&-10.926&$\pm$&  0.041& 15.457&$\pm$&  0.030&-11.194&$\pm$&  0.041& 15.463&$\pm$&  0.031& 17.8&$\pm$&  0.5&-268&$\pm$& 58&   6&$\pm$& 43\\                            
ID0019&-11.310&$\pm$&  0.040& 15.103&$\pm$&  0.031&-11.551&$\pm$&  0.040& 15.040&$\pm$&  0.031& 19.9&$\pm$&  0.7&-241&$\pm$& 57& -63&$\pm$& 44\\                            
ID0020&-10.113&$\pm$&  0.037& 13.975&$\pm$&  0.028&-10.377&$\pm$&  0.037& 13.970&$\pm$&  0.028& 18.1&$\pm$&  0.5&-264&$\pm$& 52&  -5&$\pm$& 40\\                            
ID0021&-11.011&$\pm$&  0.037& 13.883&$\pm$&  0.030&-11.253&$\pm$&  0.037& 13.822&$\pm$&  0.031& 20.0&$\pm$&  0.3&-242&$\pm$& 52& -61&$\pm$& 43\\                            
ID0022& -8.846&$\pm$&  0.035& 13.106&$\pm$&  0.024& -9.117&$\pm$&  0.035& 13.102&$\pm$&  0.025& 13.8&$\pm$&  0.5&-271&$\pm$& 49&  -4&$\pm$& 35\\                            
ID0023&  7.949&$\pm$&  0.042& 16.054&$\pm$&  0.023&  7.695&$\pm$&  0.042& 16.079&$\pm$&  0.022& 18.9&$\pm$&  0.5&-254&$\pm$& 59&  25&$\pm$& 32\\                            
ID0024&  8.492&$\pm$&  0.044& 16.573&$\pm$&  0.024&  8.262&$\pm$&  0.044& 16.594&$\pm$&  0.024& 20.5&$\pm$&  0.7&-230&$\pm$& 62&  21&$\pm$& 34\\                            
ID0025&  5.126&$\pm$&  0.038& 14.583&$\pm$&  0.016&  4.881&$\pm$&  0.038& 14.596&$\pm$&  0.015& 17.5&$\pm$&  0.5&-245&$\pm$& 54&  13&$\pm$& 22\\                            
ID0026&  4.724&$\pm$&  0.038& 14.350&$\pm$&  0.015&  4.477&$\pm$&  0.038& 14.395&$\pm$&  0.015& 19.7&$\pm$&  0.5&-247&$\pm$& 54&  45&$\pm$& 21\\                            
ID0027&  3.015&$\pm$&  0.038& 14.689&$\pm$&  0.012&  2.762&$\pm$&  0.039& 14.711&$\pm$&  0.011& 18.1&$\pm$&  0.4&-253&$\pm$& 54&  22&$\pm$& 16\\                            
ID0028&  2.606&$\pm$&  0.040& 15.448&$\pm$&  0.011&  2.364&$\pm$&  0.041& 15.467&$\pm$&  0.011& 17.6&$\pm$&  0.5&-242&$\pm$& 57&  19&$\pm$& 16\\                            
ID0029&  0.896&$\pm$&  0.038& 14.477&$\pm$&  0.009&  0.648&$\pm$&  0.038& 14.497&$\pm$&  0.009& 18.5&$\pm$&  0.3&-248&$\pm$& 54&  20&$\pm$& 13\\                            
ID0030&  2.528&$\pm$&  0.037& 14.003&$\pm$&  0.011&  2.284&$\pm$&  0.037& 14.002&$\pm$&  0.010& 19.5&$\pm$&  0.5&-244&$\pm$& 52& ---&$\pm$& 15\\                            
ID0031&  1.235&$\pm$&  0.035& 13.261&$\pm$&  0.009&  0.991&$\pm$&  0.035& 13.267&$\pm$&  0.008& 19.4&$\pm$&  0.5&-244&$\pm$& 49&   6&$\pm$& 12\\                            
ID0032& -2.163&$\pm$&  0.032& 12.380&$\pm$&  0.009& -2.420&$\pm$&  0.032& 12.377&$\pm$&  0.010& 19.6&$\pm$&  0.4&-257&$\pm$& 45&  -3&$\pm$& 13\\                            
ID0033& -5.715&$\pm$&  0.034& 12.750&$\pm$&  0.017& -5.990&$\pm$&  0.034& 12.723&$\pm$&  0.017& 18.9&$\pm$&  0.5&-275&$\pm$& 48& -27&$\pm$& 24\\                            
ID0034& 10.187&$\pm$&  0.034& 12.688&$\pm$&  0.028&  9.948&$\pm$&  0.034& 12.707&$\pm$&  0.027& 16.5&$\pm$&  0.5&-239&$\pm$& 48&  19&$\pm$& 39\\                            
ID0035& 10.304&$\pm$&  0.035& 13.084&$\pm$&  0.028& 10.039&$\pm$&  0.035& 13.088&$\pm$&  0.027& 19.1&$\pm$&  0.3&-265&$\pm$& 49&   4&$\pm$& 39\\                            
ID0036&  8.481&$\pm$&  0.034& 12.975&$\pm$&  0.024&  8.265&$\pm$&  0.034& 12.919&$\pm$&  0.023& 19.5&$\pm$&  0.3&-216&$\pm$& 48& -56&$\pm$& 33\\                            
ID0037&  8.773&$\pm$&  0.032& 11.895&$\pm$&  0.024&  8.539&$\pm$&  0.032& 11.913&$\pm$&  0.023& 18.5&$\pm$&  0.5&-234&$\pm$& 45&  18&$\pm$& 33\\                            
ID0038& 11.459&$\pm$&  0.028& 10.500&$\pm$&  0.031& 11.254&$\pm$&  0.028& 10.510&$\pm$&  0.030& 17.4&$\pm$&  0.5&-205&$\pm$& 40&  10&$\pm$& 43\\                            
ID0039&  6.865&$\pm$&  0.030& 11.321&$\pm$&  0.019&  6.626&$\pm$&  0.030& 11.318&$\pm$&  0.019& 17.9&$\pm$&  0.4&-239&$\pm$& 42&  -3&$\pm$& 27\\                            
ID0040&  5.618&$\pm$&  0.033& 12.429&$\pm$&  0.016&  5.363&$\pm$&  0.033& 12.441&$\pm$&  0.016& 17.9&$\pm$&  0.4&-255&$\pm$& 47&  12&$\pm$& 23\\                            
ID0041&  4.147&$\pm$&  0.032& 12.160&$\pm$&  0.013&  3.906&$\pm$&  0.032& 12.163&$\pm$&  0.013& 17.2&$\pm$&  0.4&-241&$\pm$& 45&   3&$\pm$& 18\\                            
ID0042&  5.329&$\pm$&  0.030& 11.286&$\pm$&  0.016&  5.088&$\pm$&  0.030& 11.269&$\pm$&  0.015& 19.1&$\pm$&  0.5&-241&$\pm$& 42& -17&$\pm$& 22\\                            
ID0043&  2.489&$\pm$&  0.031& 11.910&$\pm$&  0.010&  2.237&$\pm$&  0.031& 11.904&$\pm$&  0.009& 19.4&$\pm$&  0.6&-252&$\pm$& 44&  -6&$\pm$& 13\\                            
ID0044&  3.924&$\pm$&  0.028& 10.574&$\pm$&  0.012&  3.676&$\pm$&  0.028& 10.575&$\pm$&  0.011& 17.6&$\pm$&  0.4&-248&$\pm$& 40&   1&$\pm$& 16\\                            
ID0045&  1.229&$\pm$&  0.030& 11.519&$\pm$&  0.008&  0.976&$\pm$&  0.030& 11.525&$\pm$&  0.007& 17.6&$\pm$&  0.4&-253&$\pm$& 42&   6&$\pm$& 11\\                            
ID0046&  2.264&$\pm$&  0.025&  9.561&$\pm$&  0.008&  2.020&$\pm$&  0.025&  9.555&$\pm$&  0.008& 17.1&$\pm$&  0.4&-244&$\pm$& 35&  -6&$\pm$& 11\\                            
ID0047&  2.615&$\pm$&  0.025&  9.453&$\pm$&  0.009&  2.372&$\pm$&  0.025&  9.458&$\pm$&  0.008& 19.5&$\pm$&  0.3&-243&$\pm$& 35&   5&$\pm$& 12\\                            
ID0048&  3.465&$\pm$&  0.025&  9.642&$\pm$&  0.011&  3.222&$\pm$&  0.025&  9.652&$\pm$&  0.010& 19.2&$\pm$&  0.4&-243&$\pm$& 35&  10&$\pm$& 15\\                            
ID0049& -0.968&$\pm$&  0.027& 10.156&$\pm$&  0.007& -1.226&$\pm$&  0.027& 10.138&$\pm$&  0.007& 18.8&$\pm$&  0.4&-258&$\pm$& 38& -18&$\pm$& 10\\                            
ID0050& -1.061&$\pm$&  0.024&  9.312&$\pm$&  0.006& -1.316&$\pm$&  0.024&  9.311&$\pm$&  0.006& 18.2&$\pm$&  0.4&-255&$\pm$& 34& ---&$\pm$&  8\\                            
ID0051& -1.644&$\pm$&  0.024&  8.968&$\pm$&  0.007& -1.897&$\pm$&  0.023&  8.956&$\pm$&  0.007& 18.6&$\pm$&  0.4&-253&$\pm$& 33& -12&$\pm$& 10\\                            
ID0052& -5.516&$\pm$&  0.031& 11.827&$\pm$&  0.016& -5.800&$\pm$&  0.031& 11.769&$\pm$&  0.017& 19.8&$\pm$&  0.7&-284&$\pm$& 44& -58&$\pm$& 23\\                            
ID0053& -4.664&$\pm$&  0.030& 11.306&$\pm$&  0.014& -4.940&$\pm$&  0.030& 11.301&$\pm$&  0.015& 19.5&$\pm$&  0.3&-276&$\pm$& 42&  -5&$\pm$& 21\\                            
ID0054& -8.807&$\pm$&  0.033& 12.330&$\pm$&  0.024& -9.122&$\pm$&  0.033& 12.337&$\pm$&  0.025& 18.8&$\pm$&  0.5&-315&$\pm$& 47&   7&$\pm$& 35\\                            
ID0055&-12.998&$\pm$&  0.034& 12.774&$\pm$&  0.035&-13.254&$\pm$&  0.034& 12.739&$\pm$&  0.035& 19.3&$\pm$&  0.6&-256&$\pm$& 48& -35&$\pm$& 49\\                            
ID0056&-13.804&$\pm$&  0.033& 12.147&$\pm$&  0.037&-14.088&$\pm$&  0.033& 12.130&$\pm$&  0.038& 18.4&$\pm$&  0.5&-284&$\pm$& 47& -17&$\pm$& 53\\                            
ID0057&-14.100&$\pm$&  0.031& 11.487&$\pm$&  0.038&-14.349&$\pm$&  0.031& 11.426&$\pm$&  0.038& 19.4&$\pm$&  0.6&-249&$\pm$& 44& -61&$\pm$& 54\\                            
ID0058&-15.891&$\pm$&  0.024&  8.285&$\pm$&  0.042&-16.159&$\pm$&  0.024&  8.268&$\pm$&  0.043& 15.2&$\pm$&  0.5&-268&$\pm$& 34& -17&$\pm$& 60\\                            
ID0059& -7.046&$\pm$&  0.026&  9.665&$\pm$&  0.019& -7.302&$\pm$&  0.026&  9.656&$\pm$&  0.020& 17.4&$\pm$&  0.4&-256&$\pm$& 37&  -9&$\pm$& 28\\                            
ID0060& -7.789&$\pm$&  0.029& 10.781&$\pm$&  0.021& -8.055&$\pm$&  0.029& 10.749&$\pm$&  0.022& 19.6&$\pm$&  0.6&-266&$\pm$& 41& -32&$\pm$& 30\\                            
ID0061& -8.341&$\pm$&  0.029& 10.856&$\pm$&  0.023& -8.635&$\pm$&  0.029& 10.842&$\pm$&  0.023& 20.9&$\pm$&  0.5&-294&$\pm$& 41& -14&$\pm$& 33\\                            
ID0062& -6.989&$\pm$&  0.023&  8.738&$\pm$&  0.019& -7.252&$\pm$&  0.023&  8.722&$\pm$&  0.020& 20.0&$\pm$&  0.9\tablefootmark{d}&-263&$\pm$& 33& -16&$\pm$& 28\\                            
ID0063& -6.684&$\pm$&  0.019&  7.250&$\pm$&  0.018& -6.953&$\pm$&  0.019&  7.229&$\pm$&  0.019& 19.0&$\pm$&  0.4&-269&$\pm$& 27& -21&$\pm$& 26\\                            
ID0064& -4.377&$\pm$&  0.018&  6.647&$\pm$&  0.012& -4.627&$\pm$&  0.018&  6.640&$\pm$&  0.013& 17.9&$\pm$&  0.5&-250&$\pm$& 25&  -7&$\pm$& 18\\                            
ID0065& -4.525&$\pm$&  0.016&  6.039&$\pm$&  0.012& -4.772&$\pm$&  0.016&  6.032&$\pm$&  0.013& 17.0&$\pm$&  0.4&-247&$\pm$& 23&  -7&$\pm$& 18\\                            
ID0066& -4.525&$\pm$&  0.016&  6.039&$\pm$&  0.012& -4.772&$\pm$&  0.016&  6.032&$\pm$&  0.013& 17.0&$\pm$&  0.4&-247&$\pm$& 23&  -7&$\pm$& 18\\                            
ID0067& -1.017&$\pm$&  0.021&  7.835&$\pm$&  0.005& -1.246&$\pm$&  0.020&  7.816&$\pm$&  0.006& 18.4&$\pm$&  1.5\tablefootmark{a}&-229&$\pm$& 29& -19&$\pm$&  8\\                            
ID0068& -1.637&$\pm$&  0.019&  7.295&$\pm$&  0.006& -1.880&$\pm$&  0.019&  7.282&$\pm$&  0.007& 18.6&$\pm$&  0.5&-243&$\pm$& 27& -13&$\pm$&  9\\                            
ID0069& -1.399&$\pm$&  0.015&  5.886&$\pm$&  0.005& -1.636&$\pm$&  0.015&  5.869&$\pm$&  0.005& 18.2&$\pm$&  0.4&-237&$\pm$& 21& -17&$\pm$&  7\\                            
ID0070&  0.403&$\pm$&  0.021&  7.970&$\pm$&  0.005&  0.159&$\pm$&  0.021&  7.961&$\pm$&  0.005& 17.0&$\pm$&  0.3&-244&$\pm$& 30&  -9&$\pm$&  7\\                            
ID0071&-12.492&$\pm$&  0.026&  9.687&$\pm$&  0.033&-12.776&$\pm$&  0.027&  9.731&$\pm$&  0.034& 20.1&$\pm$&  0.4&-284&$\pm$& 37&  44&$\pm$& 47\\                            
ID0072&-11.868&$\pm$&  0.022&  7.887&$\pm$&  0.031&-12.115&$\pm$&  0.022&  7.863&$\pm$&  0.032& 19.6&$\pm$&  0.4&-247&$\pm$& 31& -24&$\pm$& 45\\                            
ID0073&-12.086&$\pm$&  0.018&  6.158&$\pm$&  0.032&-12.353&$\pm$&  0.018&  6.133&$\pm$&  0.033& 17.9&$\pm$&  0.5&-267&$\pm$& 25& -25&$\pm$& 46\\                            
ID0074&-12.515&$\pm$&  0.019&  6.636&$\pm$&  0.033&-12.817&$\pm$&  0.019&  6.630&$\pm$&  0.034& 20.0&$\pm$&  0.8\tablefootmark{d}&-302&$\pm$& 27&  -6&$\pm$& 47\\                            
ID0075&-11.278&$\pm$&  0.018&  6.521&$\pm$&  0.030&-11.558&$\pm$&  0.018&  6.522&$\pm$&  0.030& 19.1&$\pm$&  0.4&-280&$\pm$& 25&   1&$\pm$& 42\\                            
ID0076& -9.094&$\pm$&  0.020&  7.428&$\pm$&  0.024& -9.368&$\pm$&  0.020&  7.472&$\pm$&  0.025& 19.5&$\pm$&  0.4&-274&$\pm$& 28&  44&$\pm$& 35\\                            
ID0077& -8.837&$\pm$&  0.019&  7.130&$\pm$&  0.023& -9.111&$\pm$&  0.019&  7.008&$\pm$&  0.024& 19.7&$\pm$&  0.4&-274&$\pm$& 27&-122&$\pm$& 33\\                            
ID0078&-13.843&$\pm$&  0.019&  6.393&$\pm$&  0.036&-14.117&$\pm$&  0.019&  6.363&$\pm$&  0.037& 19.2&$\pm$&  0.7&-274&$\pm$& 27& -30&$\pm$& 52\\                            
ID0079&-12.928&$\pm$&  0.013&  4.072&$\pm$&  0.034&-13.151&$\pm$&  0.013&  4.010&$\pm$&  0.035& 20.0&$\pm$&  0.5&-223&$\pm$& 18& -62&$\pm$& 49\\                            
ID0080&-13.304&$\pm$&  0.013&  4.143&$\pm$&  0.035&-13.535&$\pm$&  0.013&  4.117&$\pm$&  0.036& 19.5&$\pm$&  0.4&-231&$\pm$& 18& -26&$\pm$& 50\\                            
ID0081&-13.845&$\pm$&  0.014&  4.111&$\pm$&  0.036&-14.116&$\pm$&  0.014&  4.072&$\pm$&  0.037& 18.9&$\pm$&  0.7&-271&$\pm$& 20& -39&$\pm$& 52\\                            
ID0082& -3.875&$\pm$&  0.008&  2.857&$\pm$&  0.010& -4.086&$\pm$&  0.008&  2.951&$\pm$&  0.011& 17.1&$\pm$&  0.4&-211&$\pm$& 11&  94&$\pm$& 15\\                            
ID0083& -4.401&$\pm$&  0.010&  3.594&$\pm$&  0.012& -4.649&$\pm$&  0.010&  3.580&$\pm$&  0.012& 19.8&$\pm$&  0.7&-248&$\pm$& 14& -14&$\pm$& 17\\                            
ID0084& -4.962&$\pm$&  0.010&  3.546&$\pm$&  0.013& -5.218&$\pm$&  0.010&  3.523&$\pm$&  0.014& 18.1&$\pm$&  0.4&-256&$\pm$& 14& -23&$\pm$& 19\\                            
ID0085& -5.102&$\pm$&  0.010&  3.747&$\pm$&  0.014& -5.362&$\pm$&  0.010&  3.724&$\pm$&  0.014& 18.8&$\pm$&  0.5&-260&$\pm$& 14& -23&$\pm$& 20\\                            
ID0086& -3.561&$\pm$&  0.012&  4.340&$\pm$&  0.010& -3.809&$\pm$&  0.012&  4.340&$\pm$&  0.010& 19.5&$\pm$&  0.3&-248&$\pm$& 17&   0&$\pm$& 14\\                            
ID0087& -5.510&$\pm$&  0.013&  4.835&$\pm$&  0.015& -5.763&$\pm$&  0.013&  4.822&$\pm$&  0.015& 18.5&$\pm$&  0.3&-253&$\pm$& 18& -13&$\pm$& 21\\                            
ID0088&-10.682&$\pm$&  0.007& -1.022&$\pm$&  0.028&-10.935&$\pm$&  0.007& -1.060&$\pm$&  0.029& 17.4&$\pm$&  0.5&-253&$\pm$& 10& -38&$\pm$& 40\\                            
ID0089&-11.034&$\pm$&  0.007& -0.490&$\pm$&  0.029&-11.289&$\pm$&  0.007& -0.505&$\pm$&  0.030& 19.5&$\pm$&  0.3&-255&$\pm$& 10& -15&$\pm$& 42\\                            
ID0090& -7.401&$\pm$&  0.006& -1.589&$\pm$&  0.019& -7.647&$\pm$&  0.006& -1.620&$\pm$&  0.020& 17.1&$\pm$&  0.4&-246&$\pm$&  8& -31&$\pm$& 28\\                            
ID0091& -7.058&$\pm$&  0.007& -2.340&$\pm$&  0.019& -7.304&$\pm$&  0.008& -2.367&$\pm$&  0.019& 17.2&$\pm$&  0.3&-246&$\pm$& 11& -27&$\pm$& 27\\                            
ID0092& -8.906&$\pm$&  0.008&  2.189&$\pm$&  0.023& -9.174&$\pm$&  0.008&  2.131&$\pm$&  0.024& 22.4&$\pm$&  1.8\tablefootmark{a,d}&-268&$\pm$& 11& -58&$\pm$& 33\\                            
ID0093&-11.528&$\pm$&  0.009&  2.211&$\pm$&  0.030&-11.819&$\pm$&  0.009&  2.199&$\pm$&  0.031& 19.5&$\pm$&  0.4&-291&$\pm$& 13& -12&$\pm$& 43\\                            
ID0094&  1.450&$\pm$&  0.015&  5.663&$\pm$&  0.005&  1.209&$\pm$&  0.015&  5.651&$\pm$&  0.005& 19.2&$\pm$&  0.7&-241&$\pm$& 21& -12&$\pm$&  7\\                            
ID0095&  1.188&$\pm$&  0.014&  5.164&$\pm$&  0.004&  0.925&$\pm$&  0.013&  5.152&$\pm$&  0.004& 18.4&$\pm$&  0.5&-263&$\pm$& 19& -12&$\pm$&  6\\                            
ID0096&  1.571&$\pm$&  0.013&  5.052&$\pm$&  0.005&  1.333&$\pm$&  0.013&  5.030&$\pm$&  0.005& 18.8&$\pm$&  0.3&-238&$\pm$& 18& -22&$\pm$&  7\\                            
ID0097&  1.897&$\pm$&  0.013&  4.968&$\pm$&  0.006&  1.655&$\pm$&  0.013&  4.965&$\pm$&  0.005& 17.9&$\pm$&  0.3&-242&$\pm$& 18&  -3&$\pm$&  8\\                            
ID0098&  1.885&$\pm$&  0.012&  4.628&$\pm$&  0.006&  1.644&$\pm$&  0.012&  4.631&$\pm$&  0.005& 18.4&$\pm$&  0.4&-241&$\pm$& 17&   3&$\pm$&  8\\                            
ID0099&  4.854&$\pm$&  0.024&  9.288&$\pm$&  0.014&  4.619&$\pm$&  0.024&  9.288&$\pm$&  0.013& 19.0&$\pm$&  0.5&-235&$\pm$& 34&   0&$\pm$& 19\\                            
ID0100&  4.326&$\pm$&  0.021&  7.897&$\pm$&  0.012&  4.068&$\pm$&  0.021&  7.903&$\pm$&  0.012& 19.5&$\pm$&  0.5&-258&$\pm$& 30&   6&$\pm$& 17\\                            
ID0101&  8.097&$\pm$&  0.024&  8.856&$\pm$&  0.022&  7.867&$\pm$&  0.023&  8.808&$\pm$&  0.021& 22.0&$\pm$&  0.7&-230&$\pm$& 33& -48&$\pm$& 30\\                            
ID0102& 10.651&$\pm$&  0.022&  8.107&$\pm$&  0.028& 10.427&$\pm$&  0.022&  8.128&$\pm$&  0.028& 18.1&$\pm$&  0.4&-224&$\pm$& 31&  21&$\pm$& 40\\                            
ID0103&  7.372&$\pm$&  0.016&  5.737&$\pm$&  0.020&  7.136&$\pm$&  0.016&  5.740&$\pm$&  0.019& 14.0&$\pm$&  0.4&-236&$\pm$& 23&   3&$\pm$& 28\\                            
ID0104&  6.457&$\pm$&  0.014&  5.051&$\pm$&  0.017&  6.228&$\pm$&  0.014&  5.043&$\pm$&  0.017& 17.9&$\pm$&  0.3&-229&$\pm$& 20&  -8&$\pm$& 24\\                            
ID0105&  6.426&$\pm$&  0.012&  4.379&$\pm$&  0.017&  6.192&$\pm$&  0.012&  4.389&$\pm$&  0.016& 13.8&$\pm$&  0.3&-234&$\pm$& 17&  10&$\pm$& 23\\                            
ID0106&  4.477&$\pm$&  0.011&  4.009&$\pm$&  0.012&  4.237&$\pm$&  0.011&  4.006&$\pm$&  0.011& 19.3&$\pm$&  0.4&-240&$\pm$& 16&  -3&$\pm$& 16\\                            
ID0107&  8.397&$\pm$&  0.010&  3.350&$\pm$&  0.022&  8.181&$\pm$&  0.010&  3.356&$\pm$&  0.022& 19.0&$\pm$&  0.5&-216&$\pm$& 14&   6&$\pm$& 31\\                            
ID0108&  8.364&$\pm$&  0.006&  1.566&$\pm$&  0.022&  8.141&$\pm$&  0.006&  1.563&$\pm$&  0.021& 18.5&$\pm$&  0.3&-223&$\pm$&  8&  -3&$\pm$& 30\\                            
ID0109& 11.177&$\pm$&  0.008&  1.725&$\pm$&  0.029& 10.972&$\pm$&  0.008&  1.729&$\pm$&  0.029& 19.5&$\pm$&  1.2\tablefootmark{a}&-205&$\pm$& 11&   4&$\pm$& 41\\                            
ID0110&  6.199&$\pm$&  0.005&  1.517&$\pm$&  0.016&  5.972&$\pm$&  0.005&  1.458&$\pm$&  0.016& 18.0&$\pm$&  0.6&-227&$\pm$&  7& -59&$\pm$& 23\\                            
ID0111&  7.798&$\pm$&  0.006& -1.437&$\pm$&  0.020&  7.570&$\pm$&  0.006& -1.443&$\pm$&  0.020& 16.6&$\pm$&  0.3&-228&$\pm$&  8&  -6&$\pm$& 28\\                            
ID0112&  9.992&$\pm$&  0.007& -1.069&$\pm$&  0.026&  9.774&$\pm$&  0.006& -1.074&$\pm$&  0.026& 17.6&$\pm$&  0.4&-218&$\pm$&  9&  -5&$\pm$& 37\\                            
ID0113& 10.644&$\pm$&  0.007& -1.060&$\pm$&  0.028& 10.412&$\pm$&  0.007& -1.080&$\pm$&  0.027& 19.4&$\pm$&  1.0\tablefootmark{a,d}&-232&$\pm$& 10& -20&$\pm$& 39\\                            
ID0114& 10.654&$\pm$&  0.007& -0.774&$\pm$&  0.028& 10.433&$\pm$&  0.007& -0.780&$\pm$&  0.027& 18.2&$\pm$&  0.3&-221&$\pm$& 10&  -6&$\pm$& 39\\                            
ID0115&  4.980&$\pm$&  0.007& -2.438&$\pm$&  0.013&  4.758&$\pm$&  0.007& -2.457&$\pm$&  0.013& 19.0&$\pm$&  0.3&-222&$\pm$& 10& -19&$\pm$& 18\\                            
ID0116&  9.866&$\pm$&  0.010& -2.898&$\pm$&  0.026&  9.640&$\pm$&  0.010& -2.901&$\pm$&  0.025& 18.5&$\pm$&  0.3&-226&$\pm$& 14&  -3&$\pm$& 36\\                            
ID0117&  9.696&$\pm$&  0.010& -3.045&$\pm$&  0.025&  9.480&$\pm$&  0.010& -3.045&$\pm$&  0.025& 17.4&$\pm$&  0.3&-216&$\pm$& 14&   0&$\pm$& 35\\                            
ID0118&  8.358&$\pm$&  0.012& -4.031&$\pm$&  0.022&  8.132&$\pm$&  0.012& -4.039&$\pm$&  0.021& 16.5&$\pm$&  0.3&-226&$\pm$& 17&  -8&$\pm$& 30\\                            
ID0119&  8.940&$\pm$&  0.013& -4.545&$\pm$&  0.024&  8.720&$\pm$&  0.013& -4.560&$\pm$&  0.023& 17.6&$\pm$&  0.4&-220&$\pm$& 18& -15&$\pm$& 33\\                            
ID0120&  9.528&$\pm$&  0.015& -5.240&$\pm$&  0.025&  9.311&$\pm$&  0.015& -5.260&$\pm$&  0.025& 19.5&$\pm$&  0.3&-217&$\pm$& 21& -20&$\pm$& 35\\                            
ID0121&  9.333&$\pm$&  0.014& -5.014&$\pm$&  0.025&  9.121&$\pm$&  0.014& -5.029&$\pm$&  0.024& 19.0&$\pm$&  0.3&-212&$\pm$& 20& -15&$\pm$& 35\\                            
ID0122& 10.977&$\pm$&  0.018& -6.303&$\pm$&  0.029& 10.764&$\pm$&  0.018& -6.327&$\pm$&  0.028& 18.0&$\pm$&  0.5&-213&$\pm$& 25& -24&$\pm$& 40\\                            
ID0123& 10.231&$\pm$&  0.018& -6.534&$\pm$&  0.027& 10.012&$\pm$&  0.018& -6.550&$\pm$&  0.026& 16.9&$\pm$&  0.4&-219&$\pm$& 25& -16&$\pm$& 37\\                            
ID0124& 10.000&$\pm$&  0.019& -6.989&$\pm$&  0.026&  9.786&$\pm$&  0.019& -7.007&$\pm$&  0.026& 17.3&$\pm$&  0.4&-214&$\pm$& 27& -18&$\pm$& 37\\                            
ID0125&  9.228&$\pm$&  0.017& -5.959&$\pm$&  0.024&  9.011&$\pm$&  0.017& -5.974&$\pm$&  0.024& 17.1&$\pm$&  0.3&-217&$\pm$& 24& -15&$\pm$& 34\\                            
ID0126&  7.641&$\pm$&  0.024& -9.017&$\pm$&  0.021&  7.425&$\pm$&  0.024& -9.048&$\pm$&  0.020& 17.9&$\pm$&  0.4&-216&$\pm$& 34& -31&$\pm$& 29\\                            
ID0127&  7.916&$\pm$&  0.022& -8.255&$\pm$&  0.021&  7.704&$\pm$&  0.022& -8.274&$\pm$&  0.021& 19.2&$\pm$&  0.3&-212&$\pm$& 31& -19&$\pm$& 30\\                            
ID0128&  7.315&$\pm$&  0.020& -7.610&$\pm$&  0.020&  7.088&$\pm$&  0.020& -7.621&$\pm$&  0.019& 19.2&$\pm$&  0.3&-227&$\pm$& 28& -11&$\pm$& 28\\                            
ID0129&  6.573&$\pm$&  0.025& -9.555&$\pm$&  0.018&  6.357&$\pm$&  0.025& -9.580&$\pm$&  0.018& 19.5&$\pm$&  0.4&-216&$\pm$& 35& -25&$\pm$& 25\\                            
ID0130&  5.174&$\pm$&  0.023& -8.561&$\pm$&  0.014&  4.966&$\pm$&  0.023& -8.578&$\pm$&  0.014& 19.3&$\pm$&  0.3&-208&$\pm$& 33& -17&$\pm$& 20\\                            
ID0131&  5.090&$\pm$&  0.022& -8.221&$\pm$&  0.014&  4.863&$\pm$&  0.022& -8.240&$\pm$&  0.014& 18.9&$\pm$&  0.3&-227&$\pm$& 31& -19&$\pm$& 20\\                            
ID0132&  4.854&$\pm$&  0.021& -7.864&$\pm$&  0.014&  4.647&$\pm$&  0.021& -7.903&$\pm$&  0.013& 19.8&$\pm$&  0.6&-207&$\pm$& 30& -39&$\pm$& 19\\                            
ID0133&  4.921&$\pm$&  0.019& -7.214&$\pm$&  0.014&  4.691&$\pm$&  0.019& -7.216&$\pm$&  0.013& 19.5&$\pm$&  0.3&-230&$\pm$& 27&  -2&$\pm$& 19\\                            
ID0134&  6.189&$\pm$&  0.014& -5.245&$\pm$&  0.016&  5.973&$\pm$&  0.014& -5.272&$\pm$&  0.016& 18.9&$\pm$&  0.3&-216&$\pm$& 20& -27&$\pm$& 23\\                            
ID0135&  5.003&$\pm$&  0.003& -0.052&$\pm$&  0.013&  4.759&$\pm$&  0.003& -0.075&$\pm$&  0.012& 19.5&$\pm$&  0.4&-244&$\pm$&  4& -23&$\pm$& 18\\                            
ID0136&  1.442&$\pm$&  0.018& -6.727&$\pm$&  0.005&  1.208&$\pm$&  0.018& -6.776&$\pm$&  0.005& 19.1&$\pm$&  0.7&-234&$\pm$& 25& -49&$\pm$&  7\\                            
ID0137&  1.217&$\pm$&  0.017& -6.560&$\pm$&  0.005&  0.982&$\pm$&  0.017& -6.580&$\pm$&  0.005& 22.1&$\pm$&  2.1\tablefootmark{a,d}&-235&$\pm$& 24& -20&$\pm$&  7\\                            
ID0138&  0.147&$\pm$&  0.015& -5.920&$\pm$&  0.004& -0.079&$\pm$&  0.016& -5.950&$\pm$&  0.004& 18.4&$\pm$&  0.5&-226&$\pm$& 22& -30&$\pm$&  6\\                            
ID0139& -1.573&$\pm$&  0.019& -7.348&$\pm$&  0.006& -1.785&$\pm$&  0.019& -7.386&$\pm$&  0.006& 19.0&$\pm$&  0.5&-212&$\pm$& 27& -38&$\pm$&  8\\                            
ID0140& -2.515&$\pm$&  0.022& -8.293&$\pm$&  0.008& -2.743&$\pm$&  0.022& -8.327&$\pm$&  0.009& 18.4&$\pm$&  0.3&-228&$\pm$& 31& -34&$\pm$& 12\\                            
ID0141& -3.150&$\pm$&  0.024& -9.116&$\pm$&  0.010& -3.391&$\pm$&  0.024& -9.158&$\pm$&  0.010& 19.0&$\pm$&  0.4&-241&$\pm$& 34& -42&$\pm$& 14\\                            
ID0142& -2.760&$\pm$&  0.019& -7.169&$\pm$&  0.008& -2.989&$\pm$&  0.019& -7.207&$\pm$&  0.009& 14.6&$\pm$&  0.3&-229&$\pm$& 27& -38&$\pm$& 12\\                            
ID0143& -4.665&$\pm$&  0.017& -6.336&$\pm$&  0.013& -4.901&$\pm$&  0.017& -6.363&$\pm$&  0.013& 18.7&$\pm$&  0.4&-236&$\pm$& 24& -27&$\pm$& 18\\                            
ID0144& -3.964&$\pm$&  0.016& -6.069&$\pm$&  0.011& -4.186&$\pm$&  0.016& -6.067&$\pm$&  0.012& 19.8&$\pm$&  0.3&-222&$\pm$& 23&   2&$\pm$& 16\\                            
ID0145& -5.291&$\pm$&  0.019& -7.130&$\pm$&  0.014& -5.520&$\pm$&  0.019& -7.151&$\pm$&  0.015& 20.0&$\pm$&  0.3&-229&$\pm$& 27& -21&$\pm$& 21\\                            
ID0146& -5.354&$\pm$&  0.015& -5.633&$\pm$&  0.014& -5.606&$\pm$&  0.015& -5.661&$\pm$&  0.015& 17.2&$\pm$&  0.3&-252&$\pm$& 21& -28&$\pm$& 21\\                            
ID0147& -5.459&$\pm$&  0.012& -4.415&$\pm$&  0.015& -5.702&$\pm$&  0.012& -4.441&$\pm$&  0.015& 17.2&$\pm$&  0.3&-243&$\pm$& 17& -26&$\pm$& 21\\                            
ID0148& -5.068&$\pm$&  0.011& -3.964&$\pm$&  0.013& -5.297&$\pm$&  0.011& -3.989&$\pm$&  0.014& 19.7&$\pm$&  0.4&-229&$\pm$& 16& -25&$\pm$& 19\\                            
ID0149& -6.688&$\pm$&  0.015& -5.706&$\pm$&  0.018& -6.925&$\pm$&  0.016& -5.743&$\pm$&  0.018& 19.7&$\pm$&  0.8\tablefootmark{a,d}&-237&$\pm$& 22& -37&$\pm$& 25\\                            
ID0150& -8.104&$\pm$&  0.015& -5.406&$\pm$&  0.021& -8.336&$\pm$&  0.015& -5.452&$\pm$&  0.022& 18.8&$\pm$&  0.4&-232&$\pm$& 21& -46&$\pm$& 30\\                            
ID0151&-10.863&$\pm$&  0.022& -7.955&$\pm$&  0.029&-11.105&$\pm$&  0.022& -8.008&$\pm$&  0.029& 18.6&$\pm$&  0.4&-242&$\pm$& 31& -53&$\pm$& 41\\                            
ID0152&-10.940&$\pm$&  0.018& -6.415&$\pm$&  0.029&-11.195&$\pm$&  0.018& -6.466&$\pm$&  0.030& 19.4&$\pm$&  0.5&-255&$\pm$& 25& -51&$\pm$& 42\\                            
ID0153&-10.961&$\pm$&  0.016& -5.556&$\pm$&  0.029&-11.207&$\pm$&  0.016& -5.595&$\pm$&  0.030& 19.3&$\pm$&  0.3&-246&$\pm$& 23& -39&$\pm$& 42\\                            
ID0154&-12.309&$\pm$&  0.014& -4.657&$\pm$&  0.032&-12.561&$\pm$&  0.014& -4.687&$\pm$&  0.033& 18.6&$\pm$&  0.5&-252&$\pm$& 20& -30&$\pm$& 46\\                            
ID0155&-13.460&$\pm$&  0.014& -4.197&$\pm$&  0.035&-13.710&$\pm$&  0.014& -4.235&$\pm$&  0.036& 18.1&$\pm$&  0.5&-250&$\pm$& 20& -38&$\pm$& 50\\                            
ID0156&-14.947&$\pm$&  0.014& -4.246&$\pm$&  0.039&-15.213&$\pm$&  0.014& -4.284&$\pm$&  0.040& 19.5&$\pm$&  0.6&-266&$\pm$& 20& -38&$\pm$& 56\\                            
ID0157&-14.638&$\pm$&  0.016& -4.995&$\pm$&  0.038&-14.886&$\pm$&  0.016& -5.029&$\pm$&  0.039& 18.4&$\pm$&  0.4&-248&$\pm$& 23& -34&$\pm$& 54\\                            
ID0158&-12.859&$\pm$&  0.012& -3.316&$\pm$&  0.034&-13.115&$\pm$&  0.012& -3.348&$\pm$&  0.034& 18.2&$\pm$&  0.3&-256&$\pm$& 17& -32&$\pm$& 48\\                            
ID0159&-10.131&$\pm$&  0.010& -2.930&$\pm$&  0.027&-10.395&$\pm$&  0.010& -2.905&$\pm$&  0.027& 19.6&$\pm$&  0.3&-264&$\pm$& 14&  25&$\pm$& 38\\                            
ID0160& -3.122&$\pm$&  0.016& -5.979&$\pm$&  0.009& -3.332&$\pm$&  0.016& -6.042&$\pm$&  0.009& 17.9&$\pm$&  0.5&-210&$\pm$& 23& -63&$\pm$& 13\\                            
ID0161&  9.903&$\pm$&  0.012& -3.896&$\pm$&  0.026&  9.687&$\pm$&  0.012& -3.908&$\pm$&  0.025& 19.0&$\pm$&  0.3&-216&$\pm$& 17& -12&$\pm$& 36\\                            
ID0162&  0.333&$\pm$&  0.009& -3.306&$\pm$&  0.002&\multicolumn{3}{c}{...}&\multicolumn{3}{c|}{...}& 18.5&$\pm$&  0.3&\multicolumn{3}{c}{...}&\multicolumn{3}{c}{...}\\
ID0163&  3.099&$\pm$&  0.003& -0.813&$\pm$&  0.008&\multicolumn{3}{c}{...}&\multicolumn{3}{c|}{...}& 17.9&$\pm$&  0.3&\multicolumn{3}{c}{...}&\multicolumn{3}{c}{...}\\
\hline
        HD 175742             &\multicolumn{6}{l|}{2004-07-30     }&\multicolumn{6}{l|}{2006-06-27     }&&&&&&&&&\\
\hline
ID0001& -7.334&$\pm$&  0.016&  5.982&$\pm$&  0.019& -7.641&$\pm$&  0.018&  6.505&$\pm$&  0.020& 16.5&$\pm$&  0.5&-307&$\pm$& 24& 523&$\pm$& 28\\                            
ID0003& -3.055&$\pm$&  0.018&  6.934&$\pm$&  0.009& -3.358&$\pm$&  0.020&  7.474&$\pm$&  0.010& 18.9&$\pm$&  0.5&-303&$\pm$& 27& 540&$\pm$& 13\\                            
ID0004&  2.678&$\pm$&  0.002&  0.050&$\pm$&  0.007&  2.409&$\pm$&  0.002&  0.586&$\pm$&  0.006& 15.9&$\pm$&  0.3&-269&$\pm$&  3& 536&$\pm$&  9\\                            
ID0005&  9.109&$\pm$&  0.010& -3.397&$\pm$&  0.024&  8.852&$\pm$&  0.009& -2.859&$\pm$&  0.023& 18.9&$\pm$&  0.4&-257&$\pm$& 13& 538&$\pm$& 33\\                            
ID0006& -3.039&$\pm$&  0.023& -8.884&$\pm$&  0.010& -3.313&$\pm$&  0.022& -8.375&$\pm$&  0.010& 16.8&$\pm$&  0.4&-274&$\pm$& 32& 509&$\pm$& 14\\                            
ID0007&\multicolumn{3}{c}{...}&\multicolumn{3}{c|}{...}&  8.801&$\pm$&  0.046& 17.394&$\pm$&  0.025& 18.0&$\pm$&  0.3&\multicolumn{3}{c}{...}&\multicolumn{3}{c}{...}\\
ID0008&\multicolumn{3}{c}{...}&\multicolumn{3}{c|}{...}&  3.383&$\pm$&  0.042& 16.147&$\pm$&  0.013& 18.5&$\pm$&  0.3&\multicolumn{3}{c}{...}&\multicolumn{3}{c}{...}\\
ID0009&\multicolumn{3}{c}{...}&\multicolumn{3}{c|}{...}& -2.847&$\pm$&  0.021&  7.972&$\pm$&  0.009& 19.7&$\pm$&  0.3&\multicolumn{3}{c}{...}&\multicolumn{3}{c}{...}\\
ID0010&\multicolumn{3}{c}{...}&\multicolumn{3}{c|}{...}&  3.063&$\pm$&  0.012&  4.392&$\pm$&  0.008& 18.9&$\pm$&  0.3&\multicolumn{3}{c}{...}&\multicolumn{3}{c}{...}\\
ID0011&\multicolumn{3}{c}{...}&\multicolumn{3}{c|}{...}&  2.610&$\pm$&  0.024&  9.054&$\pm$&  0.009& 20.0&$\pm$&  0.3&\multicolumn{3}{c}{...}&\multicolumn{3}{c}{...}\\
ID0012&\multicolumn{3}{c}{...}&\multicolumn{3}{c|}{...}& -8.037&$\pm$&  0.045& 17.125&$\pm$&  0.023& 19.9&$\pm$&  0.3&\multicolumn{3}{c}{...}&\multicolumn{3}{c}{...}\\
ID0013&\multicolumn{3}{c}{...}&\multicolumn{3}{c|}{...}&-11.710&$\pm$&  0.010& -2.498&$\pm$&  0.031& 19.5&$\pm$&  0.3&\multicolumn{3}{c}{...}&\multicolumn{3}{c}{...}\\
ID0014&\multicolumn{3}{c}{...}&\multicolumn{3}{c|}{...}& -3.767&$\pm$&  0.019& -7.102&$\pm$&  0.011& 19.4&$\pm$&  0.3&\multicolumn{3}{c}{...}&\multicolumn{3}{c}{...}\\
\hline
       HIP 104383             &\multicolumn{6}{l|}{2003-11-07     }&\multicolumn{6}{l|}{}&&&&&&&&&\\
\hline
ID0001&-10.192&$\pm$&  0.046& 17.479&$\pm$&  0.029&\multicolumn{3}{c}{...}&\multicolumn{3}{c|}{...}& 16.2&$\pm$&  0.3&\multicolumn{3}{c}{...}&\multicolumn{3}{c}{...}\\
ID0002&  0.401&$\pm$&  0.034& 12.871&$\pm$&  0.008&\multicolumn{3}{c}{...}&\multicolumn{3}{c|}{...}& 17.5&$\pm$&  0.3&\multicolumn{3}{c}{...}&\multicolumn{3}{c}{...}\\
ID0003&-16.661&$\pm$&  0.039& 14.241&$\pm$&  0.044&\multicolumn{3}{c}{...}&\multicolumn{3}{c|}{...}& 16.5&$\pm$&  0.3&\multicolumn{3}{c}{...}&\multicolumn{3}{c}{...}\\
ID0004& -4.384&$\pm$&  0.003& -0.608&$\pm$&  0.011&\multicolumn{3}{c}{...}&\multicolumn{3}{c|}{...}& 17.8&$\pm$&  0.3&\multicolumn{3}{c}{...}&\multicolumn{3}{c}{...}\\
ID0005& -4.876&$\pm$&  0.003& -0.504&$\pm$&  0.013&\multicolumn{3}{c}{...}&\multicolumn{3}{c|}{...}& 15.7&$\pm$&  0.3&\multicolumn{3}{c}{...}&\multicolumn{3}{c}{...}\\

\end{longtable}
\tablefoot{A list of unique sequential identifiers (prefix 'ID') has been assigned to the candidates {\bf (1)} (cf. Figs.~\ref{fig:images1}-\ref{fig:images5}). For each candidate, the separation from the central star in right ascension and declination is given for the first epoch {\bf (2, 3)} and the second epoch when present {\bf (4, 5)}. Finally, apparent $K_s$ magnitude is given for all candidates {\bf (6)} and - when two epochs are available - the change of separation {\bf (7,8)}. The change in separation reflects the stellar parallactic and proper motion since all candidates are background stars (see text and Table~\ref{tab:par_motion}). Large error bars in the magnitude measurement are ascribed to different causes. Those are indicated by letters to the column for error bars $\ge0.8$: \tablefoottext{a}{affected by background features (like diffraction spikes, ghosts, reflections)} \tablefoottext{b}{smeared in first epoch} \tablefoottext{c}{aperture photometry misses flux at edge of frame} \tablefoottext{d}{measurement affected by noise in first epoch.}
}
\end{landscape}
\end{center}}

\begin{figure}
\includegraphics[bb=0 0 800 800,trim=50 300 300 0,width=4cm]{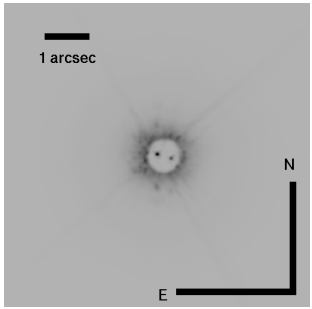}
\caption{\label{fig:HIP104383} The visual binary HIP\,104383\,A below the semi-transparent coronagraphic mask. The binary has already been characterised (\citealp{Balega04}). The full frame exposing companion candidates is shown in Fig.~\ref{fig:images4}. The magnitudes of the binary components differ by 0.45\,mag in the $\Ks$ band.}
\end{figure}

\section{Detection limits}
\label{app:det_limits}

\begin{figure*}
\center
\begin{tabular}{ccc}
\includegraphics[width=5cm]{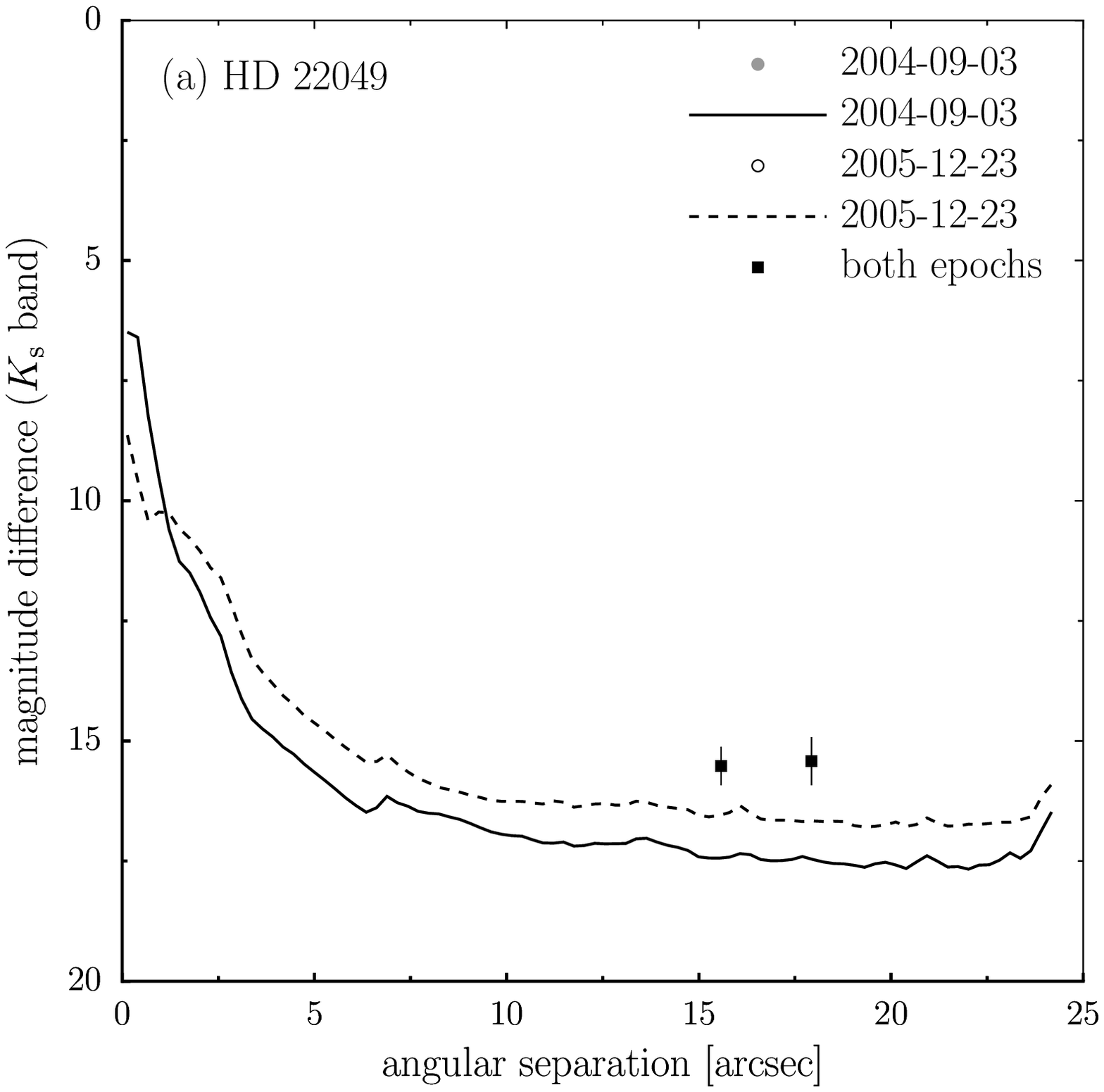}&
\includegraphics[width=5cm]{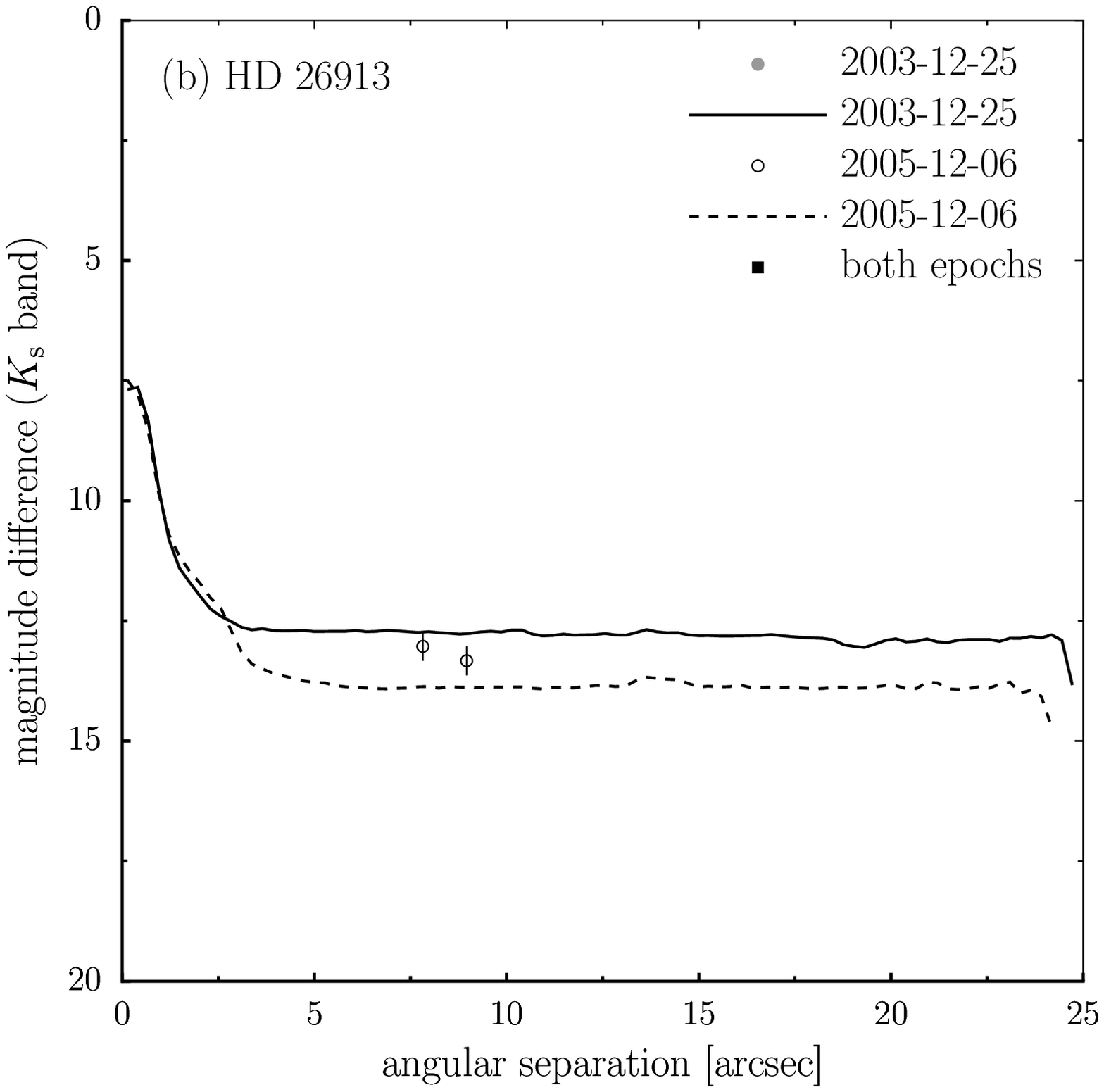}&
\includegraphics[width=5cm]{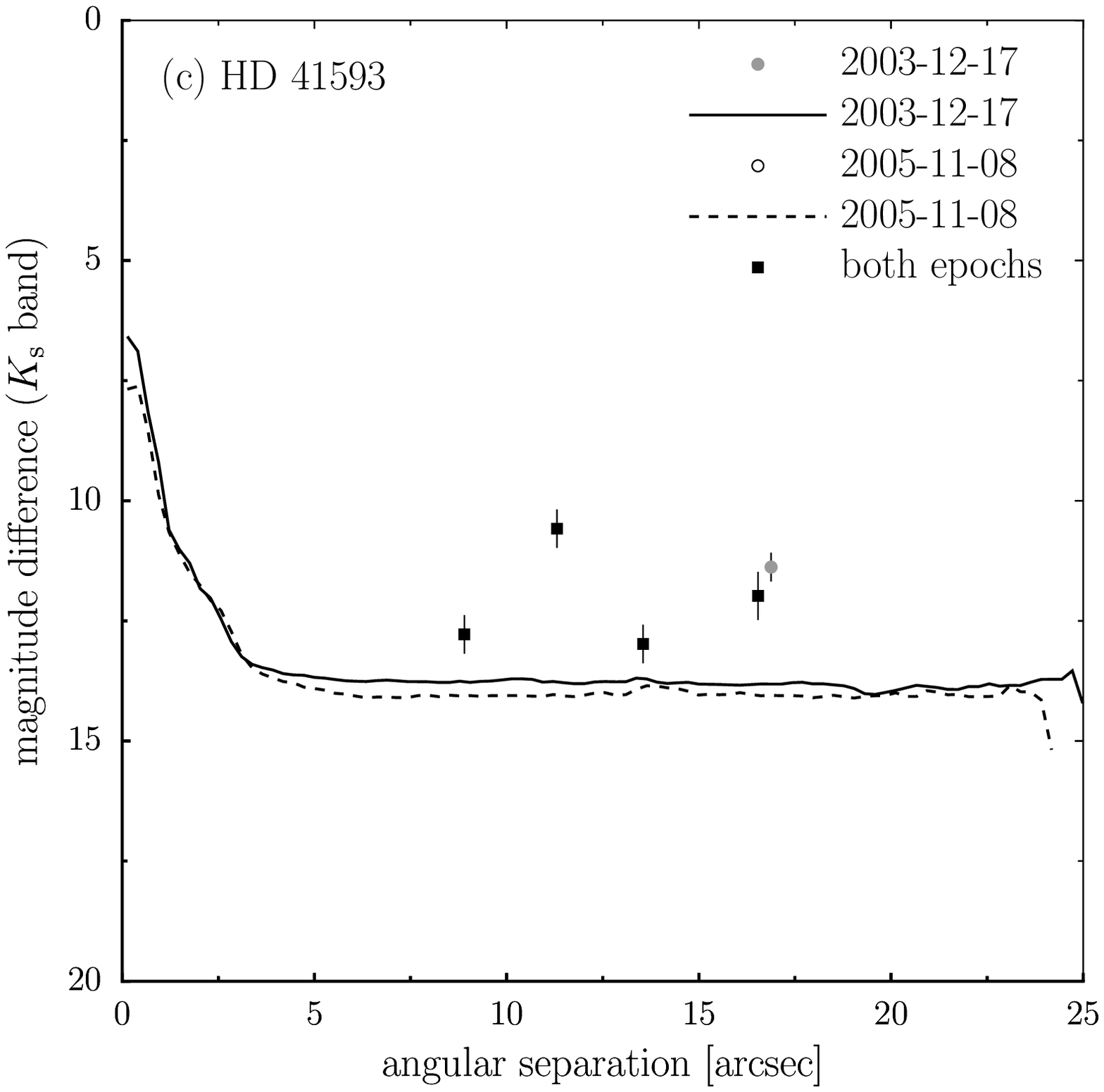}\\
\includegraphics[width=5cm]{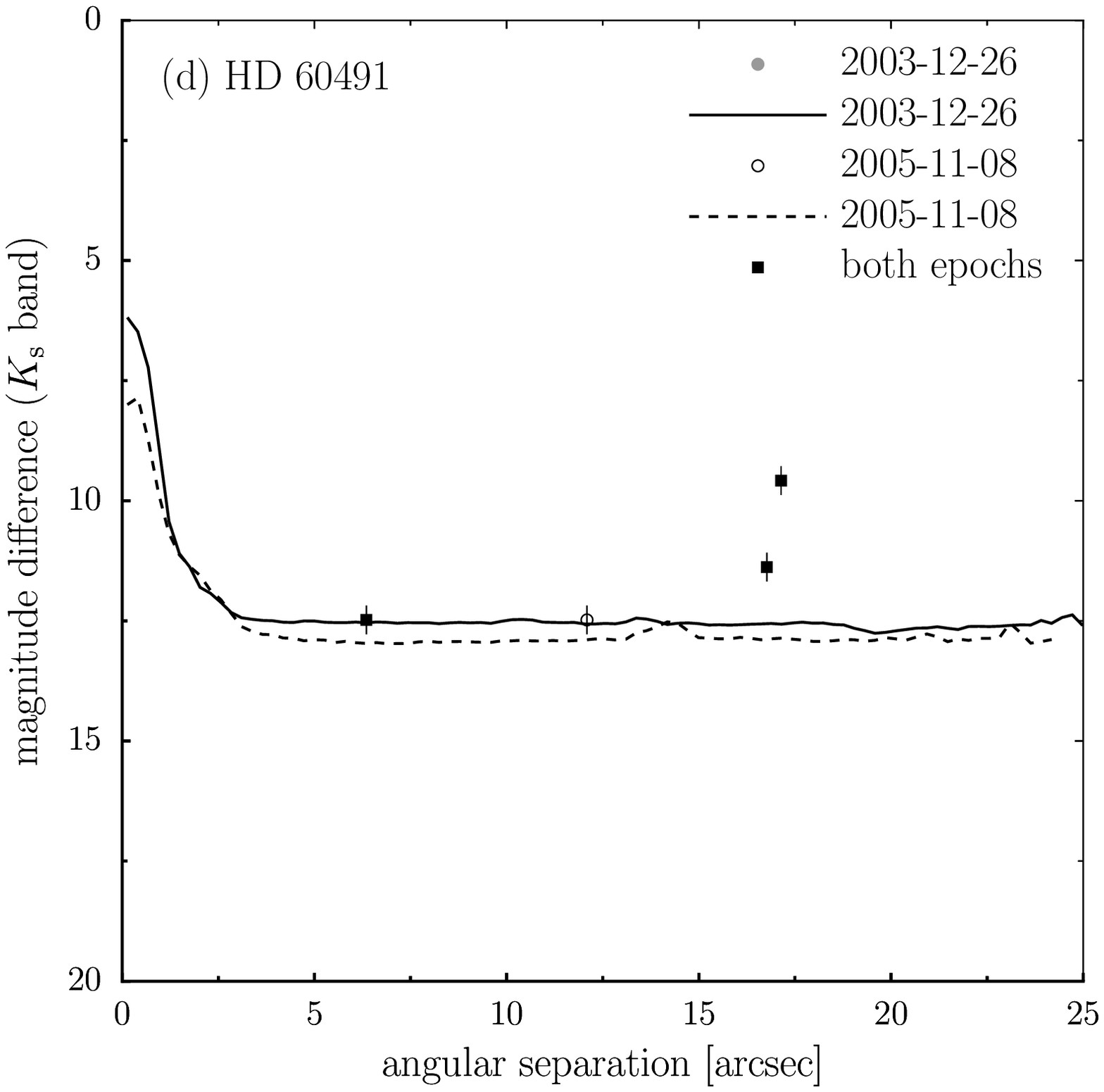}&
\includegraphics[width=5cm]{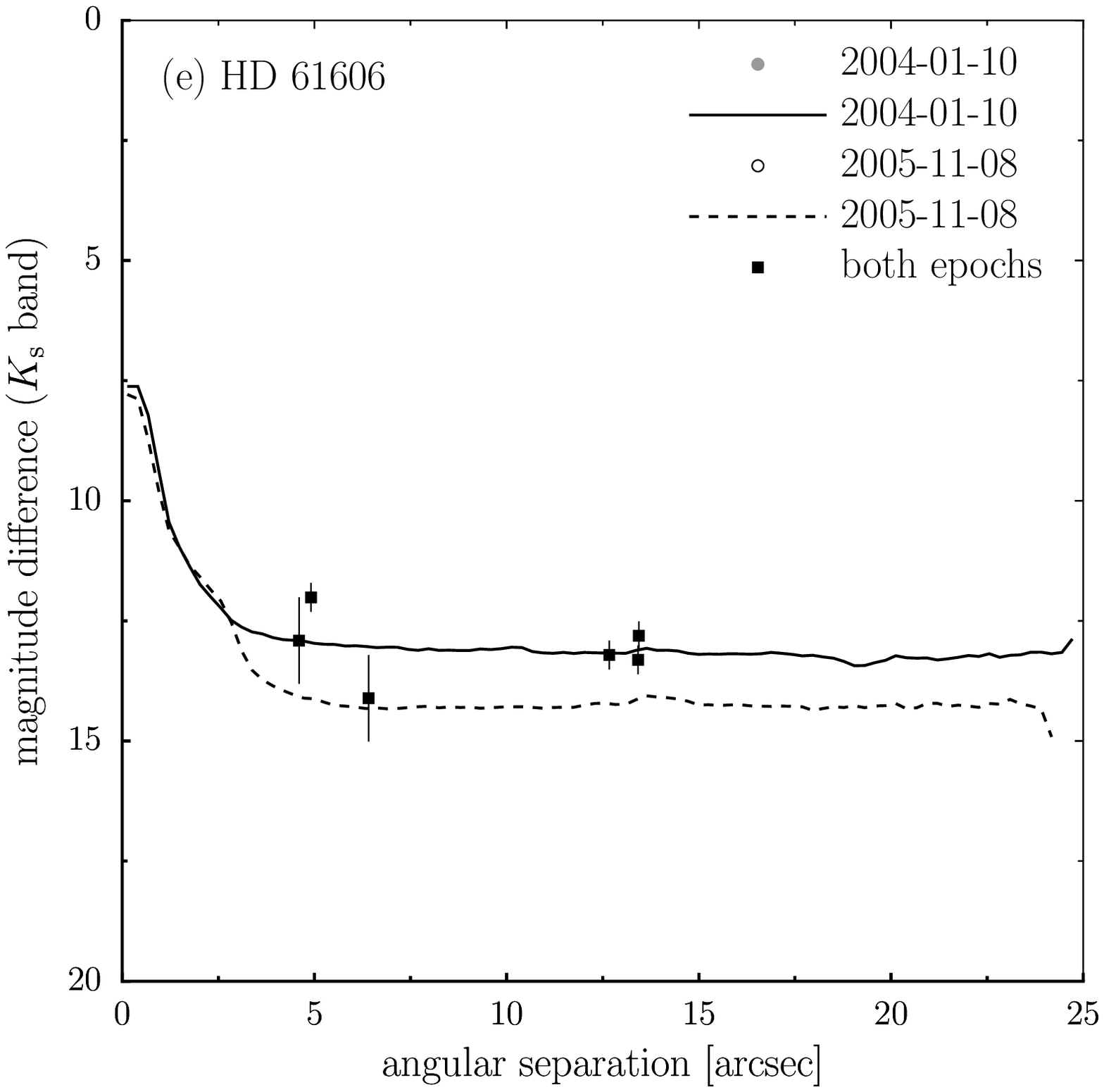}&
\includegraphics[width=5cm]{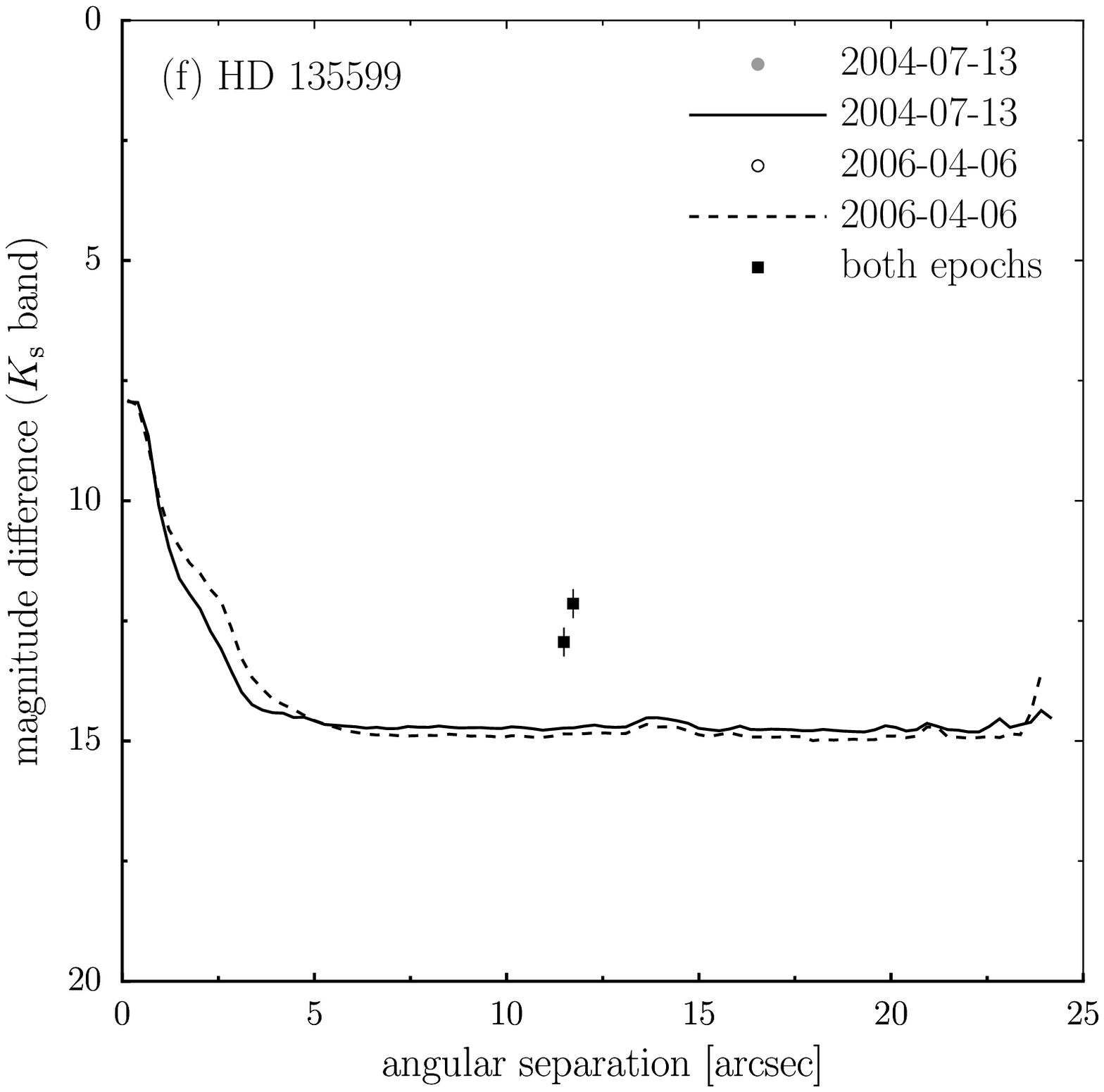}
\end{tabular}
\caption{\label{fig:candidates_2epochs1} Dynamic range curves are shown for stars with candidates and observations in two epochs: HD 22049, HD 26913, HD 41593, HD 60491, HD 61606, and HD\,135599.  The lines show the $10\,\sigma$ detection limits for the $1^\mathrm{st}$ and $2^\mathrm{nd}$ epoch as is indicated in the legend. The adoption of a $5\,\sigma$ limit to describe the visual detection of candidates would lower the curves by $2.5\,{\log}2=0.75\,$mag. Furthermore, the curves can vary by 0.3\,mag due to the uncertainty of the transmission of the coronagraph. Filled circles indicate candidates found in the $1^\mathrm{st}$ epoch, open circles those in the $2^\mathrm{nd}$ epoch, and filled squares those seen in both epochs.}
\end{figure*}

\begin{figure*}
\center
\begin{tabular}{ccc}
\includegraphics[width=5cm]{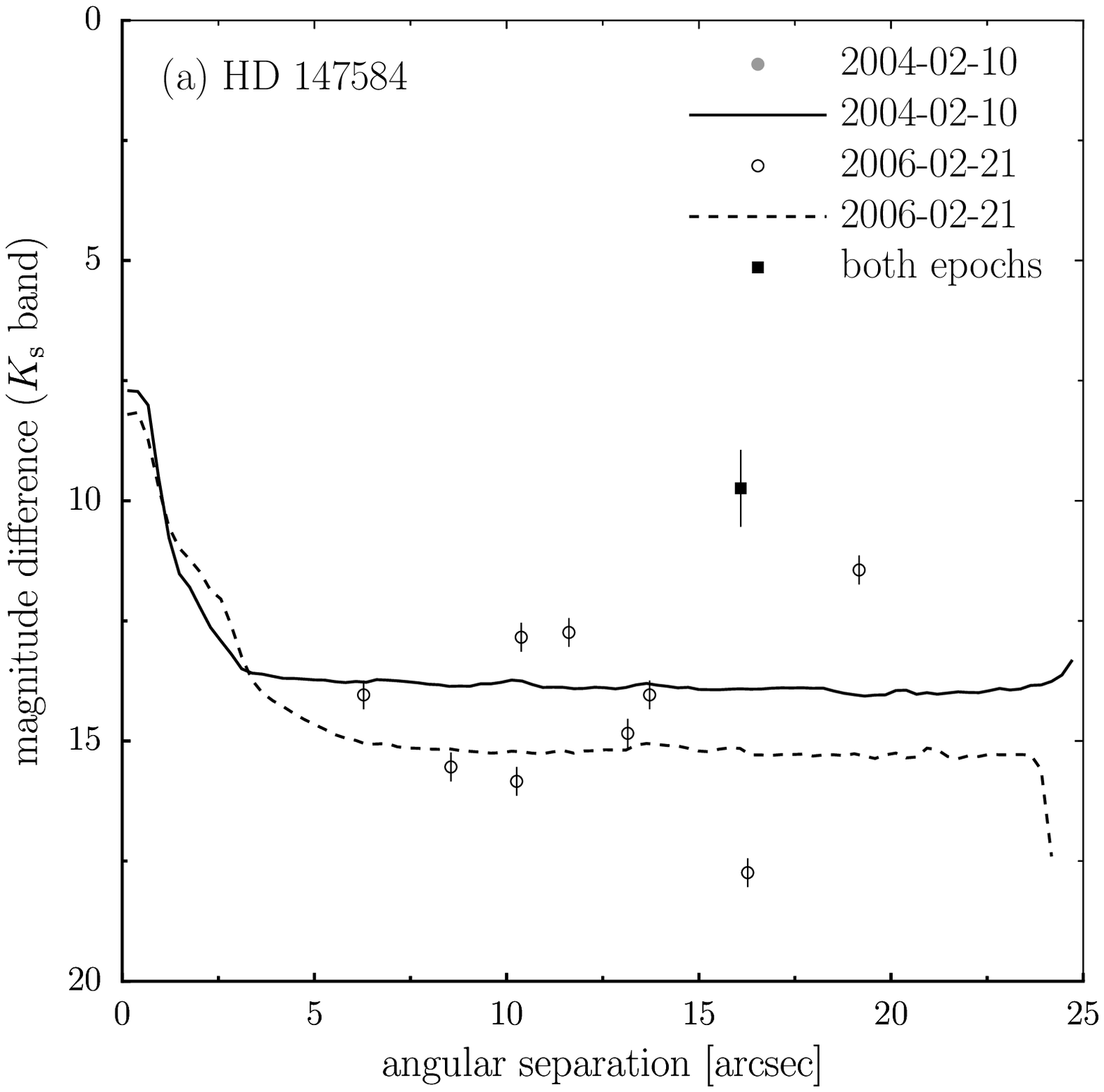}&
\includegraphics[width=5cm]{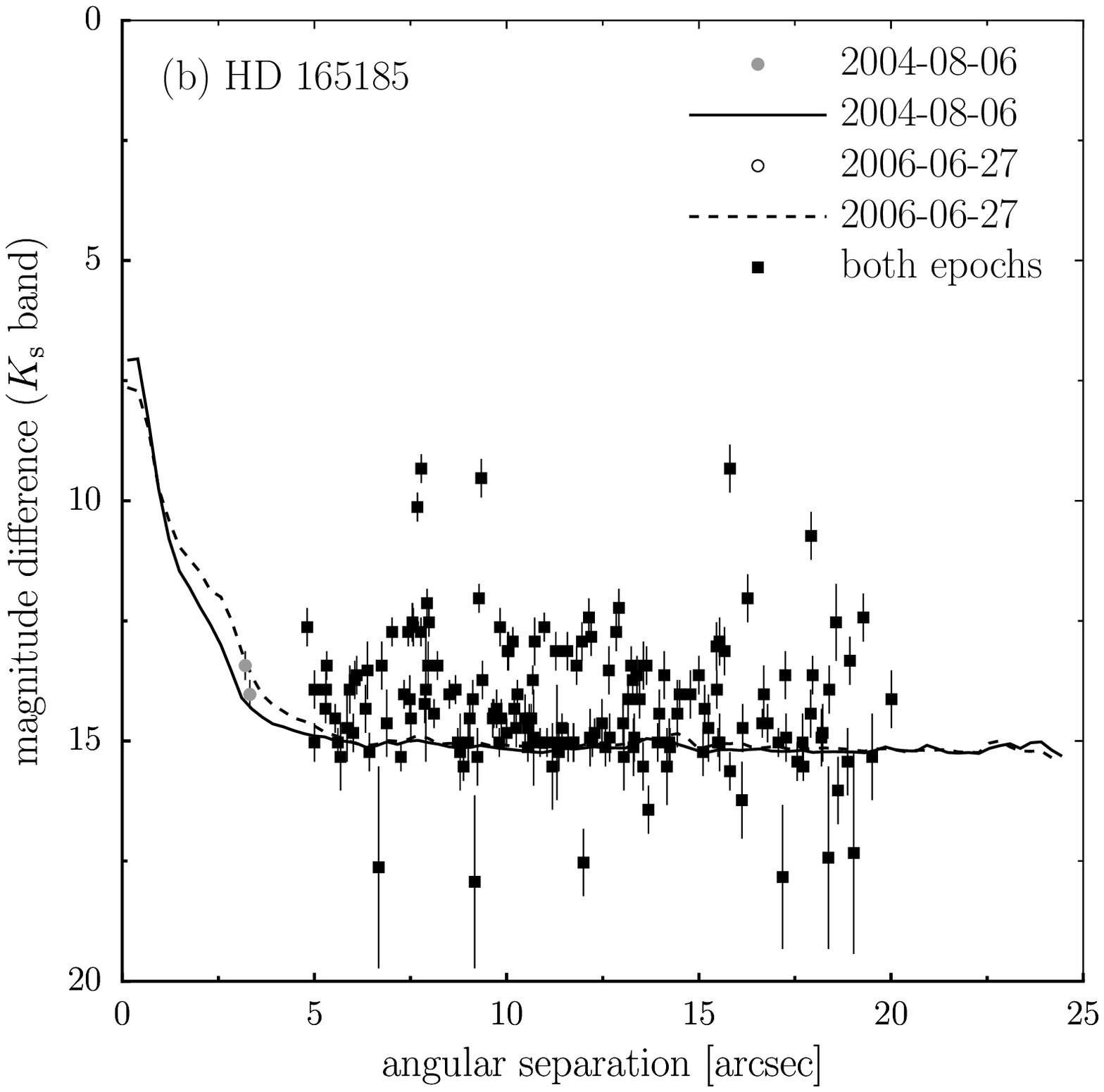}&
\includegraphics[width=5cm]{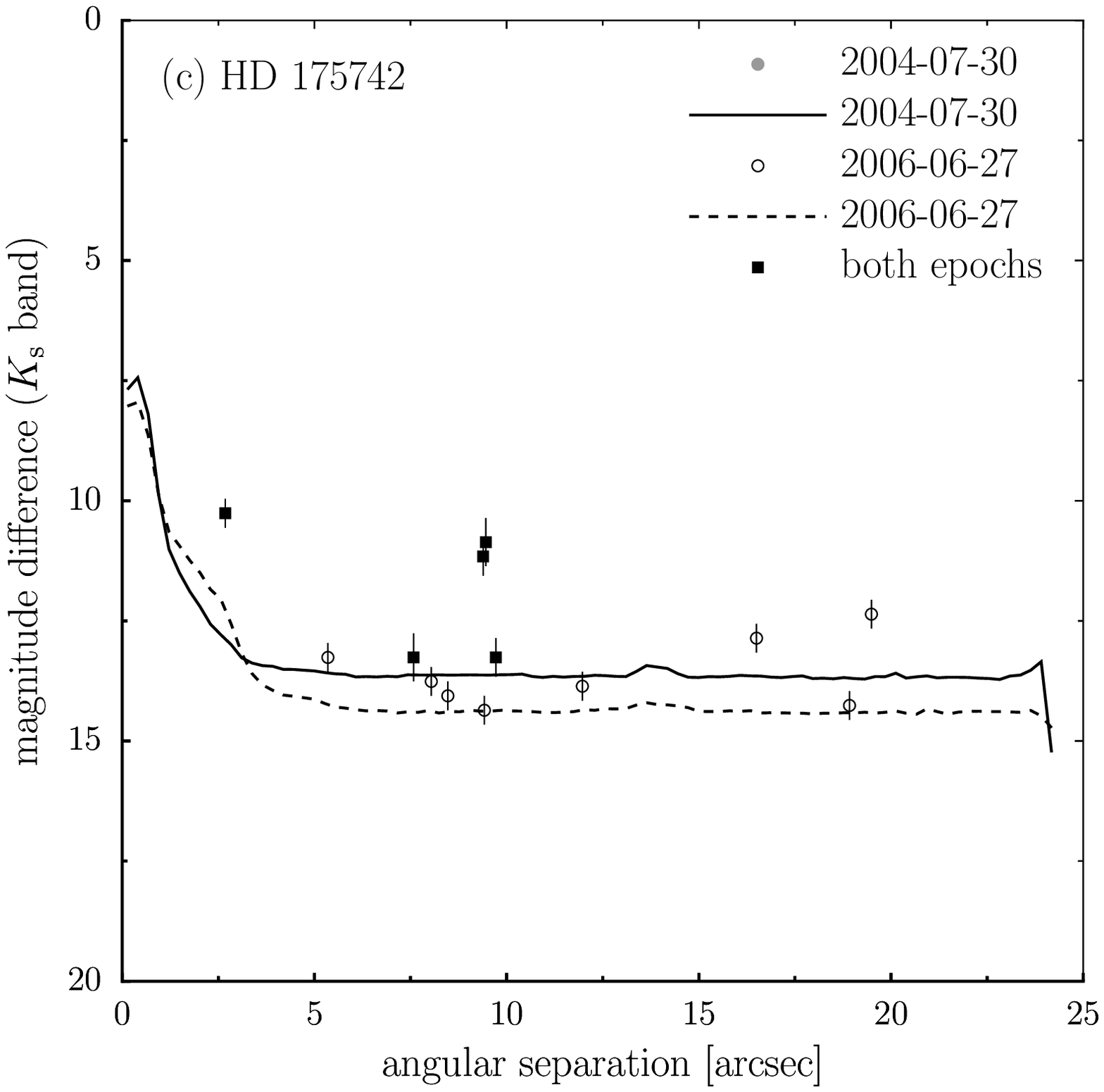}\\
\end{tabular}
\caption{\label{fig:candidates_2epochs2} Similar to Fig.~\ref{fig:candidates_2epochs1} for HD 147584, HD 165185, and HD 175742. Some data points are below the detection limits owing to unreliable magnitude measurements (discussed in the text).}
\end{figure*}

\begin{figure*}
\center
\begin{tabular}{ccc}
\includegraphics[width=5cm]{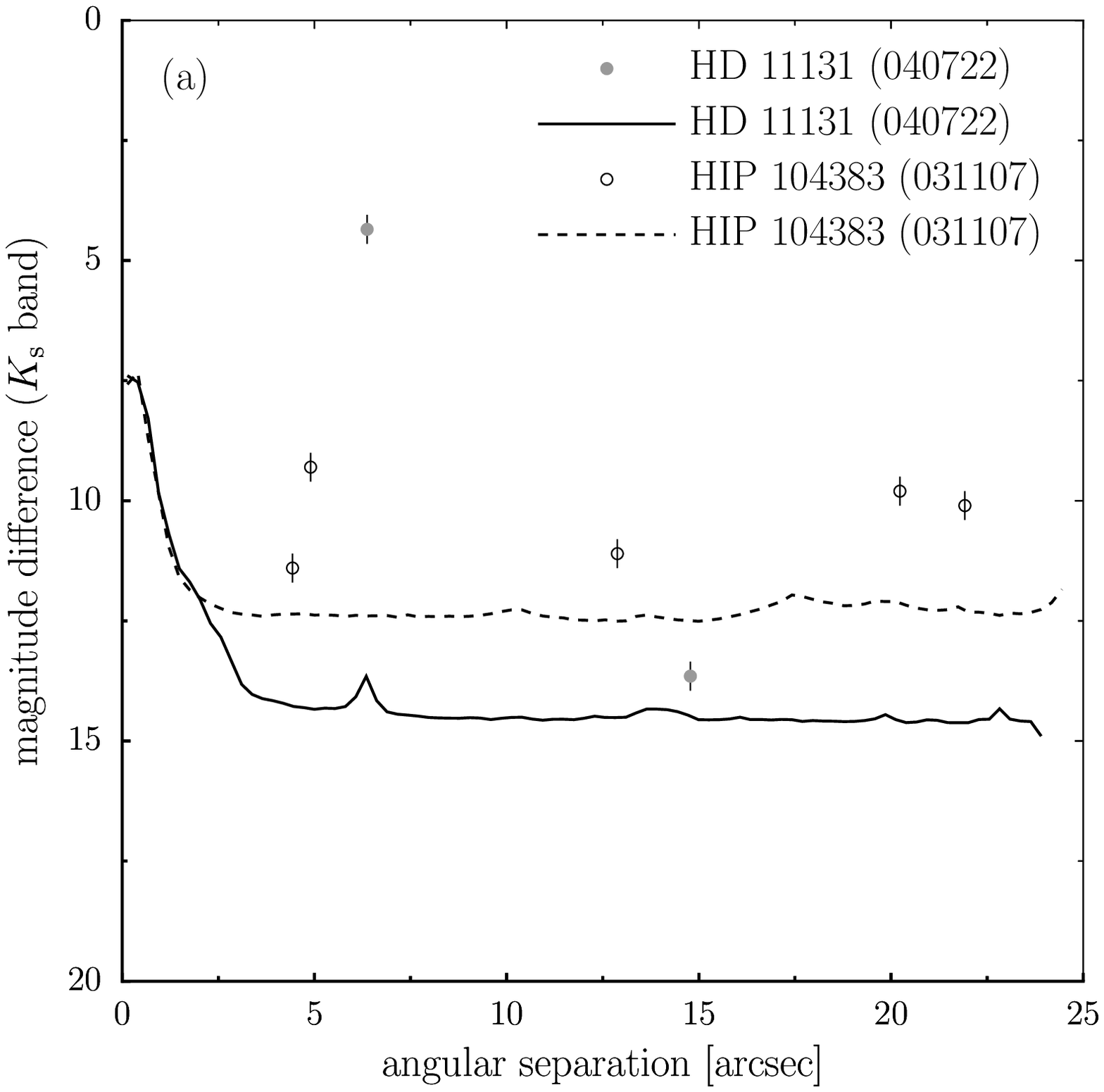}&
\includegraphics[width=5cm]{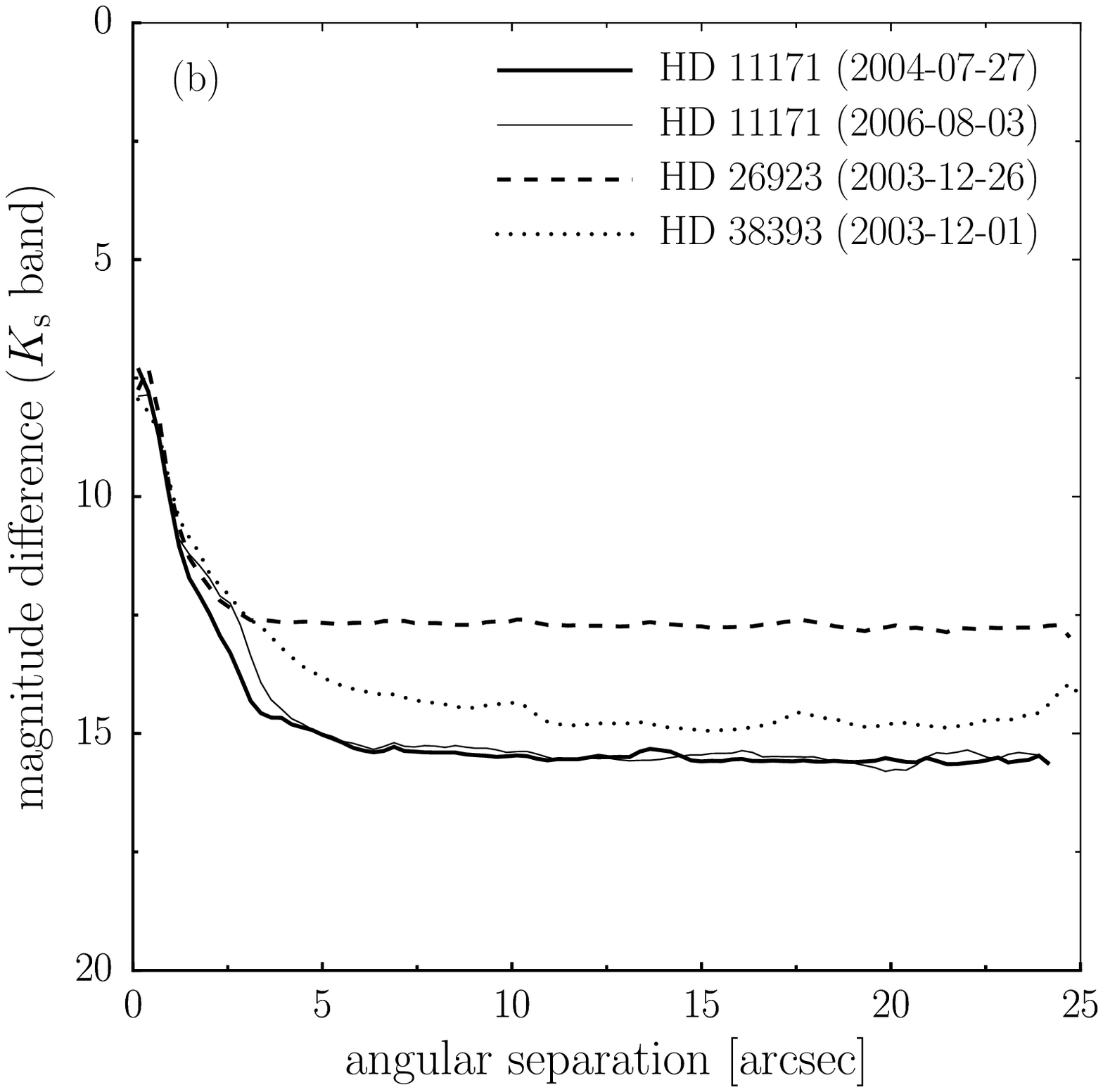}\\
\includegraphics[width=5cm]{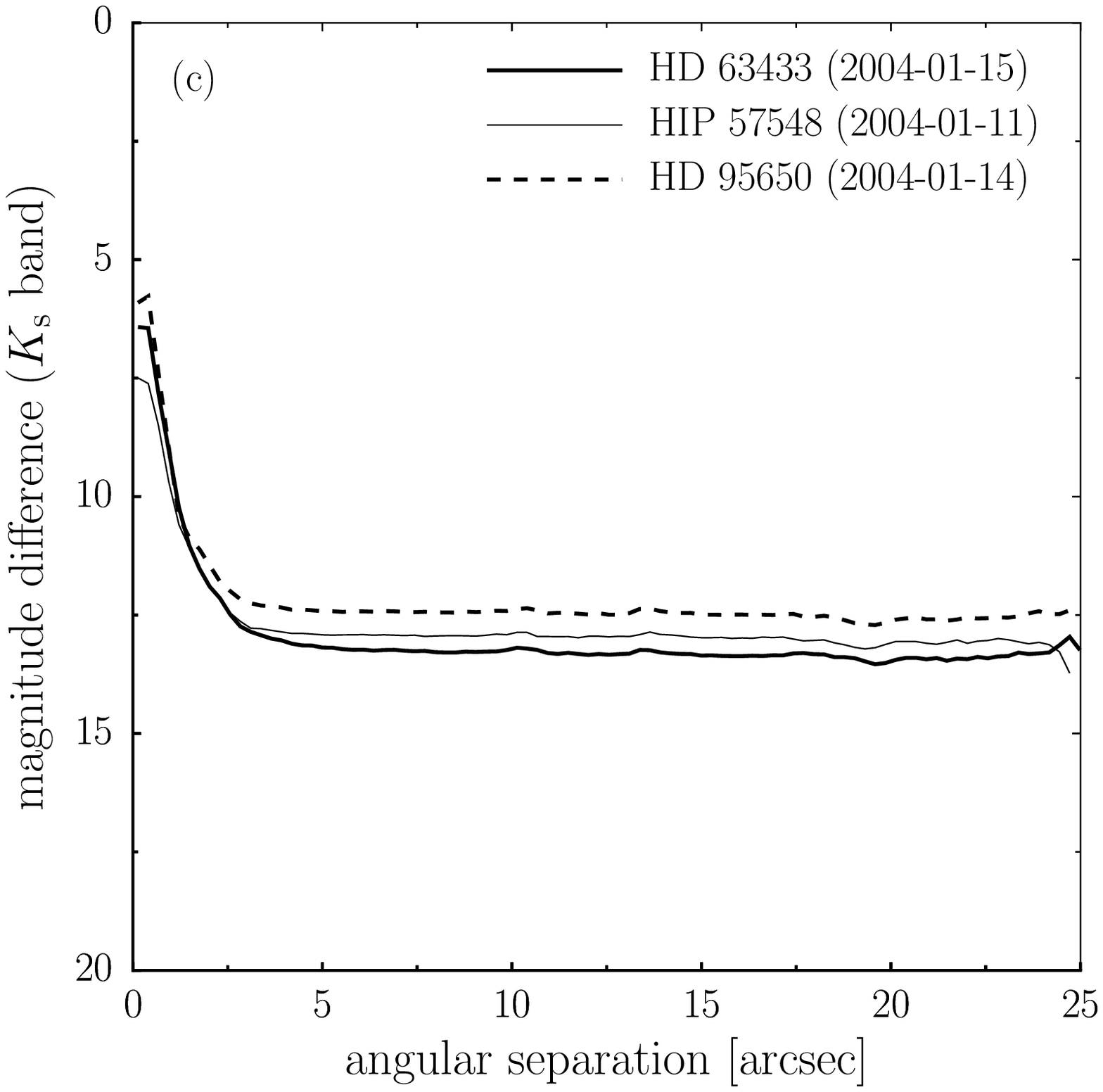}&
\includegraphics[width=5cm]{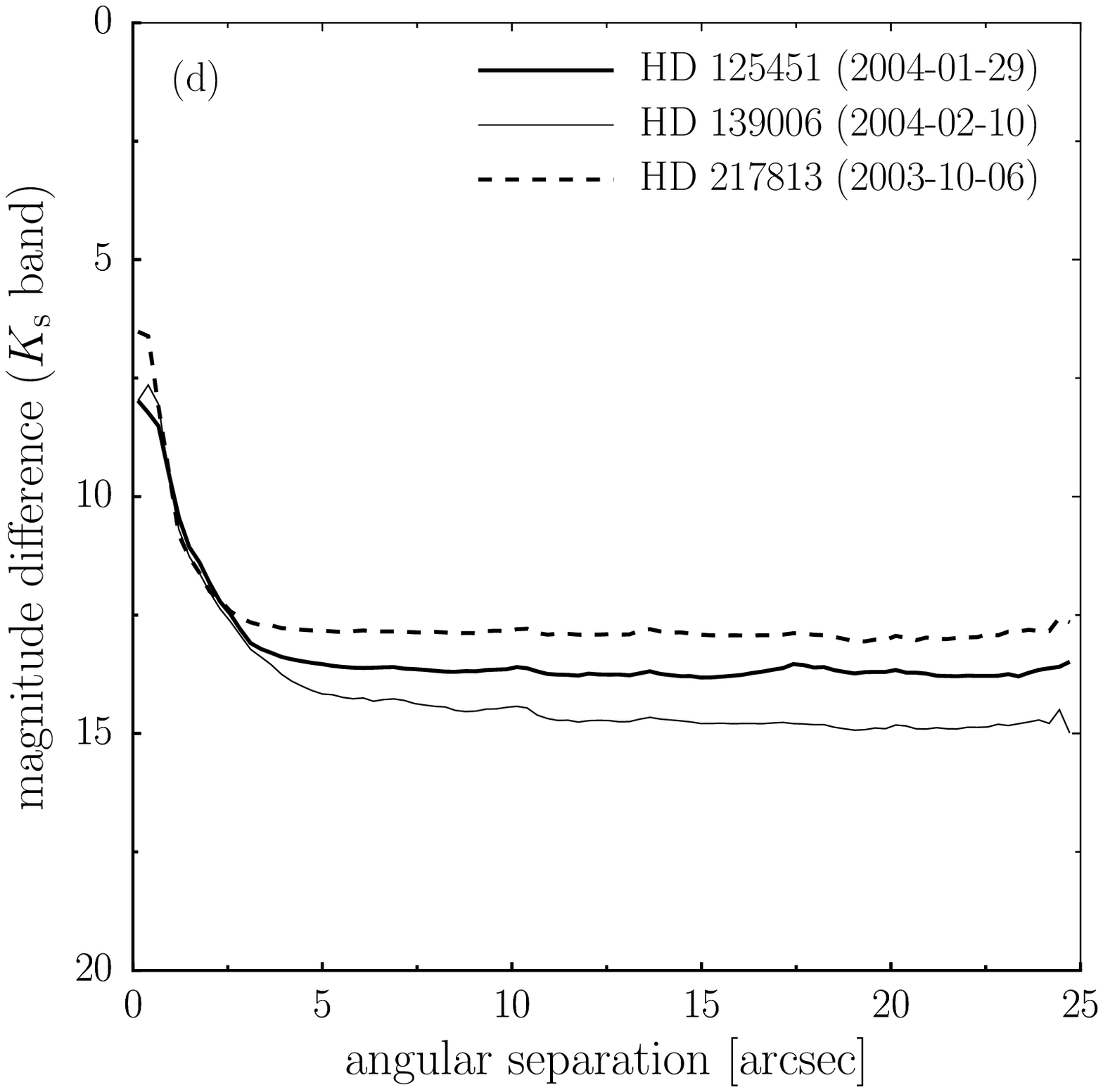}\\
\end{tabular}
\caption{\label{fig:more_detlimits}Dynamic range curves are shown for stars with candidates detected in a single epoch (HD 11131 and HIP 104383) and for all the other stars without any candidates.
}
\end{figure*}

\begin{table*}
\centering
\caption{\label{tab:det_limits2} Detection limits derived from exposures of different epochs.}
\begin{tabular}{lcrrrrrrrrrr}
\hline\hline
star&epoch&\multicolumn{1}{c}{exp. time}&\multicolumn{9}{c}{min. separation [au]}\\
&&\multicolumn{1}{c}{on target}     &\multicolumn{3}{c}{$12\,M_\mathrm{Jup}$}&\multicolumn{3}{c}{$20\,M_\mathrm{Jup}$}&\multicolumn{3}{c}{$35\,M_\mathrm{Jup}$}\\
&&\multicolumn{1}{c}{[s]}&             100\,Myr&500\,Myr&1\,Gyr&100\,Myr&500\,Myr&1\,Gyr&100\,Myr&500\,Myr&1\,Gyr\\
\hline
         HD 11131             &2004-07-22& 720&20.8   &86.7   &...    &17.6   &50.7   &96.7   &<3.2   &22.3   &34.8   \\                                                                                                              
         HD 11171             &2004-07-27& 495&27.0   &182.1  &...    &22.5   &63.9   &220.8  &14.8   &29.3   &50.3   \\                                                                                                              
         HD 11171             &2006-08-03&1680&27.7   &247.7  &...    &22.5   &75.5   &255.5  &14.5   &32.9   &66.2   \\                                                                                                              
    HD 22049\tablefootmark{a} &2004-09-03&2475&2.4    &8.7    &18.4   &1.9    &5.0    &8.8    &<0.4   &2.6    &3.8    \\                                                                                                              
         HD 22049             &2005-12-23& 720&<0.4   &10.7   &29.1   &<0.4   &7.2    &10.8   &<0.4   &0.7    &4.4    \\                                                                                                              
         HD 26913             &2003-12-25& 864&17.1   &515.8  &...    &13.8   &40.4   &516.3  &<2.8   &18.6   &27.2   \\                                                                                                              
         HD 26913             &2005-12-06& 630&16.4   &97.0   &...    &12.3   &45.5   &112.0  &<2.8   &18.0   &29.1   \\                                                                                                              
         HD 26923             &2003-12-26& 630&19.0   &...    &...    &16.0   &52.1   &...    &<2.9   &20.7   &33.2   \\                                                                                                              
    HD 38393\tablefootmark{b} &2003-12-01&1673&9.6    &96.4   &...    &7.5    &31.1   &97.8   &<1.2   &10.6   &21.0   \\                                                                                                              
    HD 41593\tablefootmark{c} &2003-12-17&1260&13.3   &60.9   &...    &10.2   &30.2   &65.5   &3.1    &14.7   &20.7   \\                                                                                                              
         HD 41593             &2005-11-08& 630&11.4   &53.8   &...    &8.3    &30.1   &56.2   &<2.1   &12.6   &20.0   \\                                                                                                              
    HD 60491\tablefootmark{d} &2003-12-26&1855&23.0   &...    &...    &19.7   &45.7   &...    &10.1   &24.5   &32.0   \\                                                                                                              
         HD 60491             &2005-11-08& 560&16.5   &...    &...    &10.5   &48.8   &...    &<3.3   &18.8   &29.7   \\                                                                                                              
         HD 61606             &2004-01-10& 560&11.0   &345.3  &...    &7.2    &25.6   &344.3  &<1.9   &12.2   &18.0   \\                                                                                                              
         HD 61606             &2005-11-08& 630&9.1    &45.6   &...    &4.1    &26.4   &46.3   &<1.9   &10.5   &16.9   \\                                                                                                              
         HD 63433             &2004-01-15& 980&21.4   &...    &...    &17.6   &46.5   &...    &11.1   &23.2   &33.2   \\                                                                                                              
        HIP 57548             &2004-01-11& 630&<0.5   &2.9    &8.2    &<0.5   &0.7    &3.0    &<0.5   &<0.5   &<0.5   \\                                                                                                              
         HD 95650             &2004-01-14& 630&7.9    &30.9   &...    &6.4    &13.7   &32.3   &<1.6   &8.6    &11.8   \\                                                                                                              
        HD 125451             &2004-01-29& 630&31.7   &...    &...    &25.7   &91.0   &...    &10.4   &35.8   &63.7   \\                                                                                                              
   HD 135599\tablefootmark{e} &2004-07-13&1080&10.8   &43.0   &...    &6.4    &22.8   &43.7   &<2.1   &11.9   &17.5   \\                                                                                                              
        HD 135599             &2006-04-06&1680&10.0   &49.8   &...    &4.6    &32.2   &50.5   &<2.1   &11.6   &19.7   \\                                                                                                              
        HD 139006             &2004-02-10& 620&52.2   &...    &...    &39.3   &...    &...    &24.7   &60.4   &140.0  \\                                                                                                              
        HD 147584             &2004-02-10& 665&11.4   &...    &...    &9.7    &25.3   &...    &<1.6   &12.4   &17.5   \\                                                                                                              
        HD 147584             &2006-02-21& 630&10.7   &47.4   &292.1  &8.1    &32.7   &49.3   &<1.6   &12.0   &23.7   \\                                                                                                              
   HD 165185\tablefootmark{f} &2004-08-06& 810&15.5   &54.6   &...    &13.0   &36.2   &56.1   &7.8    &16.8   &25.3   \\                                                                                                              
        HD 165185             &2006-06-27&1680&15.4   &65.5   &...    &12.5   &47.4   &67.2   &<2.3   &17.2   &33.6   \\                                                                                                              
        HD 175742             &2004-07-30& 540&16.4   &77.5   &...    &12.9   &33.1   &88.9   &<2.9   &17.7   &24.4   \\                                                                                                              
        HD 175742             &2006-06-27&1680&15.1   &70.0   &...    &9.0    &44.8   &71.1   &<2.9   &16.8   &26.8   \\                                                                                                              
   HD 217813\tablefootmark{g} &2003-10-06&1146&23.3   &...    &...    &19.5   &60.1   &...    &12.9   &25.0   &39.3   \\                                                                                                              
       HIP 104383             &2003-11-07& 630&16.5   &...    &...    &12.5   &35.5   &...    &<3.6   &18.5   &28.3   \\                                                                                                              

\hline
\end{tabular}\\
\tablefoot{ \textbf{(1)} Name of the target star. -- \textbf{(2)} Epoch of exposure (average if several visits have been co-added, see below). -- \textbf{(3)} On-target exposure time (total exposure time on target if several visits have been co-added). -- \textbf{(4-12)} lower limit on the separation of detectable objects for masses of 12, 20, and 35\,$M_\mathrm{Jup}$ and ages of 100\,Myr, 500\,Myr, and 1\,Gyr, respectively. The relation of masses and magnitudes  has been interpolated in the DUSTY00 and COND03 grids.
Several targets have been observed repeatedly:
\tablefoottext{a}{2, 3, and 5 Sep 2004} \tablefoottext{b}{4 and 22 Nov 2003 and 4 Jan 2004} \tablefoottext{c}{22 Nov 2003 and 11 Jan 2004} \tablefoottext{d}{22 Nov 2003, 10 and 15 Jan 2004} \tablefoottext{e}{4 and 22 Jul 2004} \tablefoottext{f}{4 Jul and 9 Sep 2004} \tablefoottext{g}{6 and 7 Oct 2003.}}
\end{table*}

\begin{table*}
\centering
\caption{\label{tab:par_motion} Apparent motion in right ascension and declination predicted for all stars with at least one candidate observed in two epochs.}
\begin{tabular}{lllrr}
\hline\hline
star&1$^\mathrm{st}$ epoch date&2$^\mathrm{nd}$ epoch date&$\Delta\alpha\cos\delta$\,[mas]&$\Delta\delta$\,[mas]\\
&yyyy mm dd&yyyy mm dd&&\\
\hline
    HD 22049                  &2004-09-03&2005-12-23&-1734.94$\,\pm\,$   0.36& -155.21$\,\pm\,$   0.28\\                                                                                                
    HD 41593                  &2003-12-17&2005-11-08& -190.06$\,\pm\,$   1.40& -193.41$\,\pm\,$   0.81\\                                                                                                
    HD 60491                  &2003-12-26&2005-11-08& -128.10$\,\pm\,$   2.43&  -72.76$\,\pm\,$   1.25\\                                                                                                
         HD 61606             &2004-01-10&2005-11-08&  185.25$\,\pm\,$   1.40& -499.95$\,\pm\,$   0.57\\                                                                                                
   HD 135599                  &2004-07-13&2006-04-06&  398.45$\,\pm\,$   1.52& -255.43$\,\pm\,$   1.09\\                                                                                                
        HD 147584             &2004-02-10&2006-02-21&  410.33$\,\pm\,$   0.51&  213.21$\,\pm\,$   0.88\\                                                                                                
   HD 165185                  &2004-08-06&2006-06-27&  234.38$\,\pm\,$   1.17&   12.16$\,\pm\,$   0.60\\                                                                                                
        HD 175742             &2004-07-30&2006-06-27&  276.77$\,\pm\,$   1.05& -537.13$\,\pm\,$   1.20\\                                                                                                

\hline
\end{tabular}
\tablefoot{Based on proper motion and stellar parallactic motion measured by {\it Hipparcos} (Tables~\ref{tab:sample_mem} and \ref{tab:sample_all}). These numbers are used to identify non-moving background stars next to the targets.}
\end{table*}

\begin{figure*}
\center
\begin{tabular}{ccc}
\includegraphics[width=5cm]{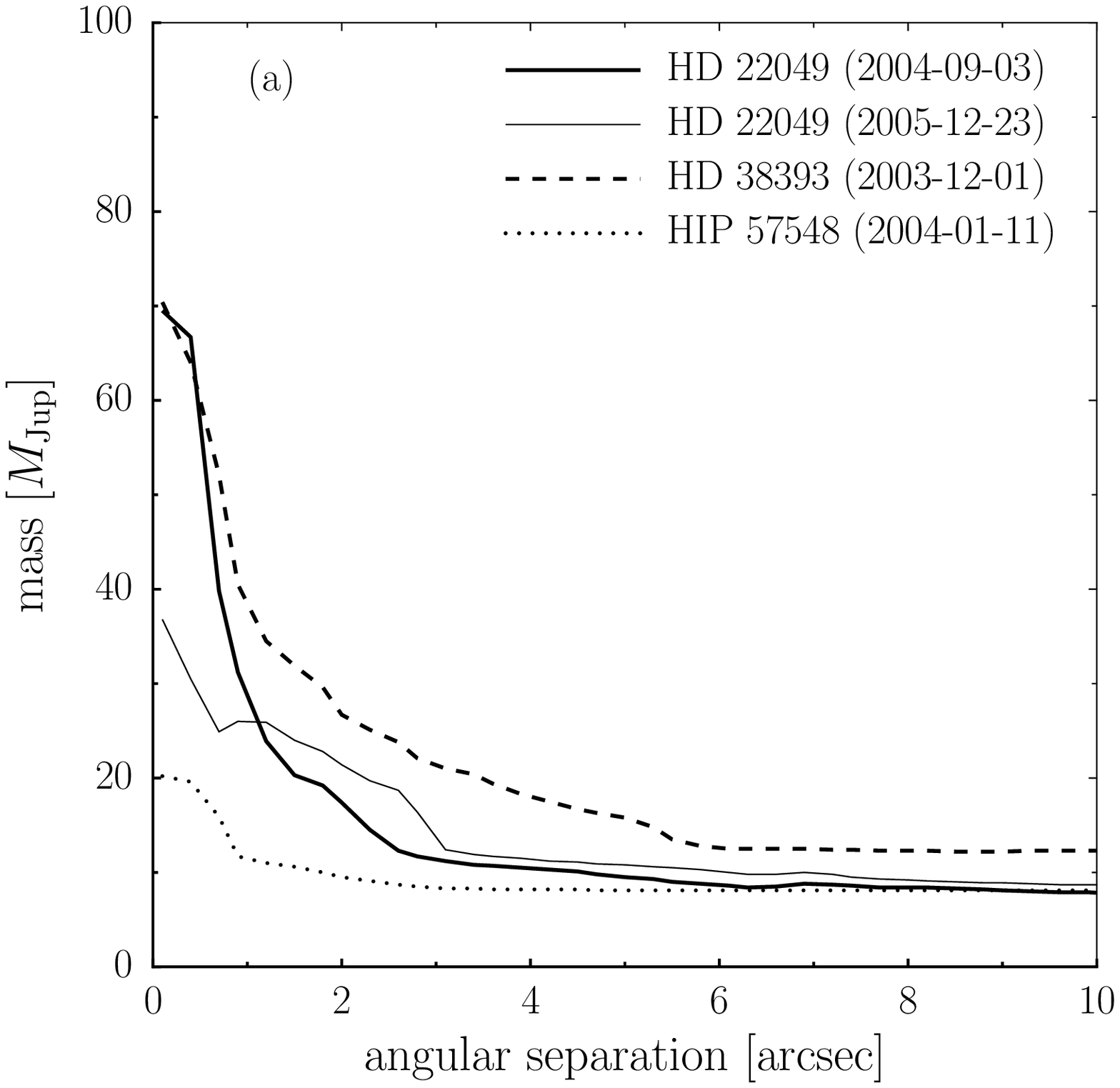}&
\includegraphics[width=5cm]{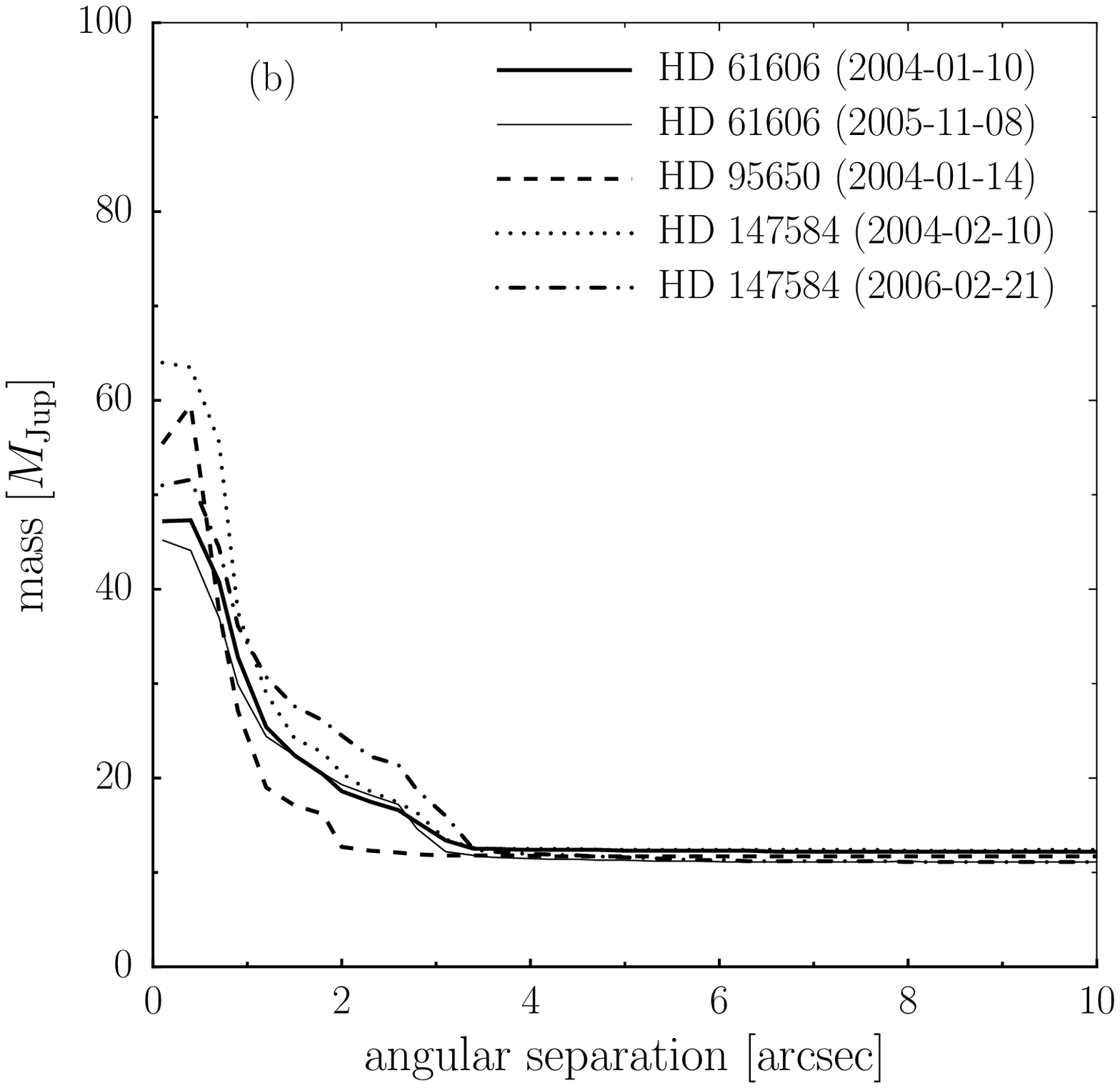}&
\includegraphics[width=5cm]{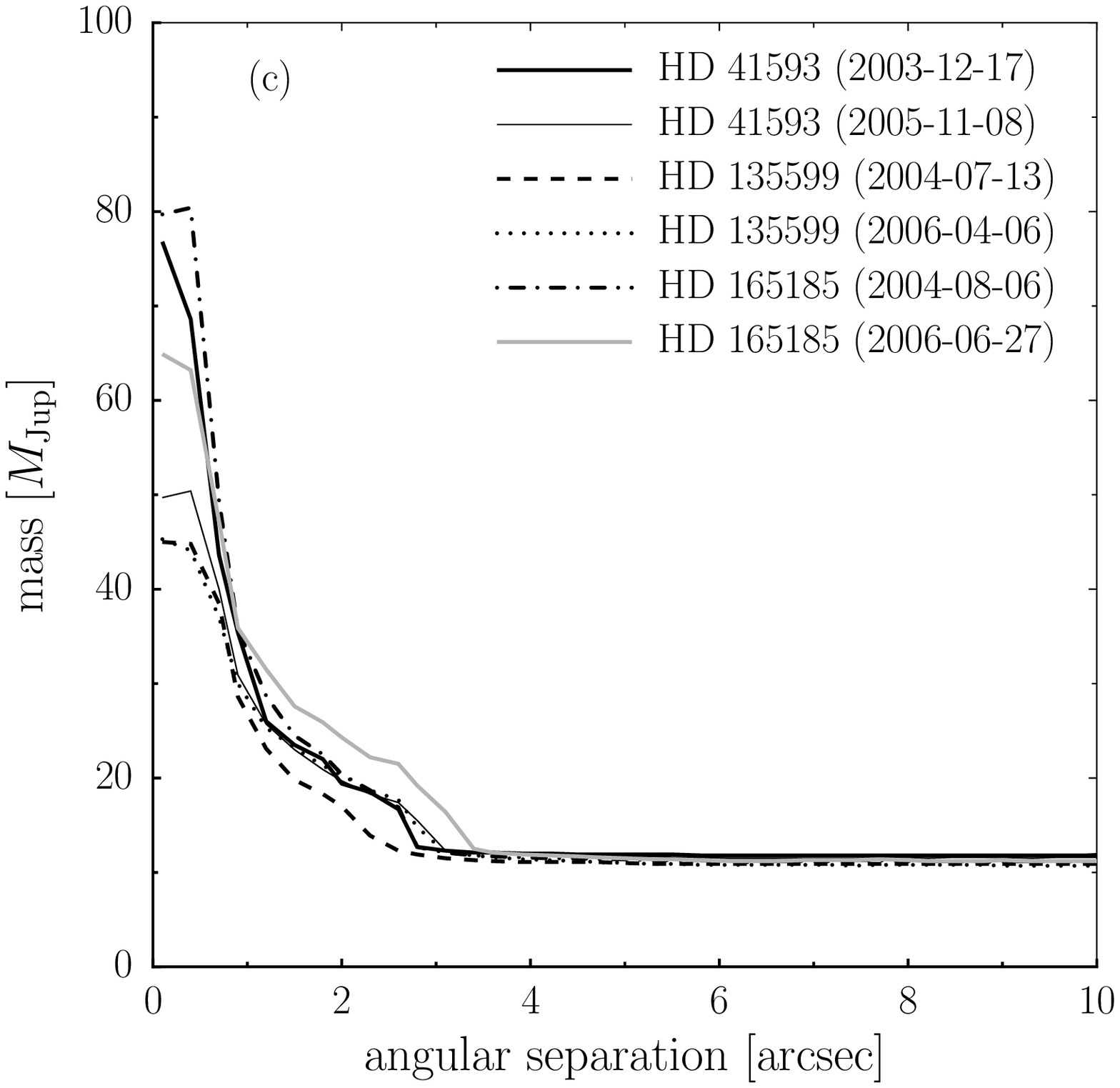}\\
\includegraphics[width=5cm]{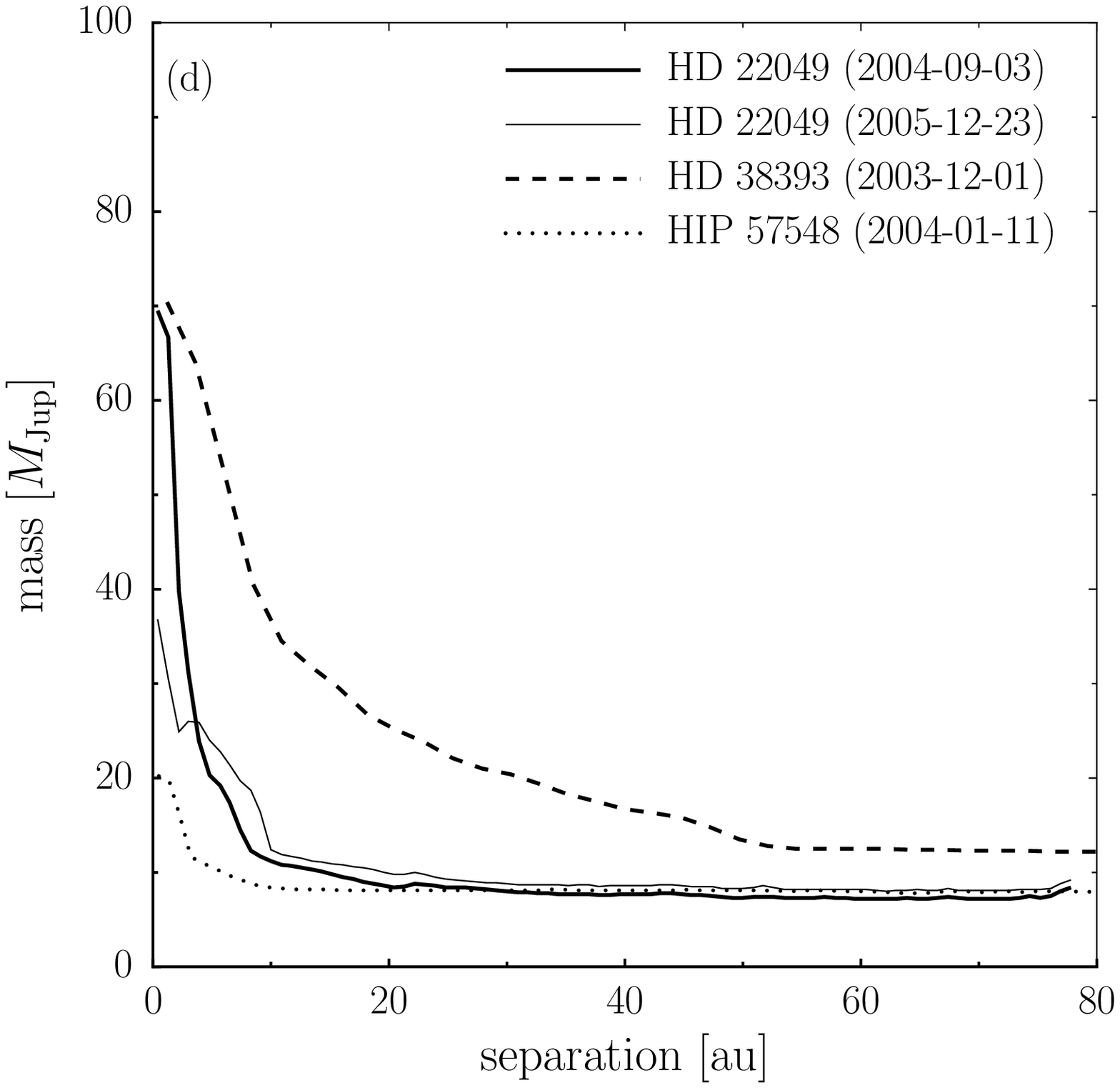}&
\includegraphics[width=5cm]{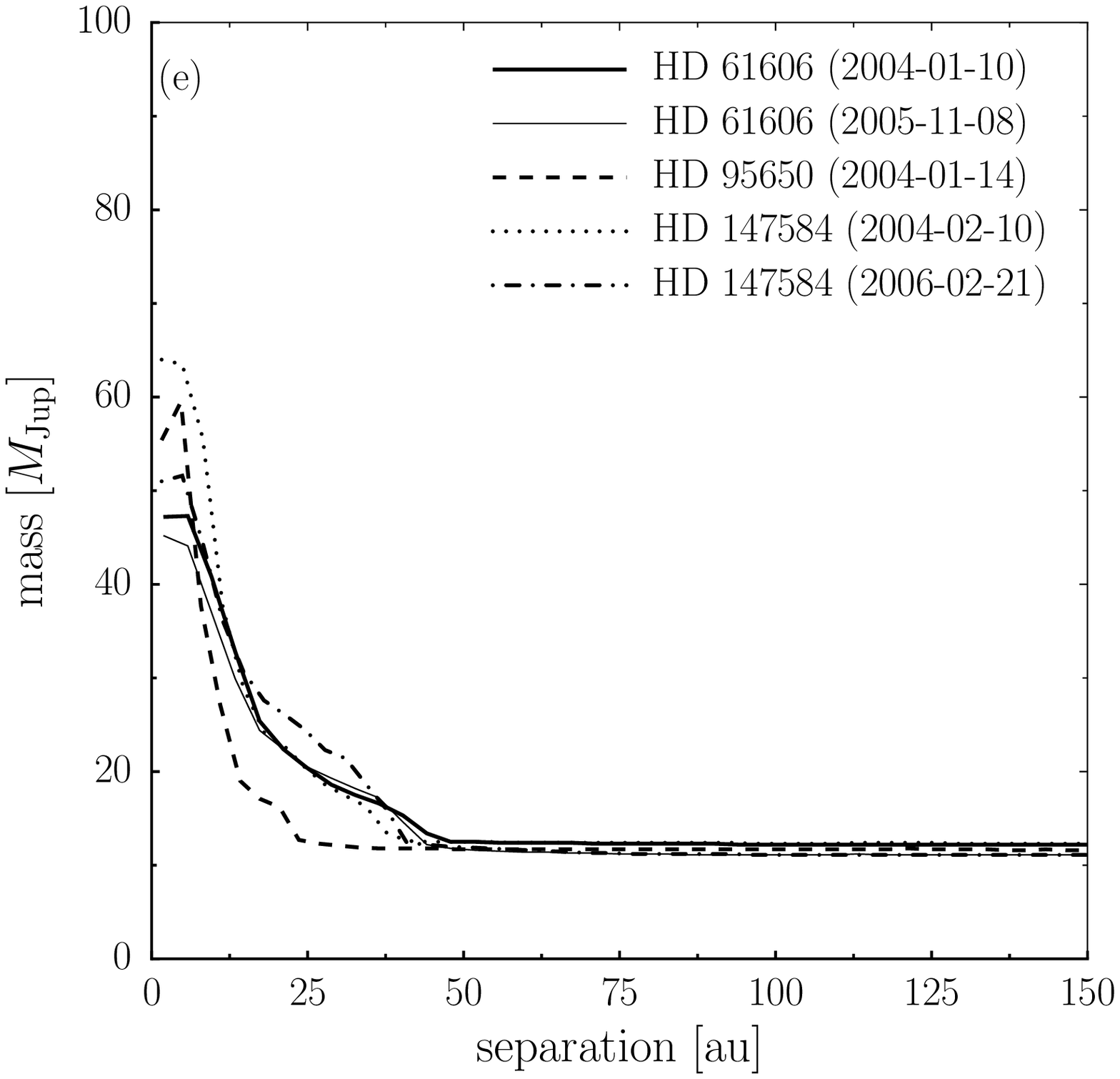}&
\includegraphics[width=5cm]{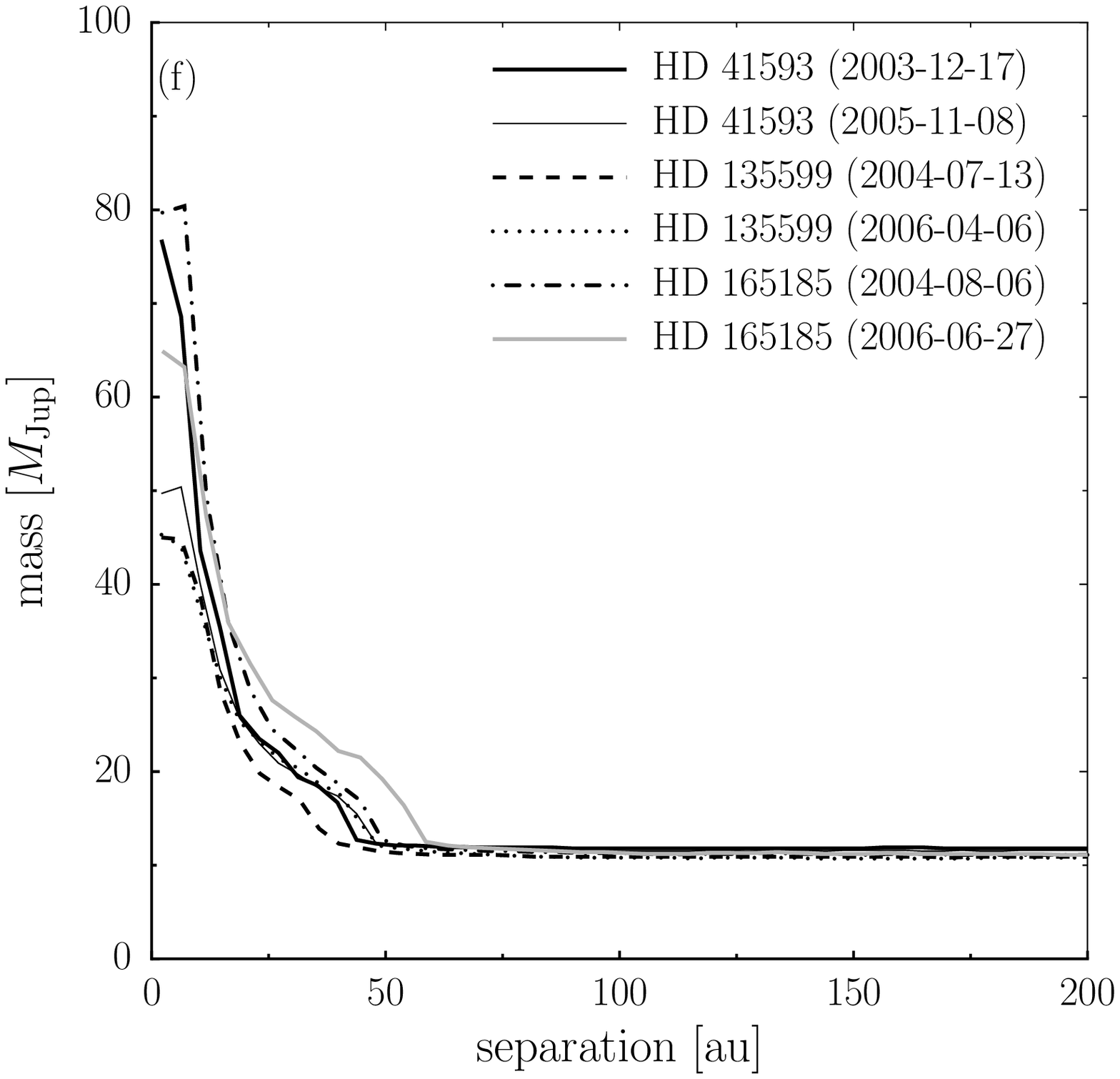}
\end{tabular}
\caption{\label{fig:mass_of_sep1}Top row: detection limits on mass as a function of angular separation. Bottom row: the same as a function of separation in linear scale (bottom row). Based on the dynamic range curves presented in Figs.~\ref{fig:curves_radec} and \ref{fig:candidates_2epochs1}-\ref{fig:more_detlimits} and on evolutionary models (DUSTY00 and COND03) assuming an age of 500\,Myr. The layout follows Figs.~\ref{fig:more_detlimits} b, c, and d.}
\end{figure*}

\begin{figure*}	
\center
\begin{tabular}{ccc}
\includegraphics[width=5cm]{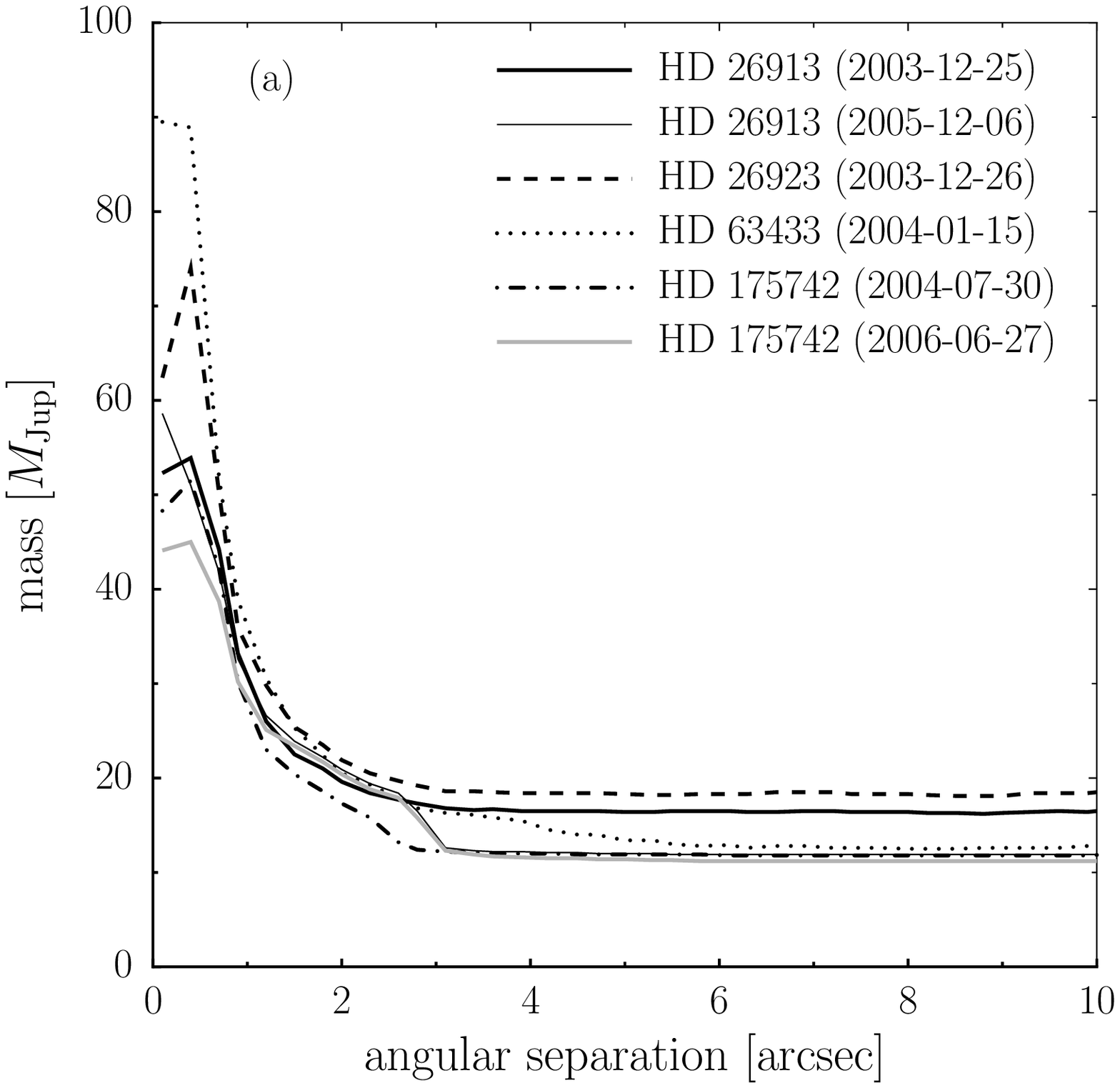}&
\includegraphics[width=5cm]{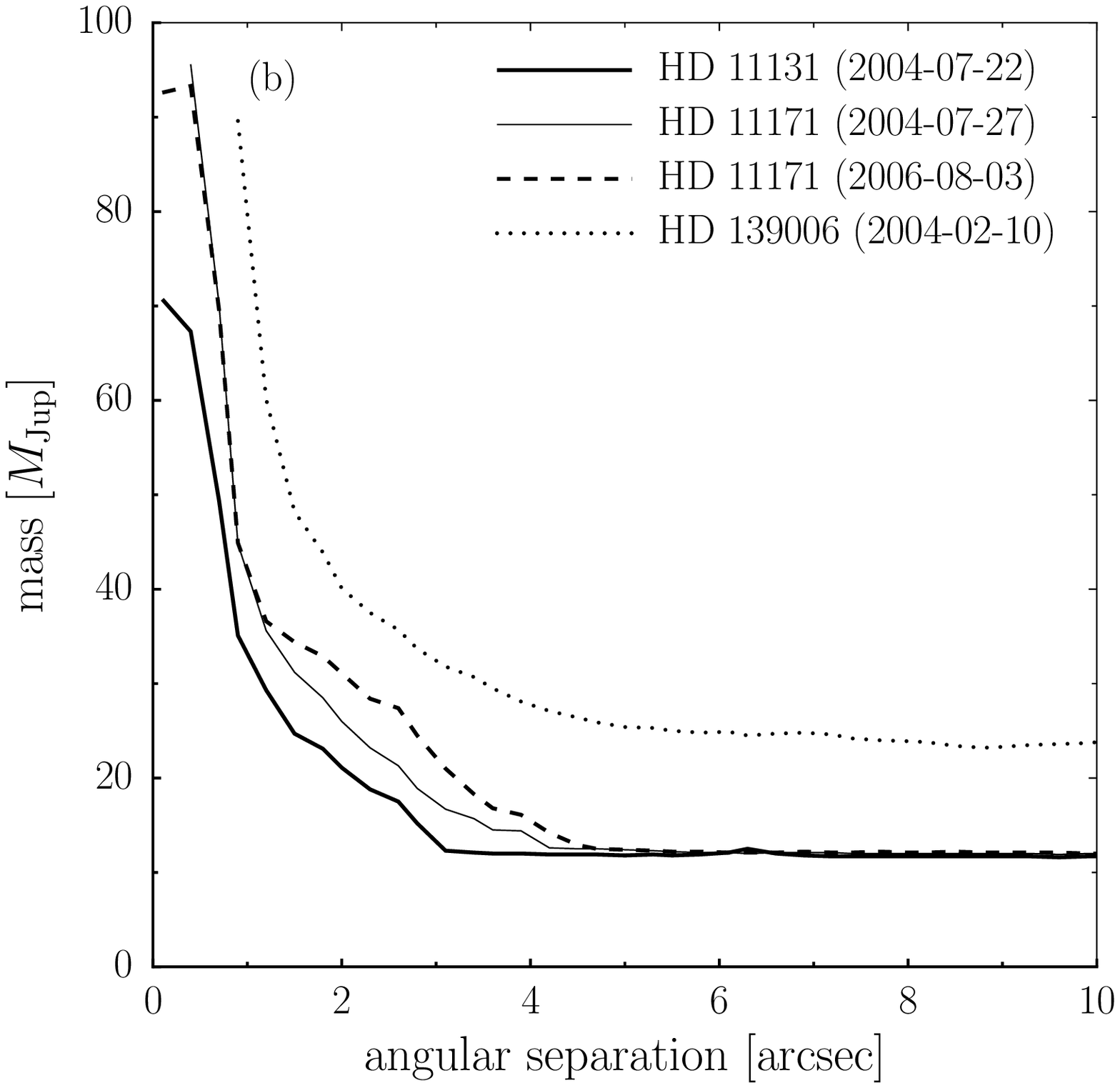}&
\includegraphics[width=5cm]{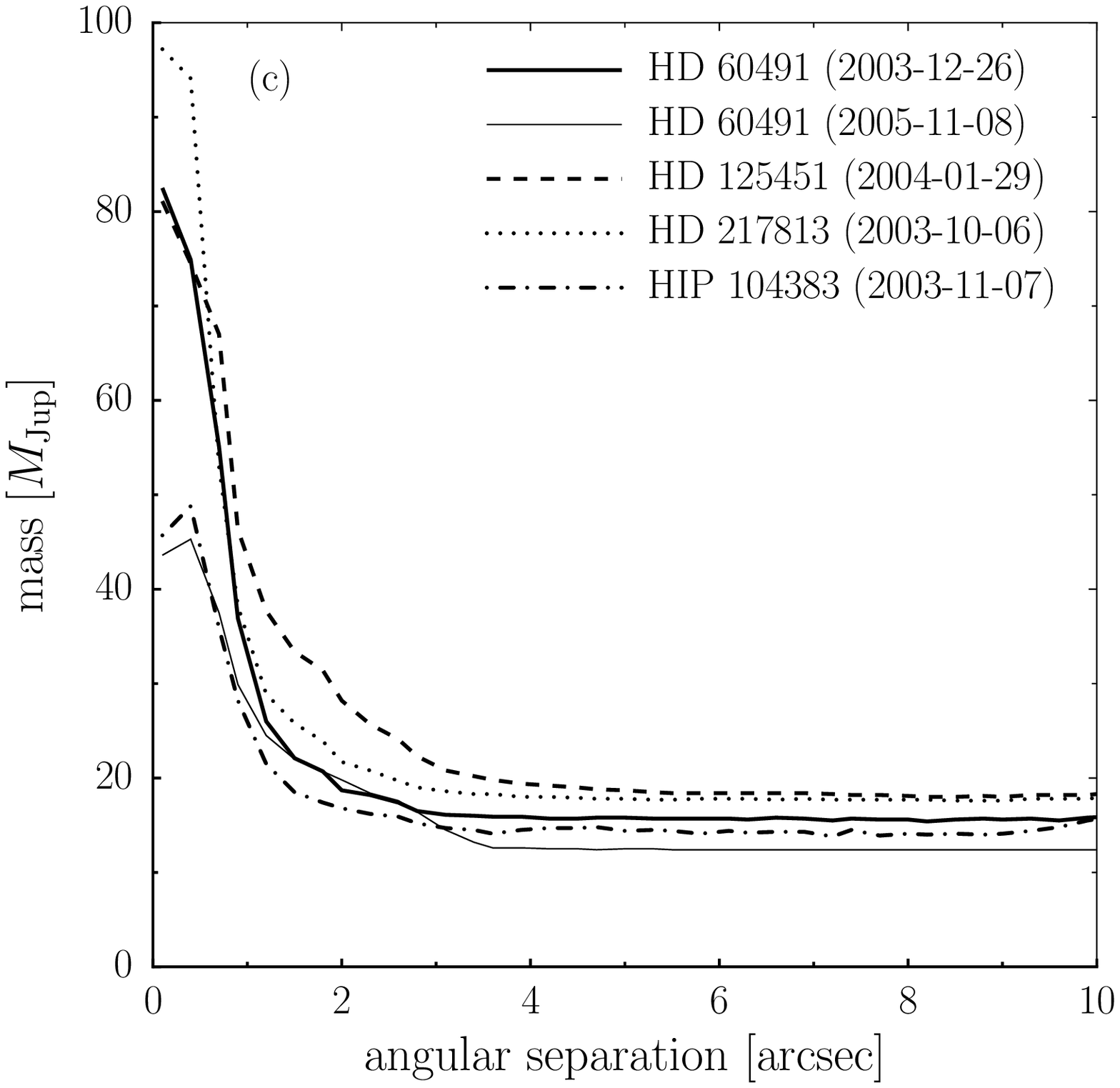}\\
\includegraphics[width=5cm]{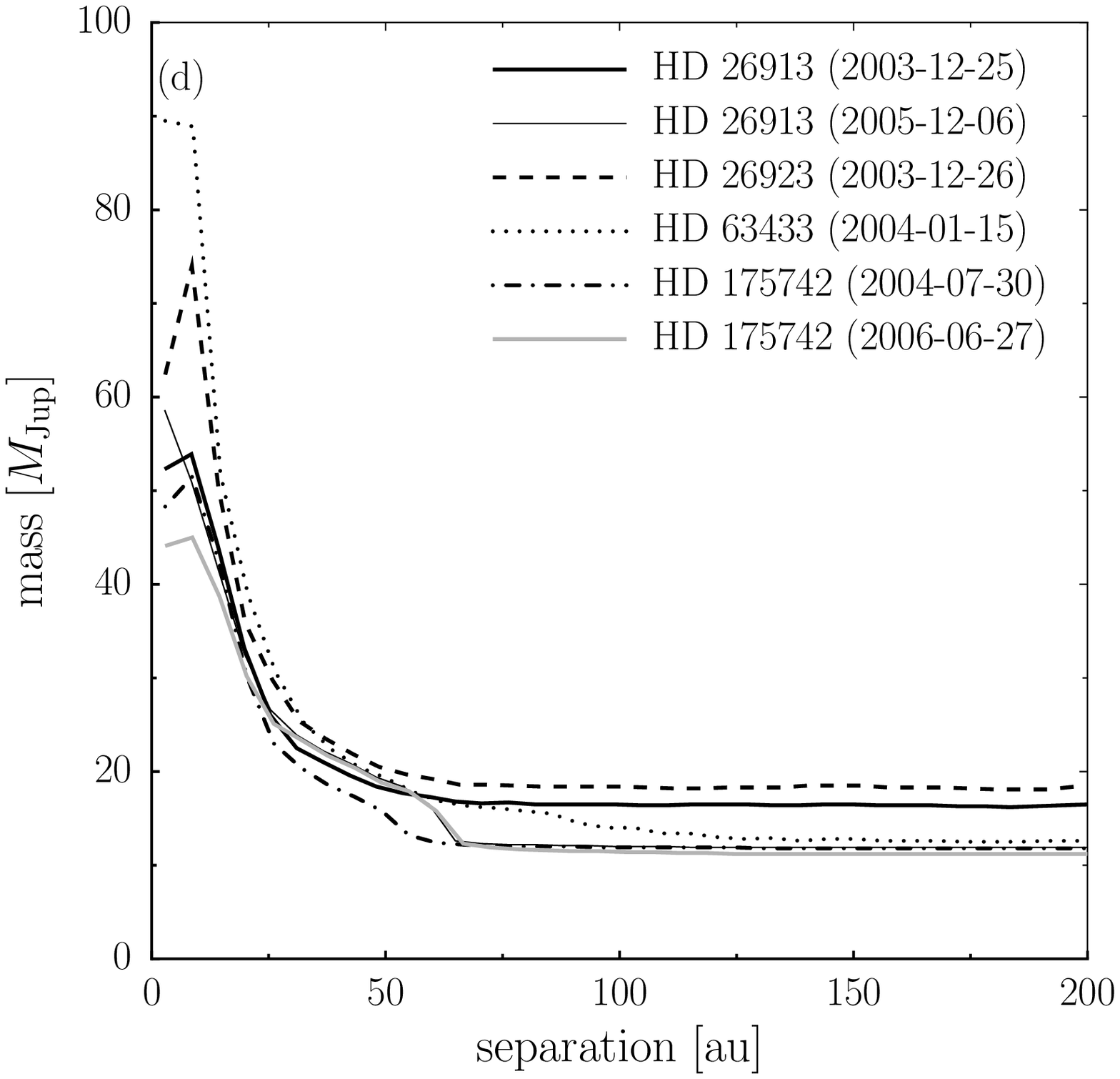}&
\includegraphics[width=5cm]{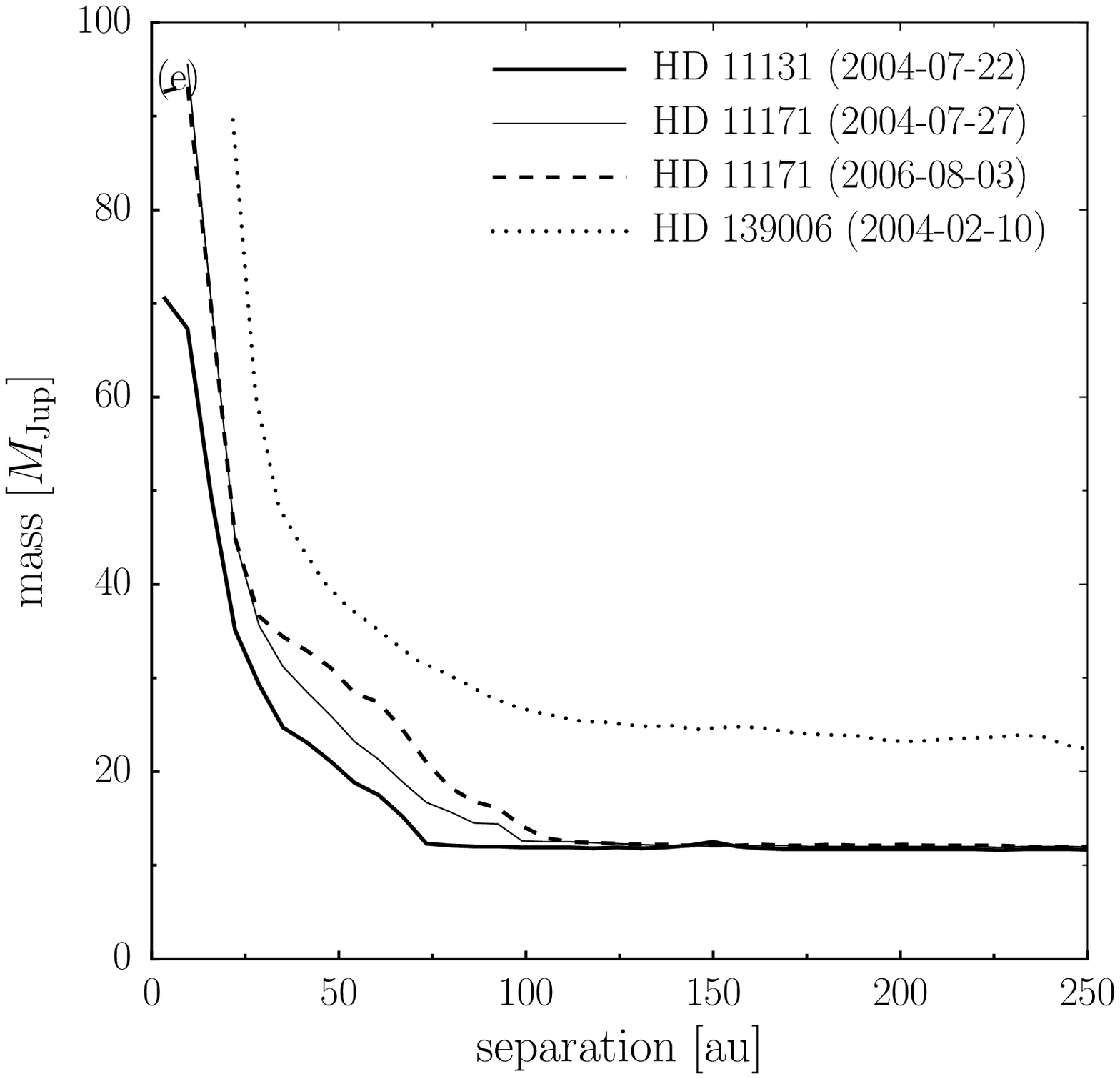}&
\includegraphics[width=5cm]{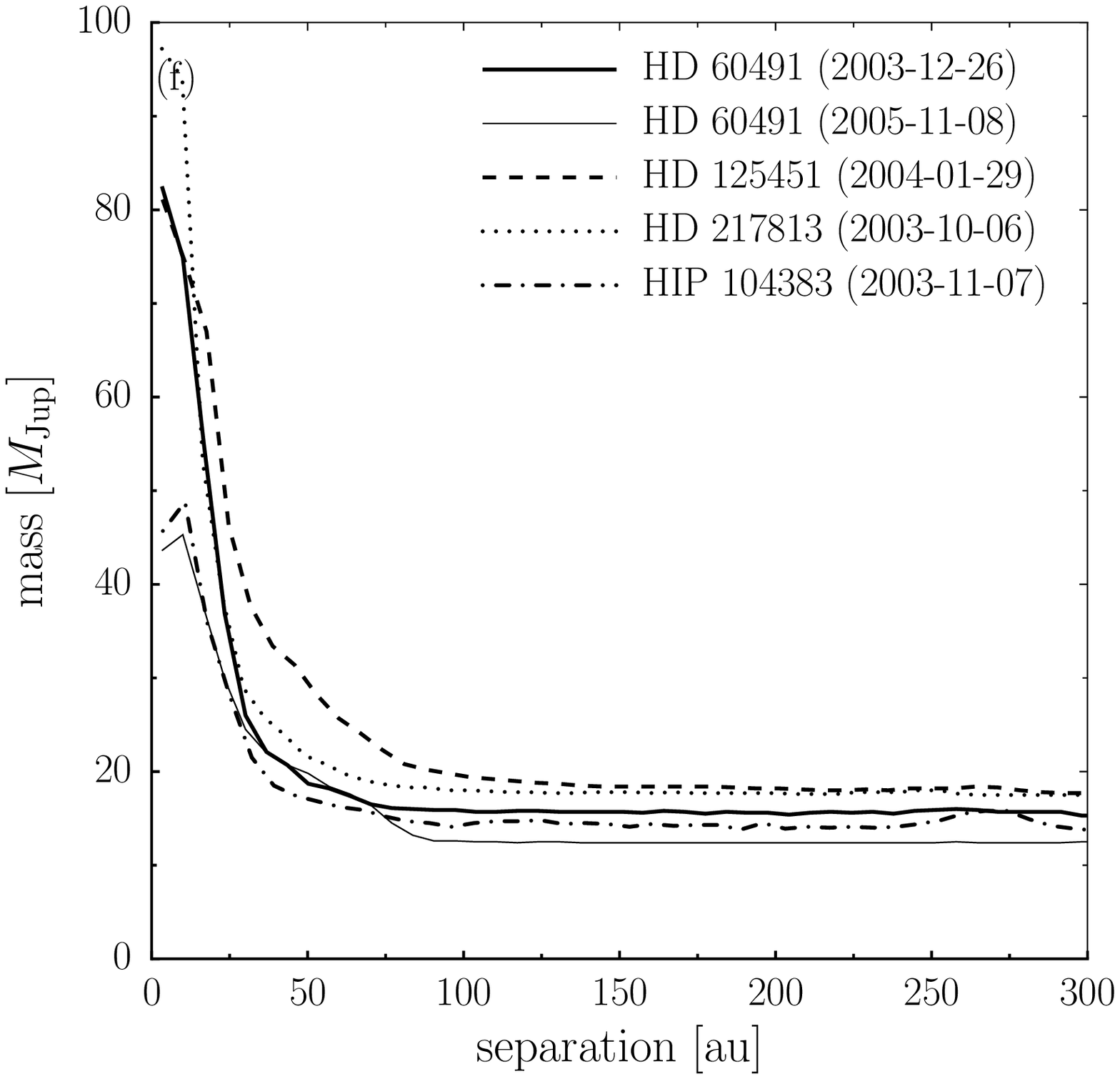}
\end{tabular}
\caption{\label{fig:mass_of_sep2} Same as in Fig.~\ref{fig:mass_of_sep1} for the remaining objects.}
\end{figure*}




\end{appendix}





\end{document}